%% file: main.tex
\documentclass[DIV=15, bibliography=totoc]{scrartcl}

\usepackage[a4paper,margin=2.25cm,bottom=2.5cm]{geometry}
\usepackage{float}
\usepackage[export]{adjustbox}
\usepackage[normalem]{ulem}
\usepackage{diagbox}
\usepackage{stmaryrd}
\usepackage{bm}
\usepackage{xstring}
\usepackage{flafter}
\usepackage{enumitem}

\input{configuration}
\input{00_frontpage}

\newcommand{\tabref}[1]{Table~\ref{#1}}
\newcommand{\figref}[1]{Figure~\ref{#1}}

\newcommand{\secref}[1]{Section~\ref{#1}}

\usepackage{xcolor}
\usepackage{comment} 

\DeclareRobustCommand{\hg}[1]{{\sethlcolor{green}\hl{#1}}}

\begin{document}

\normalem \maketitle
\normalfont\fontsize{11}{13}\selectfont

\definecolor{myblue}{rgb}{0.0, 0.0, 1.0}
\DeclareRobustCommand{\hb}[1]{{\color{myblue}#1}} 
\definecolor{mygreen}{rgb}{0.0, 0.5, 0.0}
\DeclareRobustCommand{\hg}[1]{{\color{mygreen}#1}} 

\usetikzlibrary{shapes.misc}
\tikzset{cross/.style={cross out, draw=black, minimum size=2*(#1-\pgflinewidth), inner sep=0pt, outer sep=0pt},cross/.default={1pt}}


\vspace{-1.5cm} \hrule 

\section*{Abstract}

\input{content/abstract}

\vspace{0.25cm}
\noindent \textit{Keywords:} 
implicit-explicit time integration, spectral elements, finite/spectral cell method, wave equation, immersed boundary methods
\vspace{0.25cm}


\section{Introduction}{
\label{sec:intro}
\input{content/intro}

}

\section{IMEX for Immersed Boundary Methods}
{
\label{sec:theory}

\subsection*{Critical time step analysis of cut cells}
\label{sec:motivation}
\input{content/motivation}

\subsection*{Spatial Discretization}
\label{sec:spatialDiscr}
\input{content/spatialDiscretization}

\subsection*{Partitioning of the Degrees of Freedom}
\label{sec:partitioning}
\input{content/partitioning}

\subsection*{Time Integration}
\label{sec:timeInt}
\input{content/timeIntegration}

\subsection*{Lumping of Cut Cells}
\label{sec:lumping}
\input{content/lumping}
}

\section{Numerical examples}
\label{sec:NumericExamples}
{ 
We now demonstrate fundamental principles of implicit-explicit time integration using a basic system consisting of ten spring-coupled masses before applying the immersed Newmark IMEX method to more complex examples in two and three dimensions.

\subsection*{Spring-coupled masses}{
\label{sec:SpringMasses}
\input{content/SpringMasses}
}

\subsection*{Perforated plate}{
\label{sec:bench2D}
\input{content/benchmark2D}
}

\subsection*{Three-dimensional example}{
\label{sec:bench3D}
\input{content/example3D}
}
}

\section{Conclusion}{ 
\label{sec:conclusion}
\input{content/conclusion}
}

\appendix


\section*{Acknowledgements} \label{sec:Acknowledgement}
{
We gratefully thank the Deutsche Forschungsgemeinschaft (DFG, German Research Foundation) for their support through the grants KO 4570/1-1 and RA 624/29-1. We also thank Harald Kloft and Robin D\"{o}rrie from the Institut f\"{u}r Tragwerksentwurf of the Technische Universit\"{a}t Braunschweig for providing the geometric model investigated in~\secref{sec:bench3D} which serves as a benchmark in the DFG project 414265976 TRR 277.
}


\bibliographystyle{ieeetr}


\input{main.bbl}
\end{document}

%% file: configuration.tex
\usepackage{xcolor} 
\definecolor{lightblue}{rgb}{0,0.5,1.0}
\definecolor{linkblue}{rgb}{0,0.1,0.6}
\definecolor{citegreen}{rgb}{0,0.4,0.0}
\definecolor{linkred}{rgb}{0.8,0,0.005}
\definecolor{mailviolet}{rgb}{0.3,0,0.35}
\definecolor{tumblue}{rgb}{0,0.396,0.741}
\definecolor{darkgreen}{rgb}{0,0.4,0} 
\definecolor{darkbrown}{rgb}{0.5, 0.396, 0.09}

\usepackage[hyperindex,colorlinks=true,linkcolor=linkblue,citecolor=citegreen,urlcolor=mailviolet,filecolor=linkred]{hyperref}
\usepackage[hyphenbreaks]{breakurl}
\usepackage{caption, subcaption}
\usepackage{booktabs}   
\usepackage{amsmath}
\usepackage{amssymb}
\usepackage{amsfonts}
\usepackage{lmodern}
\usepackage{graphicx}
\usepackage{listings}
\usepackage{todonotes}
\usepackage{placeins}
\usepackage{fancyhdr}
\usepackage{wrapfig}
\usepackage{mathtools}
\usepackage{url}
\usepackage{siunitx}
\usepackage[square,comma,nonamebreak,compress,numbers]{natbib}

\usepackage{soulpos}
\usepackage{ulem}
\usepackage{authblk} 

\usepackage{pgfplots}
\pgfdeclarelayer{background}
\pgfdeclarelayer{foreground}
\pgfsetlayers{background,main,foreground}

\usepackage{pgfplotstable}
\pgfplotsset{compat = newest}
\usepackage{filecontents}
\usepackage{tikz}
\usepackage{tikzscale}
\usepackage{tikz-3dplot}

\usetikzlibrary{spy}
\usetikzlibrary{intersections}
\usetikzlibrary{arrows,shapes}
\usetikzlibrary{spy}
\usetikzlibrary{backgrounds}  
\usetikzlibrary{decorations}
\usetikzlibrary{decorations.markings}
\usetikzlibrary{positioning}
\usetikzlibrary{patterns}
\usetikzlibrary{arrows}
\usetikzlibrary{arrows.meta}
\usetikzlibrary[plotmarks] 
\usetikzlibrary{calc} 
\usetikzlibrary{quotes} 
\usetikzlibrary{external} 
\usetikzlibrary{matrix}

\pgfplotscreateplotcyclelist{customCycleList}
{%
  {blue, mark=o},
  {red, mark=square},
  {darkgreen, mark=diamond},
  {cyan, mark=diamond},
  {darkbrown, mark=triangle*},
  {black, mark=pentagon},
}

\pgfplotsset{every axis/.append style= {
    cycle list name=customCycleList,
}}

\newcommand{\tensor}[1]{\IfSubStr{ABCDEFGHIJKLMNOPQRSTUVWXYZabcdefghijklmnopqrstuvwxyz}{#1}
        {\mathbf{#1}}
        {\bm{#1}}}
\renewcommand{\v}[1]{\tensor{#1}} 
\newcommand{\dv}[1]{\hat{\tensor{#1}}} 

%% file: 00_frontpage.tex
\title{Implicit-Explicit Time Integration for the Immersed Wave Equation}

\newcommand*\samethanks[1][\value{footnote}]{\footnotemark[#1]}

\author[1]{Christian Fa{\ss}bender\thanks{These two authors contributed equally to this work and share the first authorship.}}
\author[1]{Tim B\"urchner\samethanks[1]\thanks{\href{mailto:tim.buerchner@tum.de}{\texttt{tim.buerchner@tum.de}}, Corresponding author}}
\author[1]{Philipp Kopp}
\author[1, 2]{Ernst Rank}
\author[1]{Stefan Kollmannsberger}

\affil[1]{Chair of Computational Modeling and Simulation, 
	Technische Universit\"at M\"unchen 
}

\affil[2]{Institute for Advanced Study, 
	Technische Universit\"at M\"unchen 
}

\date{}
\fancypagestyle{plain}{%
	\fancyhf{}%
	\fancyfoot[L]{
		\vspace{-0.0cm} 
		\footnotesize{Preprint submitted to \textit{Computers \& Mathematics with Applications}.
		}
} }

%% file: content/abstract.tex
Efficient solution strategies for wave propagation problems with complex geometries are essential in many engineering fields. Immersed boundary methods simplify mesh generation by embedding the domain of interest into an extended domain that is easy to mesh, introducing the challenge of dealing with cells that intersect the domain boundary. We consider the finite cell method that extends the weak form from the physical into the extended domain, scaled by a small value, to stabilize badly cut cells. Combined with explicit time integration schemes, the finite cell method introduces a lower bound for the critical time step size, which is restrictive for applications that require a small amount of stabilization not to compromise the accuracy. Explicit transient analyses commonly use the spectral element method due to its natural way of obtaining diagonal mass matrices through nodal lumping. The resulting method is very efficient since nodal lumping renders the solution of the equation systems trivial while not reducing the accuracy or the critical time step size. The combination of the spectral element and finite cell methods is called the spectral cell method. Unfortunately, a direct application of nodal lumping in the spectral cell method is impossible due to the special quadrature necessary to treat the discontinuous integrand inside the cut cells. The existing approaches to lump the mass matrices of cut cells significantly reduce the accuracy of the approximation.

We analyze an implicit-explicit (IMEX) time integration method to exploit the advantages of the nodal lumping scheme for uncut cells on one side and the unconditional stability of implicit time integration schemes for cut cells on the other. In this hybrid, immersed Newmark IMEX approach, we use explicit second-order central differences to integrate the uncut degrees of freedom that lead to a diagonal block in the mass matrix and an implicit trapezoidal Newmark method to integrate the remaining degrees of freedom (those supported by at least one cut cell). The immersed Newmark IMEX approach preserves the high-order convergence rates and the geometric flexibility of the finite cell method while retaining the efficiency of the nodal lumping scheme for uncut cells without compromising its critical time step size. We analyze a simple system of spring-coupled masses to highlight some of the essential characteristics of Newmark IMEX time integration. We then solve the scalar wave equation on two- and three-dimensional examples with significant geometric complexity to show that our approach is more efficient than state-of-the-art time integration schemes when comparing accuracy and runtime. While we focus on the finite and spectral cell methods, we expect other immersed methods to benefit equally from Newmark IMEX time integration.

%% file: content/intro.tex
Various technical and scientific applications depend on efficient methods to gain information about the propagation of waves. The most important fields include mechanical and civil engineering, medical engineering, and geophysics. For example, structural health monitoring~(SHM) and non-destructive testing~(NDT) require efficient and accurate simulations of waves propagating in plate-like structures or solids. In guided wave tomography~\cite{Pavlopoulou2012, Huthwaite2013, Rao2016, Zimmermann2021}, Lamb waves are excited in thin plates for damage or corrosion detection. With its origins in seismic exploration~\cite{Lailly1983, Tarantola1984, Fichtner2011}, the full waveform inversion~(FWI) employs simulated wave fields for the reconstruction of material models and anomalies in mechanical, civil, and medical engineering~\cite{Pratt2007, Sandhu2017, Guasch2020, Rao2020, BKK22, BKK23}. The underlying mathematical models of these applications, among others, are most commonly one-, two-, or three-dimensional scalar or elastic wave equations.

Many combinations of spatial and temporal discretization methods exist to approximate wave equations. Popular spatial discretizations are the finite difference method~(FDM)~\cite{Kelly1976, Virieux1986} and variants of the finite element method~(FEM)~\cite{Bao1998, Hughes2012}, in particular, its $p$-extension~\cite{Duester2017} and isogeometric analysis~(IGA)~\cite{Cottrell2006, Hughes2014} that uses B-Splines to generate approximations with higher continuity. For solutions with sufficient regularity, such high-order approaches benefit from higher convergence rates and a reduced dispersion error, enabling the solution of large-scale problems. A discussion of $p$- and $hp$-methods and their applications in electromagnetism can be found in \cite{Demkowicz2006}. Typical time integration approaches are basic finite difference schemes~\cite{FEMBook}, the Newmark family~\cite{N59}, and Runge-Kutta methods \cite{B62, DORMAND198019}. Particularly efficient approaches combine explicit time integration schemes with spatial discretizations that generate a diagonal mass matrix, as the solution of the equation systems of a time step becomes trivial. The spectral element method~(SEM)
defines Lagrange polynomials of order $p$ on each element of a conforming mesh and combines them to $C^0$ continuous finite element basis functions. The SEM is predominantly used for explicit time integration, where lumping techniques diagonalize the mass matrix and thus render the solution of the equation system trivial. The widely adopted ``nodal lumping'' technique uses Gauss-Lobatto-Legendre~(GLL) points as Lagrange nodes in the shape function definition and then again in the numerical quadrature of the finite element integrals to obtain a diagonal mass matrix due to the Kronecker delta property of the Lagrange shape functions. Nodal lumping does not suffer from a reduced accuracy compared to the consistent variant~\cite{Komatitsch2002, Cohen2002}. The standard IGA does not naturally lump the mass matrix; however, it gains efficiency by exploiting the elevated continuity of the solution space across element interfaces to obtain higher accuracy with a reduced number of degrees of freedom (dofs)~\cite{HCB05, IGA}. Finding methods that diagonalize the mass matrix in IGA without losing much accuracy remains challenging. For instance, classical row-sum lumping leads to a breakdown in $p$-convergence~\cite{Cottrell2006}, rendering it unsuitable for wave simulations. The most promising lumping approaches currently use approximate dual basis functions~\cite{Nguyen2023, Held2023}.

Wave propagation problems often benefit from an automatic way of incorporating highly complex geometries and different types of geometric representations into the finite element simulation when a manual mesh generation is tedious or unfeasible. One possible remedy is to embed the computational geometry, called the physical domain, into an extended domain with a simple shape that can be meshed easily. Later, the influence of this newly added, fictitious domain onto the solution in the original, physical domain is eliminated during the assembly. The major challenges of such immersed boundary methods are the stabilization and numerical quadrature of cut elements (i.e., elements that intersect the domain boundary), the imposition of boundary conditions, and the solution of the (often still) ill-conditioned equation system. The finite cell method (FCM) is an immersed boundary method that retains the definiteness of the problem by defining the original physical model also in the fictitious domain but multiplied with a small number $\alpha_f$ to minimize the introduced model error~\cite{Duester2008, Duester2017}. The extended domain is often chosen as the bounding box of the geometry and is meshed with a Cartesian grid whose finite elements are then called finite cells. The quadrature of the finite cell integrals on the cut cells is challenging as their discontinuous nature renders standard Gauss-Legendre rules unsuitable. A robust (yet expensive) strategy is to create a spacetree-based subdivision of the cut cells that is refined towards the domain boundary and distribute Gauss-Legendre points on each of these newly created integration cells. Alternatively, the computation of custom weights for a fixed set of quadrature points via moment fitting can drastically reduce the assembly effort~\cite{Joulaian2016}.

The combination of the spectral element and finite cell methods is called the spectral cell method (SCM). Like the SEM, the SCM uses Lagrange polynomials with GLL nodes; however, the standard quadrature methods to address the discontinuity of the finite cell integrals introduce non-diagonal blocks in the mass matrix. Any additional quadrature point that is not a Lagrange node (GLL point) breaks the natural lumping as it adds off-diagonal terms due to the interaction of multiple nonzero basis functions. We call this approach the \textit{consistent SCM}~\cite{Duczek2014}. The \textit{HRZ-lumped SCM}~\cite{Joulaian2014, Duczek2015a} uses HRZ lumping to diagonalize the mass matrices of cut cells. Alternatively, a moment fitting with fixed GLL points recovers the nodal lumping by computing custom weights for the GLL points~\cite{NAC22, Nicoli2023}. Unfortunately, lumped SCM methods suffer from reduced accuracy~\cite{K21}, and we observe a significantly restricted critical time step size in the presence of badly cut cells. %
As a result, the cell with the lowest volume fraction dominates the stability behavior of the entire system. As shown in~\cite{Eisentraeger2023}, eigenvalue stabilization can mitigate the severe limitation of the critical time step size. The development of explicit versions of the isogeometric finite cell analysis (IGA-FCM)~\cite{Schillinger2012, Rank2012} that combine IGA with the FCM faces similar challenges. IGA-FCM has been applied to various transient problems~\cite{Leidinger2019, Messmer2022, Voet2023, BKK23}, and while the critical time step of its lumped version appears to be independent of the cut cells, the reduction in accuracy is severe~\cite{Messmer2022, Voet2023}. An in-depth discussion of the critical time step size of immersed boundary methods, particularly in the context of IGA-FCM, can be found in~\cite{SBD23}.

A similar restriction on the overall time step size exists in dynamic fluid-structure and soil-structure problems~\cite{Belytschko1979}. This limitation arises because the structure mesh introduces a stability limit significantly lower than the medium mesh when using explicit time integration. To address this challenge, Belytschko and Mullen~\cite{BM78} and Hughes and Liu~\cite{H78} suggest integrating the part of the mesh belonging to the structure implicitly in time and the part of the mesh belonging to the fluid or soil explicitly. This combination results in an implicit-explicit (IMEX) time integration scheme. The analysis in~\cite{H78, H78_stability} shows that the IMEX integration approach inherits the stability characteristics of the explicit subsystem while preserving the accuracy of both individual methods. We transfer this scheme to systems discretized using the SCM. The part of the system corresponding to cut cells --- hence strongly restricting the stability and contributing to a non-diagonal block in the global mass matrix --- is integrated implicitly over time using the trapezoidal Newmark method~\cite{Bathe}. In contrast, the part of the system corresponding to uncut cells --- thus contributing to a diagonal block in the global mass matrix --- is integrated explicitly using the second-order central difference method~(CDM)~\cite{Hughes2012}. The proposed approach, denoted as immersed Newmark IMEX method, preserves the accuracy of the consistent SCM, eliminates the dependency of the time step size on the cut cells, and exploits the efficiency of explicit time stepping for the diagonal block of the mass matrix belonging to uncut cells.

The central aspects of our paper are as follows.
\begin{enumerate}[noitemsep,topsep=3pt,parsep=3pt,partopsep=0pt]
	\item We discuss the Newmark IMEX time integration and examine its stability and accuracy on a simplified system of ten spring-coupled masses.
	\item We apply this approach to the spectral cell method, where we integrate the dofs of cut cells implicitly and the remaining dofs explicitly. We evaluate the efficiency and accuracy on a two-dimensional perforated plate and compare it to conventional time integration schemes.
	\item We demonstrate the applicability of the immersed Newmark IMEX method to large-scale problems using a complex three-dimensional example. Our approach performs better, even when dealing with an unfavorable ratio between the implicit and explicit subsystems. 
\end{enumerate}

The remainder of the paper is structured as follows. In~\secref{sec:theory}, we start with a discussion of the critical time step size to motivate the development of our immersed Newmark IMEX approach. We then introduce the scalar wave equation and outline its spatial and temporal discretization. \secref{sec:NumericExamples} applies the proposed approach to examples of increasing complexity. First, we demonstrate the convergence and stability characteristics of the implicit-explicit time integration scheme using a system of spring-coupled masses. Second, we compare the efficiency of the immersed Newmark IMEX method to conventional time integration schemes on a two-dimensional example. Third, we validate the improved performance on a three-dimensional problem. Finally, \secref{sec:conclusion} concludes our paper and highlights some of the limitations.

%% file: content/motivation.tex
Explicit time integration is computationally efficient but comes with the drawback of conditional stability. A critical time step size $\Delta t_\text{crit}$ defines the stability limit, beyond which the time integration becomes unstable, and the solution grows infinitely~\cite{Hughes2012}. In the context of the spectral cell method~(SCM), the critical time step size of the global system is predominantly determined by cells intersected by the boundaries of the geometry, even if the number of cut cells is small. For the explicit second-order central difference method~(CDM), the systems's critical time step size $t_\text{crit}$ is equal to~\cite{Hughes2012}
\begin{equation}
	\Delta t_\text{crit} = \frac{2}{\omega_\text{max}}\text{,}
\end{equation}
where $\omega_\text{max}$ is the highest eigenfrequency of the system. For a spatially discretized system, we obtain
\begin{equation}
  \omega_\text{max} = \sqrt{\lambda_\text{max}(\v{K}, \v{M})}
\end{equation}
by computing the largest eigenvalue $\lambda_\text{max}$ of the generalized eigenproblem corresponding to the stiffness matrix~$\v{K}$ and mass matrix~$\v{M}$. 

\begin{figure}[H]
	\centering
        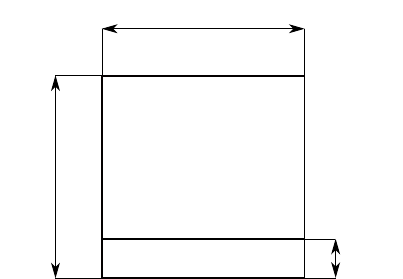
        \caption{Cell cut by a horizontal boundary}
        \label{fig:fillRatio}
\end{figure}

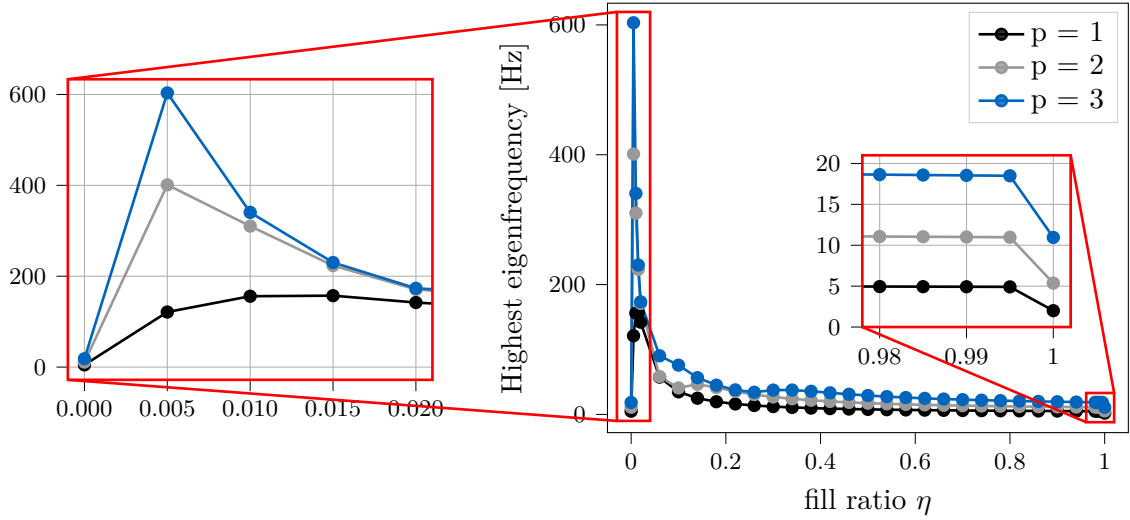
\begin{figure}[t!]
	\centering
	\input{figures/fancyFillRatios.tex} 
	\caption{Highest eigenfrequency for different fill ratios $\eta$ and polynomial degrees $p$}
	\label{fig:fillRatioHZ}
\end{figure}

When dealing with cut cells in the SCM, the maximum eigenfrequency of one cut cell depends on the ratio between its physical and fictitious parts. \figref{fig:fillRatio} shows a motivating example that we use to analyze the behavior of the eigenfrequency for badly cut cells.
A horizontal boundary with varying heights $d_\text{p}$ cuts a single quadrilateral cell. Below the cut, the material's density and wave speed are constant, i.e., $\rho_\text{p} = c_\text{p} = 1$. Above the boundary, the material is fictitious, and its density is scaled to $\rho_\text{f} = 10^{-6}$, while the wave speed remains untouched, i.e., $c_\text{f} = 1$. The fill ratio of the cell, depending on the cut position, is $\eta = \frac{d_\text{p}}{l_\text{y}}$.
The maximum eigenfrequency of the cell is computed using spectral basis functions of polynomial orders $p = 1, 2, $ and $3$, and a quadtree of depth $13$ for the integration of the mass and stiffness matrices. \figref{fig:fillRatioHZ} plots the maximum eigenfrequencies of the cells compared to their fill ratios. Cells with smaller fill ratios feature higher maximum eigenfrequencies, which leads to a decisive restriction on the critical time step size. This effect is more pronounced the higher the polynomial degree of a cell.

In addition to the cut ratio, factors such as the cut's shape also significantly affect the maximum eigenfrequency of a cut cell. A detailed discussion regarding critical time step sizes of immersed methods can be found in~\cite{SBD23}. Implicit-explicit time integration overcomes this limitation of the SCM by integrating cut cells implicitly in time.

%% file: figures/fillRatios.pdf_tex
\begingroup%
  \makeatletter%
  \providecommand\color[2][]{%
    \errmessage{(Inkscape) Color is used for the text in Inkscape, but the package 'color.sty' is not loaded}%
    \renewcommand\color[2][]{}%
  }%
  \providecommand\transparent[1]{%
    \errmessage{(Inkscape) Transparency is used (non-zero) for the text in Inkscape, but the package 'transparent.sty' is not loaded}%
    \renewcommand\transparent[1]{}%
  }%
  \providecommand\rotatebox[2]{#2}%
  \newcommand*\fsize{\dimexpr\f@size pt\relax}%
  \newcommand*\lineheight[1]{\fontsize{\fsize}{#1\fsize}\selectfont}%
  \ifx\svgwidth\undefined%
    \setlength{\unitlength}{193.36234867bp}%
    \ifx\svgscale\undefined%
      \relax%
    \else%
      \setlength{\unitlength}{\unitlength * \real{\svgscale}}%
    \fi%
  \else%
    \setlength{\unitlength}{\svgwidth}%
  \fi%
  \global\let\svgwidth\undefined%
  \global\let\svgscale\undefined%
  \makeatother%
  \begin{picture}(1,0.69144602)%
    \lineheight{1}%
    \setlength\tabcolsep{0pt}%
    \put(0,0){\includegraphics[width=\unitlength,page=1]{figures/fillRatios.pdf}}%
    \put(0.91075896,0.03164443){\color[rgb]{0,0,0}\makebox(0,0)[t]{\lineheight{1.25}\smash{\begin{tabular}[t]{c}$d_p$\end{tabular}}}}%
    \put(0.51043405,0.64429523){\color[rgb]{0,0,0}\makebox(0,0)[t]{\lineheight{1.25}\smash{\begin{tabular}[t]{c}$l_x$\end{tabular}}}}%
    \put(0.0768473,0.23841391){\color[rgb]{0,0,0}\makebox(0,0)[t]{\lineheight{1.25}\smash{\begin{tabular}[t]{c}$l_y$\end{tabular}}}}%
  \end{picture}%
\endgroup%

%% file: figures/fancyFillRatios.tex
\begin{tikzpicture}
	
	\definecolor{darkcyan0101189}{RGB}{0,101,189}
	\definecolor{darkgray153}{RGB}{153,153,153}
	\definecolor{darkgray176}{RGB}{176,176,176}
	\definecolor{lightgray204}{RGB}{204,204,204}
	\pgfplotsset{every tick label/.append style={font=\footnotesize}}
	\begin{axis}[
		scaled ticks=false,
		xmin=0,
		ymin=0,
		xlabel=In,
		ylabel=out,
		name=mainAx,
		legend cell align={left},
		legend style={fill opacity=0.8, draw opacity=1, text opacity=1, draw=lightgray204},
		tick align=outside,
		tick pos=left,
		xlabel={fill ratio $\eta$},
		xmin=-0.05, xmax=1.05,
		xtick style={color=black},
		y grid style={darkgray176},
		ylabel={Highest eigenfrequency [Hz]},
		ymin=-28.0684673209088, ymax=633.437813739085,
		ytick style={color=black}
		]
		
		\addplot [line width=1pt, black, mark=*, mark size=2, mark options={solid}]
		table {%
			0 4.89897948556724
			0.005 121.168838718037
			0.01 155.934143125144
			0.015 157.197881461798
			0.02 142.103130304999
			0.06 57.3606877514231
			0.1 34.7633579572768
			0.14 24.968063949803
			0.18 19.5502188127825
			0.22 16.1182440813372
			0.26 13.7650084967748
			0.3 12.0553902936978
			0.34 10.7602997429505
			0.38 9.751785248248
			0.42 8.94600834971271
			0.46 8.28881112524952
			0.5 7.74592951220602
			0.54 7.29076552356884
			0.58 6.9043102141407
			0.62 6.57402738951362
			0.66 6.28890803420217
			0.7 6.04060313991709
			0.74 5.82360272141913
			0.78 5.63253485570473
			0.82 5.4631664322825
			0.86 5.31277056457676
			0.9 5.17822427317554
			0.94 5.05773993231199
			0.98 4.94927179969145
			0.985 4.9364595183604
			0.99 4.92380847565658
			0.995 4.91131600264697
			1 2
		};
		\addlegendentry{p = 1}
		\addplot [line width=1pt, darkgray153, mark=*, mark size=2, mark options={solid}]
		table {%
			0 10.9544511501228
			0.005 401.107603716093
			0.01 310.200453864134
			0.015 223.412309880532
			0.02 170.946763188762
			0.06 58.2496609766696
			0.1 40.8526334971884
			0.14 46.436611990197
			0.18 41.445007528221
			0.22 35.4142555434202
			0.26 30.5738267632384
			0.3 26.878550186589
			0.34 24.031662470951
			0.38 21.79219482169
			0.42 19.9965662234014
			0.46 18.5322462062958
			0.5 17.3189287792258
			0.54 16.3012491770218
			0.58 15.4386504771568
			0.62 14.6997642573364
			0.66 14.0619142392212
			0.7 13.5074937561359
			0.74 13.0219424865886
			0.78 12.594444081613
			0.82 12.2162463896145
			0.86 11.8794439862448
			0.9 11.5791014431098
			0.94 11.3096233005532
			0.98 11.0667128492045
			0.985 11.0380661904899
			0.99 11.0099704544251
			0.995 10.9820364396147
			1 5.3665631459995
		};
		\addlegendentry{p = 2}
		\addplot [line width=1pt, darkcyan0101189, mark=*, mark size=2, mark options={solid}]
		table {%
			0 18.4458614598009
			0.005 603.369346418176
			0.01 340.450022004595
			0.015 230.158127975808
			0.02 173.393993648284
			0.06 90.3032923048559
			0.1 75.8706003373924
			0.14 56.6155189781937
			0.18 45.047122118541
			0.22 37.7649660429095
			0.26 34.160041619721
			0.3 37.4723317421237
			0.34 37.5590887798691
			0.38 35.6444979002863
			0.42 33.2839517552746
			0.46 31.0564526591679
			0.5 29.1027230842569
			0.54 27.4238583032831
			0.58 25.9852647566168
			0.62 24.747973336725
			0.66 23.6761611251191
			0.7 22.743422872567
			0.74 21.926485275773
			0.78 21.2070348728645
			0.82 20.5700958096182
			0.86 20.0035775637166
			0.9 19.4975329102817
			0.94 19.0436317951994
			0.98 18.6350183376731
			0.985 18.5867796059976
			0.99 18.5391479140802
			0.995 18.4923114709928
			1 10.9544511501033
		};
		\addlegendentry{p = 3}
		
		\coordinate (c1) at (axis cs:0.961, -12); 
		\coordinate (c2) at (axis cs:0.97, 30);
		\coordinate (c5) at (axis cs:1.02, -10);
		\coordinate (c6) at (axis cs:1.02, 33);
		\draw[draw=red,line width=1pt] (c1) rectangle (c6);		
		
		\coordinate (c3) at (axis cs:-0.03, -10);
		\coordinate (c4) at (axis cs:-0.03, 620);
		\coordinate (c7) at (axis cs:0.05, -10);
		\coordinate (c8) at (axis cs:0.04, 620);
		\draw[draw=red,line width=1pt] (c3) rectangle (c8);		
	\end{axis}

	\begin{axis}[ 
		legend cell align={left},
		legend style={fill opacity=0.8, draw opacity=1, text opacity=1, draw=lightgray204},
		tick align=outside,
		tick pos=left,
		scale=0.4,
		x grid style={darkgray176},
		xmajorgrids,
		xmin=0.978, xmax=1.002,
		xminorgrids,
		xtick style={color=black},
		y grid style={darkgray176},
		ymajorgrids,
		at={($(mainAx.south east)+(-3.5cm,1.4cm)$)},
		ymin=-0.0, ymax=21,
		yminorgrids,
		ytick style={color=black},
		name=rightAx
		]
		\addplot [line width=1pt, black, mark=*, mark size=2, mark options={solid}]
		table {%
			0.94 5.05773993231199
			0.98 4.94927179969145
			0.985 4.9364595183604
			0.99 4.92380847565658
			0.995 4.91131600264697
			1 2
		};
		\addplot [line width=1pt, darkgray153, mark=*, mark size=2, mark options={solid}]
		table {%
			0.94 11.3096233005532
			0.98 11.0667128492045
			0.985 11.0380661904899
			0.99 11.0099704544251
			0.995 10.9820364396147
			1 5.3665631459995
		};

		\addplot [line width=1pt, darkcyan0101189, mark=*, mark size=2, mark options={solid}]
		table {%
			0.94 19.0436317951994
			0.98 18.6350183376731
			0.985 18.5867796059976
			0.99 18.5391479140802
			0.995 18.4923114709928
			1 10.9544511501033
		};

	\end{axis} 
	
	\begin{axis}[ 
		legend cell align={left},
		legend style={fill opacity=0.8, draw opacity=1, text opacity=1, draw=lightgray204},
		tick align=outside,
		tick pos=left,
		scale=0.7,
		x grid style={darkgray176},
		xmajorgrids,
		xmin=-0.001, xmax=0.021,
		xminorgrids,
		xtick style={color=black},
		scaled x ticks=false,
		xticklabel=\pgfkeys{/pgf/number format/.cd,fixed,precision=3,zerofill}\pgfmathprintnumber{\tick},
		y grid style={darkgray176},
		ymajorgrids,
		at={($(mainAx.south west)+(-7.1cm,0.7cm)$)},
		ymin=-28.0684673209088, ymax=633.437813739085,
		yminorgrids,
		ytick style={color=black},
		name=leftAx
		]
		\addplot [line width=1pt, black, mark=*, mark size=2, mark options={solid}]
		table {%
			0 4.89897948556724
			0.005 121.168838718037
			0.01 155.934143125144
			0.015 157.197881461798
			0.02 142.103130304999
			0.06 57.3606877514231
			0.1 34.7633579572768
			0.14 24.968063949803
			0.18 19.5502188127825
		};

		\addplot [line width=1pt, darkgray153, mark=*, mark size=2, mark options={solid}]
		table {%
			0 10.9544511501228
			0.005 401.107603716093
			0.01 310.200453864134
			0.015 223.412309880532
			0.02 170.946763188762
			0.06 58.2496609766696
			0.1 40.8526334971884
			0.14 46.436611990197
			0.18 41.445007528221
		};

		\addplot [line width=1pt, darkcyan0101189, mark=*, mark size=2, mark options={solid}]
		table {%
			0 18.4458614598009
			0.005 603.369346418176
			0.01 340.450022004595
			0.015 230.158127975808
			0.02 173.393993648284
			0.06 90.3032923048559
			0.1 75.8706003373924
			0.14 56.6155189781937
			0.18 45.047122118541
		};

		\end{axis}

	\draw [draw=red, line width=1pt] (c1) -- (rightAx.south west);
	\draw [ draw=red, line width=1pt] (c6) -- (rightAx.north east);
	\draw [ draw=red, line width=1pt] (c3) -- (leftAx.south west);
	\draw [ draw=red, line width=1pt] (c4) -- (leftAx.north west);
	\draw[draw=red, line width=1pt] (rightAx.south west) rectangle (rightAx.north east);	
	\draw[draw=red, line width=1pt] (leftAx.south west) rectangle (leftAx.north east);			
	
\end{tikzpicture}

%% file: content/spatialDiscretization.tex
For a given spatial dimension $d = 1, 2,$ or $3$, the scalar wave equation is defined on a domain $\Omega \subset \mathbb{R}^d$ for the time interval $\mathcal{T} = \left[0, T\right]$, where $T$ is the simulation duration. By introducing the wave field $u(\v{x}, t))$, its acceleration $\ddot{u}(\v{x}, t))$, the external force term $f(\v{x}, t))$, the heterogeneous density $\rho(\v{x})$ and wave speed $c(\v{x})$, we obtain the following partial differential equation (PDE):
\begin{equation}
	\rho(\v{x}) \ddot{u}(x, t) - \nabla \cdot (\rho(\v{x}) c^2(\v{x}) \nabla u(\v{x}, t)) = f(\v{x}, t), \;
	\v{x} \in \Omega, \quad t \in \mathcal{T}\text{.}
	\label{eq:scalarWaveEq}
\end{equation}
The initial conditions are $u(\v{x}, 0) = u_\text{0}(\v{x})$ and $\dot{u}(\v{x}, 0) = \dot{u}_\text{0}(\v{x})$ on $\v{x} \in \Omega$. The Dirichlet boundary conditions are $u(\v{x}, t) = g(\v{x}, t)$ on $\v{x} \in \Gamma_\text{D}$, and the Neumann boundary conditions are ${\v{n} \cdot \nabla u(\v{x}, t) = h(\v{x}, t)}$ on $\v{x} \in \Gamma_\text{N}$, where $\partial \Omega = \Gamma = \Gamma_\text{N} \cup \Gamma_\text{D}$ and $\Gamma_\text{N} \cap \Gamma_\text{D} = \emptyset$ is satisfied. The unit normal vector $\v{n}$ on the boundary $\partial \Omega$ points outwards. 

\begin{figure}[H]
    \centering
	\subfloat[Physical domain \label{fig:physDomain}]{\includegraphics[width=0.3\textwidth]{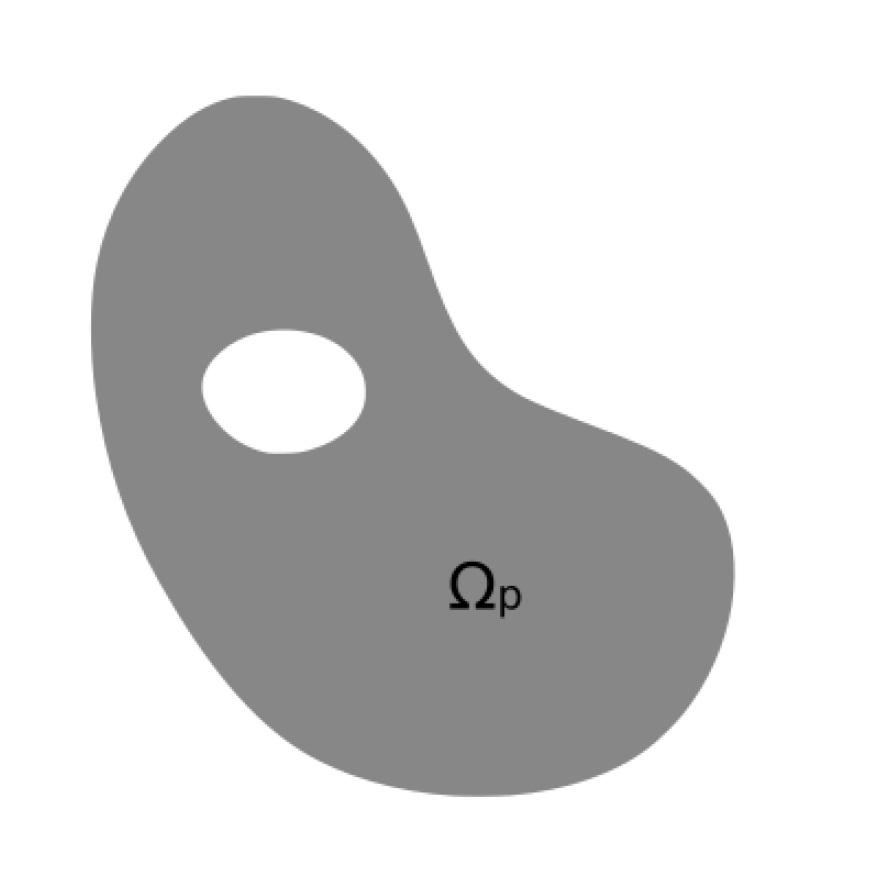}}
	\hfill
	\subfloat[Meshed embedding domain \label{fig:embeddedDomain}]{\includegraphics[width=0.3\textwidth]{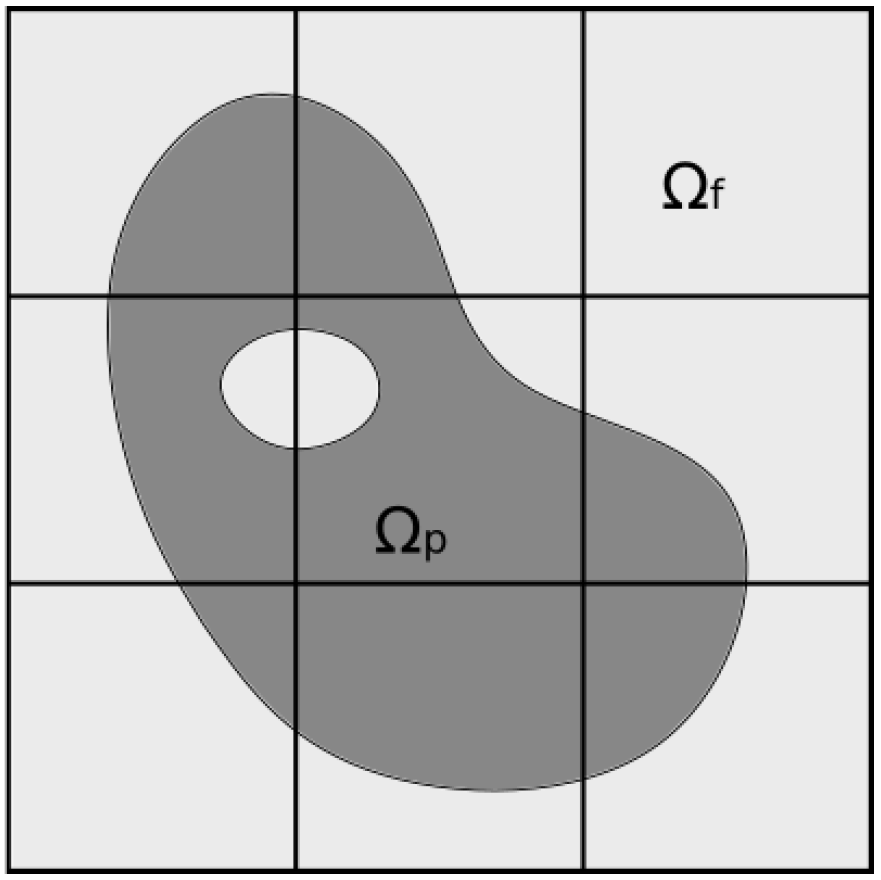}}
	\hfill
	\subfloat[Integration mesh defined by a quadtree \label{fig:fcmQuadtree}]{\includegraphics[width=0.3\textwidth]{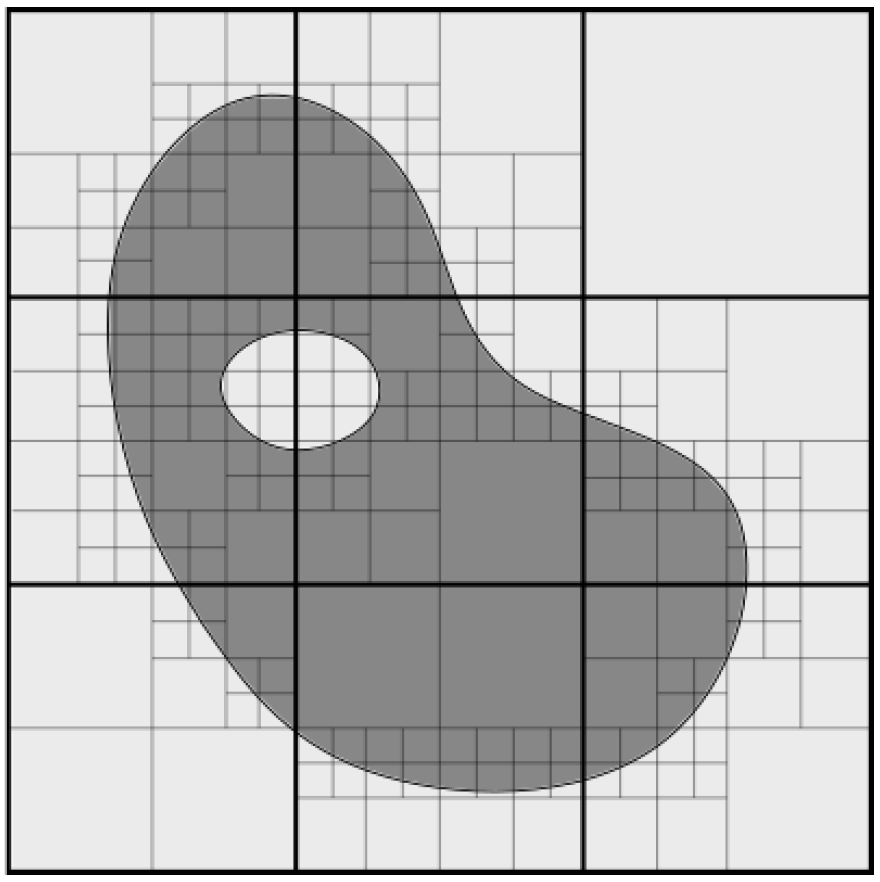}}
	\caption{Principles of the FCM/SCM~\cite{K21}}
	\label{fig:FCM}
\end{figure}

The finite element discretization introduces $n^\text{dof}$ basis functions $N_i(\v{x})$ and their corresponding time-dependent coefficients $\hat{u}_i(t)$, for $i = 1, 2, \ldots, n^\text{dof}$ to approximate the wave field as
\begin{equation}
    u(\v{x}, t) \approx \sum_{i = 1}^{n^\text{dof}} N_i(\v{x}) \hat{u}_i(t) \text{.}
\end{equation}
The hat notation indicates the coefficients of a discretized quantity. An in-depth introduction of the finite element method (FEM) for transient problems can be found in~\cite{Hughes2012}. Like other finite element methods, the SEM subdivides the computational domain into a set of elements that support tensor-products of Lagrange polynomials of degree $p$. These shape functions are connected with those of adjacent elements to form the basis functions $N_i(\v{x})$. The $p + 1$ Lagrange polynomials in the reference interval~$[-1, 1]$ are defined as follows
\begin{equation}
    N_i^\text{Lag,p}(\xi) = \prod_{j = 1, j \neq i}^{p + 1} \frac{\xi - \xi_j}{\xi_i - \xi_j} \quad \text{for } i = 1, 2, \ldots, p + 1 \text{.}
\end{equation}
We choose Gauß-Lobatto-Legendre (GLL) points for the definition of $N_i^\text{Lag,p}$ to obtain a diagonal mass matrix. We refer to~\cite{Komatitsch2002, Cohen2002, Fichtner2011} for a discussion of GLL points combined with the SEM. The $p + 1$ GLL points of one element are given by
\begin{equation}
    \xi_j = \begin{cases}
		-1 \qquad &\text{if } j = 1 \\
		\xi_{0, j-1}^{\text{Lo}, p-1} \qquad &\text{if } 2 \leq j < p + 1 \\
		1 \qquad &\text{if } j = p+1 \\
	\end{cases}.
\end{equation}
where $\xi_{0, k}^{\text{Lo}, p-1}$ with $k = 1, 2, \ldots, p-1$ are the roots of Lobatto polynomials of order $p-1$. Combining the SEM basis functions with an immersed boundary approach, in particular, the finite cell method~(FCM)~\cite{Duester2008, Duester2017}, leads to the spectral cell method (SCM)~\cite{Duczek2014, Joulaian2014}, a non-boundary-conforming adaption of the SEM. In both FCM and SCM, the physical domain $\Omega_\text{p}$ is embedded within a larger computational domain $\Omega$, which is of simple geometric shape and can be readily meshed using a Cartesian grid. As illustrated in \figref{fig:FCM}, the computational domain $\Omega$ is subdivided into the physical domain $\Omega_\text{p}$ and the fictitious domain $\Omega_\text{f}$, such that 
\begin{equation}
	\Omega_\text{p} \cup \Omega_\text{f} = \Omega \text{.}
\end{equation}
In the context of FCM, the elements generated by the Cartesian grid are referred to as cells, and those cells that are intersected by the boundaries of the physical domain are termed cut cells.
We introduce an indicator function $\alpha(\v{x})$ to recover the original geometry and differentiate between the physical domain $\Omega_\text{p}$ and the fictitious domain $\Omega_\text{f}$. Within the fictitious domain, $\alpha(\v{x})$ is set to a value close to \textit{zero}, while it remains \textit{one} in the physical domain
\begin{equation}
	\alpha(\v{x}) = \begin{cases}
		1 \qquad &\v{x} \in \Omega_\text{p} \\
		\alpha_\text{f} = 10^{-\beta} \qquad &\v{x} \in \Omega_\text{f}
	\end{cases}.
\end{equation}
In practice, $\beta$ is commonly assigned some value between five and eight. According to~\cite{Duczek2014, Joulaian2014}, the entire weak form of the scalar wave equation is multiplied by $\alpha(\v{x})$. This scaling is investigated in~\cite{BKK22} is equivalent to scaling the system's density, while keeping the wave speed untouched, i.e.,
\begin{equation}
    \rho^*(\v{x}) = \alpha(\v{x}) \rho(\v{x}) \text{.}
\end{equation}
This scaling turns out to be most favourable for treating voids within the computational domain. A generalization to the elastic wave equation and a detailed mathematical analysis is given in~\cite{Rabinovich2023}. The entries of the corresponding mass matrix $\v{M}$, stiffness matrix $\v{K}$ and time-dependent force vector $\v{f}(t)$ are obtained by
\begin{alignat}{2}
	M_{ij} &= \int_{\Omega} \alpha(\v{x}) \rho(\v{x}) N_i(\v{x}) N_j(\v{x}) d\Omega,	\label{eq:Mglobal} \\
	K_{ij} &= \int_{\Omega} \alpha(\v{x}) \rho(\v{x}) c(\v{x})^2 (\nabla N_i(\v{x}))^T \nabla N_j(\v{x}) d\Omega, \label{eq:K} \\
	\hat{f}_i(t) &= \int_{\Omega} \alpha(\v{x}) f(\v{x}, t) N_i(\v{x}) d\Omega \text{.} 	\label{eq:f}
\end{alignat}
This spatial discretization yields the second-order system
\begin{equation}
	\v{M} \ddot{\dv{u}}(t) + \v{K} \dv{u}(t) = \dv{f}(t) \text{,}
	\label{eq:discretizedWaveEquation}
\end{equation}
where the time-dependent coefficient vector $\dv{u}(t)$ contains the unknowns of the system. 
In wave propagation applications, it is often possible to split the force term into a spatial contribution and a temporal excitation
\begin{equation}
    \dv{f}(t) = f_\text{t}(t) \cdot \dv{f}_{\text{x}}.
\end{equation}
The spatial force vector $\dv{f}_{\text{x}}$ entails the geometric information of the excitation, e.g., the source positions. The term $f_\text{t}(t)$ contains the excitation signal in time.

In the finite and spectral cell methods, the boundaries of the physical domain are recovered by accurately computing the discontinuous integrands in Equations \eqref{eq:Mglobal}, \eqref{eq:K} and \eqref{eq:f}. There are several approaches for integrating cut cells, among them are: space-trees~\cite{Duester2008, Peto2020}, smart octrees~\cite{Kudela2016}, and moment-fitting~\cite{Joulaian2016, NAC22}. We use space-trees (i.e. binary-, quad-, and octrees in one, two, and three dimension) that are refined towards the domain boundary for the integration of cut cells. \figref{fig:fcmQuadtree} shows an exemplary integration mesh for a quadtree of depth $3$. The leaves of the spacetree are used to accumulate Gauss-Legendre points close to the domain boundary. 

In the SCM, the choice of a cell's quadrature rule depends on whether a cell is cut or not. Cells not intersected by the boundaries of the physical domain (from now on referred to as uncut cells) are integrated using the Gauß-Lobatto quadrature with GLL points as integration points. The latter causes the element mass matrices to be diagonal for uncut elements, since in each uncut element at all GLL points exactly one Lagrange basis function is unequal to zero. However, if a cell is cut, we use a space-tree integration scheme. Since the diagonality of the element mass matrix is lost on cut elements, we prefer to distribute the more accurate Gauß-Legendre quadrature on the spacetree integration mesh. The element stiffness matrices and the load vectors are integrated using Gauß-Legendre quadrature regardless of whether the cell is cut or uncut.

%% file: content/partitioning.tex
\begin{figure}[b!]
\centering
	\subfloat[Mass matrix $\v{M}$ \label{fig:MStructure}]{\input{figures/MStructure.tex}}
	\hfill
	\subfloat[Stiffness matrix $\v{K}$ \label{fig:KStructure}]{\input{figures/KStructure.tex}}
	\caption{Typical matrix structures for a consistent SCM discretization}
\end{figure}
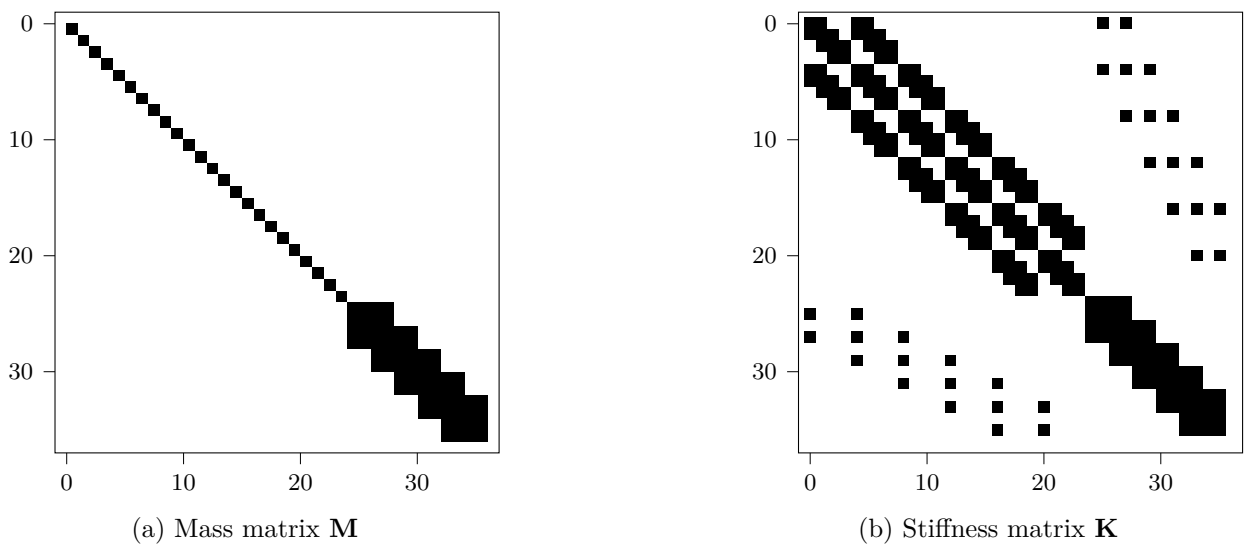

The degrees of freedom of basis functions supported exclusively in uncut cells contribute to exactly one entry in the diagonal mass matrix block $\v{M}^\text{dd}$. Conversely, dofs with support in at least one cut cell contribute to a non-diagonal block of the mass matrix $\v{M}^\text{cc}$. We partition the mass matrix into diagonal and a non-diagonal blocks
\begin{equation}
	\label{eq:MStructure}
	\v{M} = \begin{bmatrix} \v{M}^\text{dd} & \v{0} \\ \v{0} & \v{M}^\text{cc} \end{bmatrix} \text{.}
\end{equation}
In the following, the dofs, the stiffness matrix, and the force vector are partitioned based on this distinction. The off-diagonal blocks in the mass matrix equal $\v{0}$ and do not introduce any coupling. Sub-vectors or sub-matrices corresponding to the dofs with the diagonal mass matrix block are denoted by the superscript $\star^\text{d}$ for `diagonal', while all other dofs corresponding to the non-diagonal block of the mass matrix are denoted by $\star^\text{c}$ for `cut'. \figref{fig:MStructure} illustrates an exemplary structure of an obtained global mass and stiffness matrix. 
The stiffness matrix $\v{K}$ couples diagonal and cut dofs, as demonstrated in~\figref{fig:KStructure}. Nevertheless, it is possible to decompose the stiffness matrix into
\begin{equation}
	\label{eq:KStructure}
	\v{K} = \begin{bmatrix} \v{K}^\text{dd} & \v{K}^\text{dc} \\ \v{K}^\text{cd} & \v{K}^\text{cc} 	\end{bmatrix}.
\end{equation}
Given the symmetric nature of the stiffness matrix, $\v{K}^\text{dc} = (\v{K}^\text{cd})^T$ holds. For future use, the sub-matrices $\v{K}^\text{d}$ and $\v{K}^\text{c}$ are introduced
\begin{alignat}{2}
	\v{K}^\text{d} &= \begin{bmatrix} \v{K}^\text{dd} & \v{K}^\text{dc} 	\end{bmatrix} 	\label{eq:Kd} \\
	\v{K}^\text{c} &= \begin{bmatrix} \v{K}^\text{cd} & \v{K}^\text{cc} 	\end{bmatrix}. 	\label{eq:Kc}
\end{alignat}
The spatial force vector $\dv{f}$ is partitioned accordingly
\begin{equation}
	\dv{f}_\text{x} = \begin{bmatrix} \bigl(\dv{f}_\text{x}^\text{d}\bigr)^T, & \bigl(\dv{f}_\text{x}^\text{c}\bigr)^T \end{bmatrix}^T \text{.}
	\label{eq:f_xStructure}
\end{equation}

%% file: figures/MStructure.tex
\begin{tikzpicture}

\definecolor{darkgray176}{RGB}{176,176,176}
\pgfplotsset{every tick label/.append style={font=\footnotesize}}
\pgfplotsset{every label/.append style={font=\footnotesize}}

\begin{axis}[
tick align= outside,
x grid style={darkgray176},
xmin=-1, xmax=37,
xtick pos=bottom,
xtick style={color=black},
y grid style={darkgray176},
ymin=-1, ymax=37,
ytick pos=bottom,
y dir = reverse,
ytick style={color=black},
width=0.45*\textwidth,
height=0.45*\textwidth,
]
\addplot graphics [includegraphics cmd=\pgfimage,xmin=0, xmax=36, ymin=0, ymax=36] {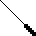};
\end{axis}

\end{tikzpicture}

%% file: figures/KStructure.tex
\begin{tikzpicture}

\definecolor{darkgray176}{RGB}{176,176,176}
\pgfplotsset{every tick label/.append style={font=\footnotesize}}
\pgfplotsset{every label/.append style={font=\footnotesize}}

\begin{axis}[
tick align= outside,
x grid style={darkgray176},
xmin=-1, xmax=37,
xtick pos=bottom,
xtick style={color=black},
y grid style={darkgray176},
ymin=-1, ymax=37,
ytick pos=bottom,
y dir = reverse,
ytick style={color=black},
width=0.45*\textwidth,
height=0.45*\textwidth,
]
\addplot graphics [includegraphics cmd=\pgfimage,xmin=-0.5, xmax=35.5, ymin=35.5, ymax=-0.5] {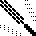};
\end{axis}

\end{tikzpicture}

%% file: content/timeIntegration.tex
\begin{figure}[t!]
	\centering
	\resizebox{0.8\textwidth}{!}{\def\svgwidth{1\textwidth}\input{figures/NewmarkImplicitOverleaf.pdf_tex}}
	\caption{Flowchart for the trapezoidal Newmark method adapted from~\cite{GR14}}
	\label{fig:implicitNewmark}
\end{figure}
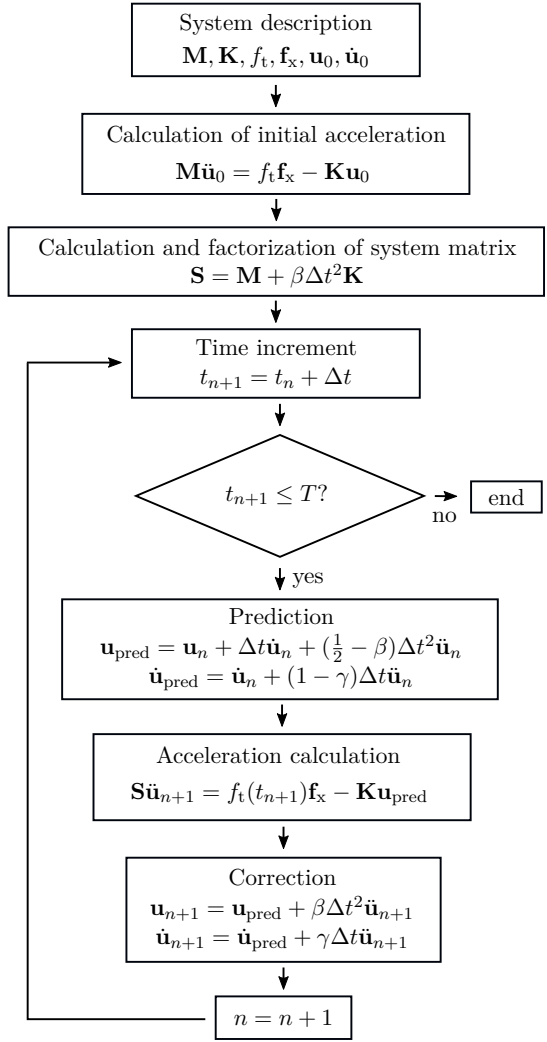

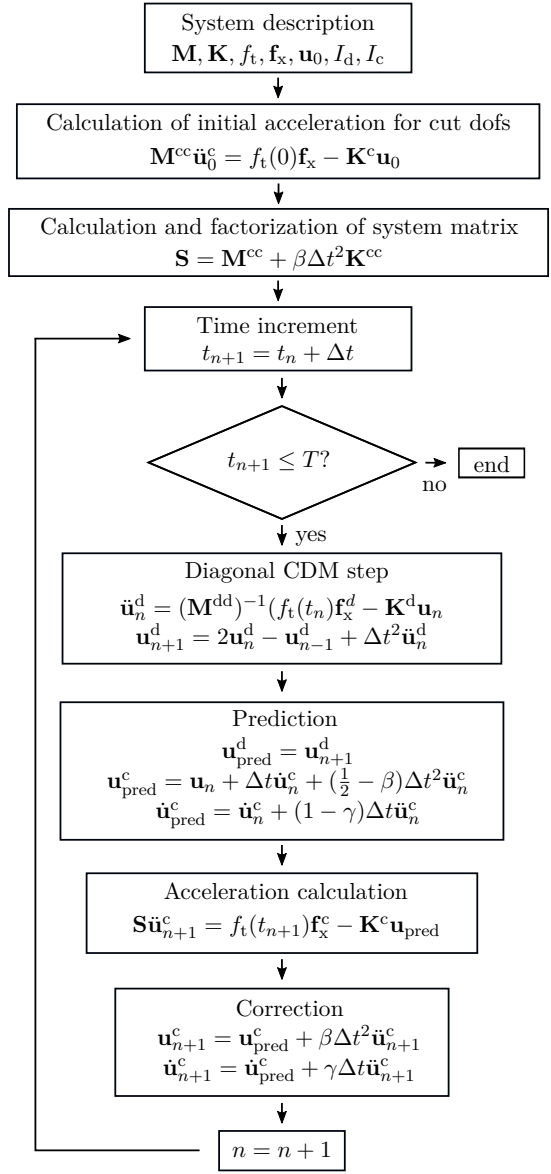
\begin{figure}[t!]
	\centering
	\resizebox{0.8\textwidth}{!}{\def\svgwidth{1\textwidth}\input{figures/CDMTrapImexOverleaf.pdf_tex}}
	\caption{Flowchart for the Newmark IMEX method}
	\label{fig:TrapCDMIMEX}
\end{figure}

Since numerical time integration exclusively deals with discretized quantities, the $\hat{\star}$ notation is omitted in this section.
The primary advantage of explicit methods is their potentially higher computational efficiency per time step compared to implicit methods. On the other hand, they are only conditionally stable. However, as long as the time step size limitation is not too restrictive, explicit methods, particularly the CDM, offer a good trade-off between accuracy and computational cost~\cite{Hughes2012}. The following derivations closely follow~\cite{K21}. Newmark's time integration formulas~\cite{N59} are well established and equal
\begin{alignat}{2}
	\v{u}_{n+1} &= \v{u}_n + \Delta t \dot{\v{u}}_n + \Delta t^2 ((\frac{1}{2} - \beta) \ddot{\v{u}}_n + \beta \ddot{\v{u}}_{n+1}) \label{eq:Newmark1} \\
	\dot{\v{u}}_{n+1} &= \dot{\v{u}}_n + \Delta t ((1-\gamma) \ddot{\v{u}}_n + \gamma \ddot{\v{u}}_{n+1}) \text{.}\label{eq:Newmark2}
\end{alignat}
The two parameters $\beta$ and $\gamma$ serve to define the weights for the acceleration calculation, $\beta$ for the displacement estimation, and $\gamma$ for the velocity estimation, respectively. Depending on the choice of these parameters, the method is either explicit or implicit. The explicit CDM, is obtained by setting $\beta = 0$ and $\gamma = \frac{1}{2}$. The formulas then simplify to 
\begin{alignat}{2}
	\v{u}_{n+1} &= \v{u}_n + \Delta t \dot{\v{u}}_n + \frac{\Delta t^2}{2} \ddot{\v{u}}_n \\
	\dot{\v{u}}_{n+1} &= \dot{\v{u}}_n + \frac{\Delta t}{2}(\ddot{\v{u}}_n + \ddot{\v{u}}_{n+1}) \text{.}
\end{alignat}
The CDM is second-order accurate and exhibits no numerical dissipation~\cite{H78}. In the context of wave propagation, one can often omit the computation of the velocities~$\dot{\v{u}}$. As a consequence, the velocity can be eliminated by substitution, resulting in a two-step scheme, where
\begin{equation}
	\v{u}_{n+1} = 2 \v{u}_n - \v{u}_{n-1} + \Delta t^2 \ddot{\v{u}}_n \text{.}
	\label{eq:CDMstep}
\end{equation}
is referred to as the CDM sum step. The accelerations $\ddot{\v{u}}$ are computed from the spatially discretized equations of motion
\begin{equation}
	\v{M} \ddot{\v{u}}_n = f_\text{t}(t_n) \v{f}_\text{x} - \v{K} \v{u}_n \text{.}
	\label{eq:CDMacc}
\end{equation}
The computation of the acceleration together with the CDM sum step will hereafter be called CDM step. Since the mass matrix remains constant, it can be factorized once outside the time integration loop, e.g., with a Cholesky-decomposition. Inside the loop, the system of equations is solved using backward and forward substitution. For a diagonal mass matrix, this factorization is omitted since the system of equations can directly be solved by element-wise inversion of the diagonal mass matrix. The CDM is extensively used, e.g., in~\cite{Joulaian2014, Messmer2022, BKK22, BKK23}.

The CDM is highly efficient for uncut cells due to its moderate time step size constraint and the low computational cost per time step. However, the situation changes in the presence of cut cells, especially for small cuts. The very small critical time step size for explicit integration of cut cells significantly slows down the overall computation process. One way to overcome this limitation is to introduce substepping on cut cells, commonly referred to as leap-frog method~\cite{LB82, NAC22}. By introducing a ratio $m$, the cut cells are integrated in time with a finer time step size $\Delta t_{\text{fine}}$:
\begin{equation}
    m = \frac{\Delta t_{\text{coarse}}}{\Delta t_{\text{fine}}} \text{.}
\end{equation}
The parameter $\Delta t_{\text{coarse}}$ denotes the coarse time step size of the overall system. In this way, only $\Delta t_{\text{fine}}$ must satisfy the highly restrictive stability constraints for CDM of cut cells. Implementations details of the leap-frog method applied to the SCM can be found in~\cite{NAC22, meineMA}.

Another way to overcome the restrictive time step size limitation of explicit methods in the presence of cut cells is to use unconditionally stable implicit time integration schemes. When using implicit time integration, larger time step sizes can compensate the higher computational costs per time step. The increased computational costs arises from the necessity of solving a system of equations in each time step. One of the most frequently employed implicit methods is the Newmark trapezoidal method~\cite{Bathe}, which we now briefly discuss as a starting point for the suggested implicit explicit time integration scheme. 

The Newmark trapezoidal method is unconditionally stable, non-dissipative, and second-order accurate~\cite{GR14, Hughes2012}. It can be implemented as a predictor-corrector scheme and obtained from the general Newmark formulas (equations \eqref{eq:Newmark1} to \eqref{eq:Newmark2}) by setting $\beta = \frac{1}{4}$ and $\gamma = \frac{1}{2}$. \figref{fig:implicitNewmark} illustrates the procedure of the method. 

The implicit-explicit Newmark method composed of the implicit trapezoidal Newmark method and the explicit CDM combines the advantages of both methods in the context of the spectral cell method. Firstly, we employ the CDM for the uncut dofs associated with the diagonal mass matrix block. The acceleration of these diagonal dofs can directly be computed using Equation~\eqref{eq:CDMacc}, without the need to solve a system of equations. Subsequently, by incorporating this acceleration to the CDM sum step (Equation \eqref{eq:CDMstep}), we obtain the solution for the uncut dofs at the next time step. Secondly, as the cut dofs correspond to a non-diagonal block in the mass matrix and may cause a very small critical time step size, we integrate them in time using the implicit trapezoidal Newmark method. To predict the displacement and velocity in the trapezoidal Newmark method, we utilize the solution of the previous CDM step for the uncut dofs. In the remainder of this paper, the combination of both algorithms is referred to as the Newmark IMEX method. Its flowchart is depicted in~\figref{fig:TrapCDMIMEX}. The symbols $I_\text{d}$ and $I_\text{c}$ denote the index sets of the diagonal and cut dofs, respectively. Furthermore, note that an update in a dof-subgroup, e.g., $\v{u}^\text{d}$, is automatically assembled into the full vector $\v{u}$ at the corresponding indices $I_\text{d}$. \\


%% file: figures/NewmarkImplicitOverleaf.pdf_tex
\begingroup%
  \makeatletter%
  \providecommand\color[2][]{%
    \errmessage{(Inkscape) Color is used for the text in Inkscape, but the package 'color.sty' is not loaded}%
    \renewcommand\color[2][]{}%
  }%
  \providecommand\transparent[1]{%
    \errmessage{(Inkscape) Transparency is used (non-zero) for the text in Inkscape, but the package 'transparent.sty' is not loaded}%
    \renewcommand\transparent[1]{}%
  }%
  \providecommand\rotatebox[2]{#2}%
  \newcommand*\fsize{\dimexpr\f@size pt\relax}%
  \newcommand*\lineheight[1]{\fontsize{\fsize}{#1\fsize}\selectfont}%
  \ifx\svgwidth\undefined%
    \setlength{\unitlength}{734.00872709bp}%
    \ifx\svgscale\undefined%
      \relax%
    \else%
      \setlength{\unitlength}{\unitlength * \real{\svgscale}}%
    \fi%
  \else%
    \setlength{\unitlength}{\svgwidth}%
  \fi%
  \global\let\svgwidth\undefined%
  \global\let\svgscale\undefined%
  \makeatother%
  \begin{picture}(1,1.0418905)%
    \lineheight{1}%
    \setlength\tabcolsep{0pt}%
    \put(0.49489978,0.98100547){\color[rgb]{0,0,0}\makebox(0,0)[t]{\lineheight{1.25}\smash{\begin{tabular}[t]{c}$\v{M}, \v{K}, f_\text{t}, \v{f}_\text{x}, \v{u}_0, \dot{\v{u}}_0$\end{tabular}}}}%
    \put(0.49508605,1.01460261){\color[rgb]{0,0,0}\makebox(0,0)[t]{\lineheight{1.25}\smash{\begin{tabular}[t]{c}System description\end{tabular}}}}%
    \put(0.49192453,0.86497991){\color[rgb]{0,0,0}\makebox(0,0)[t]{\lineheight{1.25}\smash{\begin{tabular}[t]{c}$\v{M} \ddot{\v{u}}_0 = f_\text{t} \v{f}_\text{x} - \v{K} \v{u}_0$\\\end{tabular}}}}%
    \put(0.49537473,0.9014813){\color[rgb]{0,0,0}\makebox(0,0)[t]{\lineheight{1.25}\smash{\begin{tabular}[t]{c}Calculation of  initial acceleration\end{tabular}}}}%
    \put(0.49484392,0.65873711){\color[rgb]{0,0,0}\makebox(0,0)[t]{\lineheight{1.25}\smash{\begin{tabular}[t]{c}$t_{n+1} = t_{n} + \Delta t$\end{tabular}}}}%
    \put(0.49537473,0.68754915){\color[rgb]{0,0,0}\makebox(0,0)[t]{\lineheight{1.25}\smash{\begin{tabular}[t]{c}Time increment\end{tabular}}}}%
    \put(0.50000466,0.38720836){\color[rgb]{0,0,0}\makebox(0,0)[t]{\lineheight{1.25}\smash{\begin{tabular}[t]{c}$\v{u}_{\text{pred}} = \v{u}_n + \Delta t \dot{\v{u}}_n + (\frac{1}{2} - \beta) \Delta t^2 \ddot{\v{u}}_n  $\\$\dot{\v{u}}_{\text{pred}} = \dot{\v{u}}_n + (1-\gamma) \Delta t \ddot{\v{u}}_n $\\\end{tabular}}}}%
    \put(0.50053554,0.41754721){\color[rgb]{0,0,0}\makebox(0,0)[t]{\lineheight{1.25}\smash{\begin{tabular}[t]{c}Prediction\end{tabular}}}}%
    \put(0.49830201,0.24243387){\color[rgb]{0,0,0}\makebox(0,0)[t]{\lineheight{1.25}\smash{\begin{tabular}[t]{c}$\v{S} \ddot{\v{u}}_{n+1} = f_\text{t}(t_{n+1}) \v{f}_\text{x} - \v{K} \v{u}_{\text{pred}}$ \end{tabular}}}}%
    \put(0.49743902,0.27790166){\color[rgb]{0,0,0}\makebox(0,0)[t]{\lineheight{1.25}\smash{\begin{tabular}[t]{c}Acceleration calculation\end{tabular}}}}%
    \put(0,0){\includegraphics[width=\unitlength,page=1]{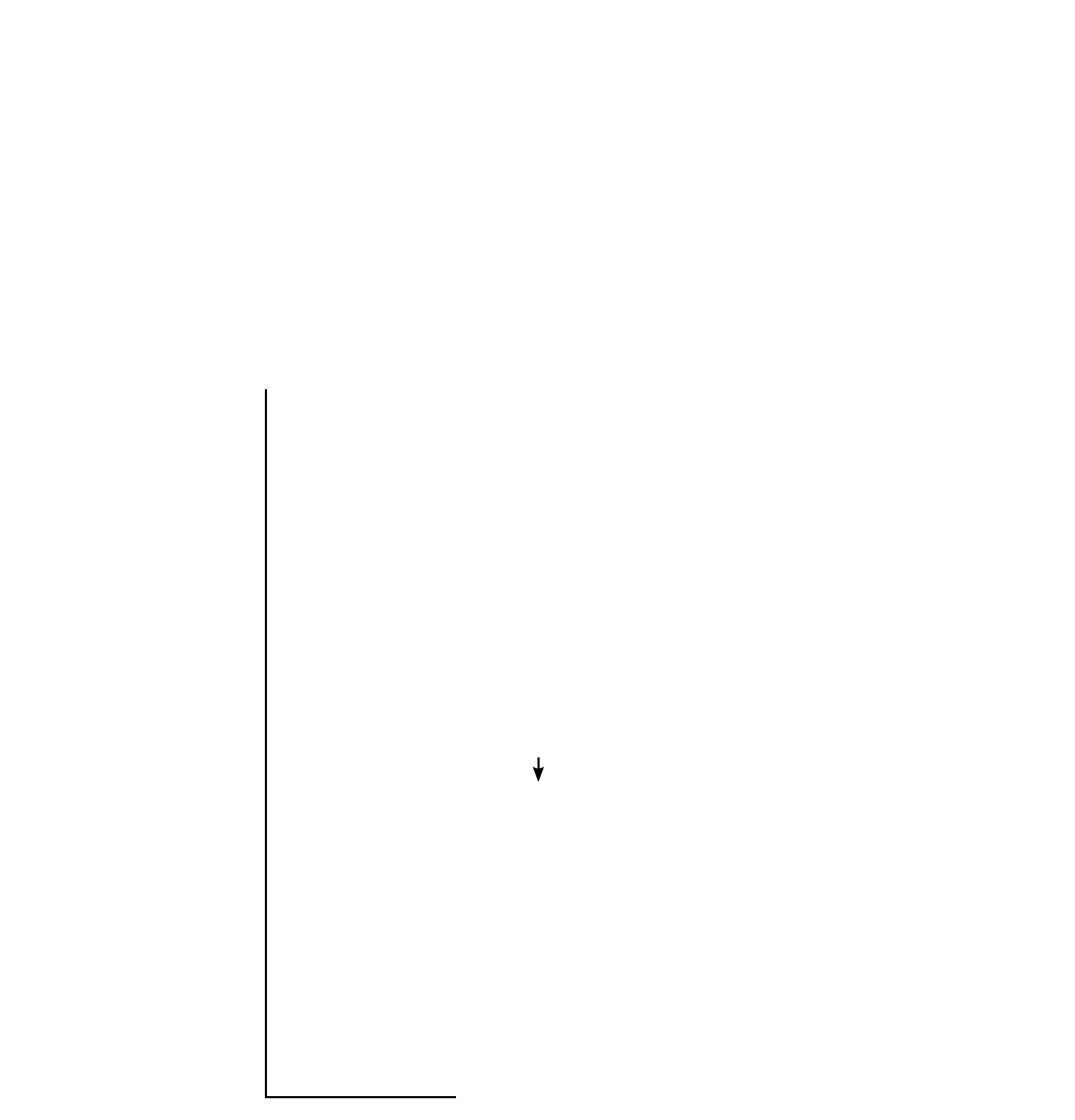}}%
    \put(0.50103688,0.12530594){\color[rgb]{0,0,0}\makebox(0,0)[t]{\lineheight{1.25}\smash{\begin{tabular}[t]{c}$\v{u}_{n+1} = \v{u}_{\text{pred}} + \beta \Delta t^2 \ddot{\v{u}}_{n+1}$\\$\dot{\v{u}}_{n+1} = \dot{\v{u}}_{\text{pred}} + \gamma \Delta t \ddot{\v{u}}_{n+1}$\end{tabular}}}}%
    \put(0.50156773,0.15596212){\color[rgb]{0,0,0}\makebox(0,0)[t]{\lineheight{1.25}\smash{\begin{tabular}[t]{c}Correction\end{tabular}}}}%
    \put(0,0){\includegraphics[width=\unitlength,page=2]{figures/NewmarkImplicit.pdf}}%
    \put(0.49601433,0.76002982){\color[rgb]{0,0,0}\makebox(0,0)[t]{\lineheight{1.25}\smash{\begin{tabular}[t]{c}$\v{S} = \v{M} + \beta \Delta t^2 \v{K}$\end{tabular}}}}%
    \put(0.49654515,0.7888419){\color[rgb]{0,0,0}\makebox(0,0)[t]{\lineheight{1.25}\smash{\begin{tabular}[t]{c}Calculation and factorization of system matrix\end{tabular}}}}%
    \put(0,0){\includegraphics[width=\unitlength,page=3]{figures/NewmarkImplicit.pdf}}%
    \put(0.49583879,0.54169768){\color[rgb]{0,0,0}\makebox(0,0)[t]{\lineheight{1.25}\smash{\begin{tabular}[t]{c}$t_{n+1} \leq T?$\\\end{tabular}}}}%
    \put(0,0){\includegraphics[width=\unitlength,page=4]{figures/NewmarkImplicit.pdf}}%
    \put(0.50249653,0.01529944){\color[rgb]{0,0,0}\makebox(0,0)[t]{\lineheight{1.25}\smash{\begin{tabular}[t]{c}$n = n+1$\\\end{tabular}}}}%
    \put(0,0){\includegraphics[width=\unitlength,page=5]{figures/NewmarkImplicit.pdf}}%
    \put(0.51139239,0.45937692){\color[rgb]{0,0,0}\makebox(0,0)[lt]{\lineheight{1.25}\smash{\begin{tabular}[t]{l}yes\end{tabular}}}}%
    \put(0,0){\includegraphics[width=\unitlength,page=6]{figures/NewmarkImplicit.pdf}}%
    \put(0.65073612,0.51981309){\color[rgb]{0,0,0}\makebox(0,0)[lt]{\lineheight{1.25}\smash{\begin{tabular}[t]{l}no\end{tabular}}}}%
    \put(0.725052,0.537107){\color[rgb]{0,0,0}\makebox(0,0)[t]{\lineheight{1.25}\smash{\begin{tabular}[t]{c}\text{end}\\\end{tabular}}}}%
    \put(0,0){\includegraphics[width=\unitlength,page=7]{figures/NewmarkImplicit.pdf}}%
  \end{picture}%
\endgroup%

%% file: figures/CDMTrapImexOverleaf.pdf_tex
\begingroup%
  \makeatletter%
  \providecommand\color[2][]{%
    \errmessage{(Inkscape) Color is used for the text in Inkscape, but the package 'color.sty' is not loaded}%
    \renewcommand\color[2][]{}%
  }%
  \providecommand\transparent[1]{%
    \errmessage{(Inkscape) Transparency is used (non-zero) for the text in Inkscape, but the package 'transparent.sty' is not loaded}%
    \renewcommand\transparent[1]{}%
  }%
  \providecommand\rotatebox[2]{#2}%
  \newcommand*\fsize{\dimexpr\f@size pt\relax}%
  \newcommand*\lineheight[1]{\fontsize{\fsize}{#1\fsize}\selectfont}%
  \ifx\svgwidth\undefined%
    \setlength{\unitlength}{792.33298045bp}%
    \ifx\svgscale\undefined%
      \relax%
    \else%
      \setlength{\unitlength}{\unitlength * \real{\svgscale}}%
    \fi%
  \else%
    \setlength{\unitlength}{\svgwidth}%
  \fi%
  \global\let\svgwidth\undefined%
  \global\let\svgscale\undefined%
  \makeatother%
  \begin{picture}(1,1.17383345)%
    \lineheight{1}%
    \setlength\tabcolsep{0pt}%
    \put(0.49017901,1.11539974){\color[rgb]{0,0,0}\makebox(0,0)[t]{\lineheight{1.25}\smash{\begin{tabular}[t]{c}$\v{M}, \v{K}, f_\text{t}, \v{f}_\text{x}, \v{u}_0, I_\text{d}, I_\text{c}$\end{tabular}}}}%
    \put(0.49035174,1.14652377){\color[rgb]{0,0,0}\makebox(0,0)[t]{\lineheight{1.25}\smash{\begin{tabular}[t]{c}System description\end{tabular}}}}%
    \put(0.48742289,1.01170568){\color[rgb]{0,0,0}\makebox(0,0)[t]{\lineheight{1.25}\smash{\begin{tabular}[t]{c}$\v{M}^\text{cc} \ddot{\v{u}}_0^\text{c} = f_\text{t}(0) \v{f}_\text{x} - \v{K}^\text{c} \v{u}_0$\\\end{tabular}}}}%
    \put(0.49061901,1.04552017){\color[rgb]{0,0,0}\makebox(0,0)[t]{\lineheight{1.25}\smash{\begin{tabular}[t]{c}Calculation of  initial acceleration for cut dofs \end{tabular}}}}%
    \put(0.49017923,0.81530355){\color[rgb]{0,0,0}\makebox(0,0)[t]{\lineheight{1.25}\smash{\begin{tabular}[t]{c}$t_{n+1} = t_{n} + \Delta t$\end{tabular}}}}%
    \put(0.48946018,0.84199472){\color[rgb]{0,0,0}\makebox(0,0)[t]{\lineheight{1.25}\smash{\begin{tabular}[t]{c}Time increment\end{tabular}}}}%
    \put(0.5000043,0.41523317){\color[rgb]{0,0,0}\makebox(0,0)[t]{\lineheight{1.25}\smash{\begin{tabular}[t]{c}$\v{u}_\text{pred}^\text{d} = \v{u}^\text{d}_{n+1}$\\$\v{u}_\text{pred}^\text{c} = \v{u}_n + \Delta t \dot{\v{u}}_n^\text{c} + (\frac{1}{2} - \beta) \Delta t^2 \ddot{\v{u}}_n^\text{c}  $\\$\dot{\v{u}}_\text{pred}^\text{c} = \dot{\v{u}}_n^\text{c} + (1-\gamma) \Delta t \ddot{\v{u}}_n^\text{c} $\\\end{tabular}}}}%
    \put(0.49698417,0.44716348){\color[rgb]{0,0,0}\makebox(0,0)[t]{\lineheight{1.25}\smash{\begin{tabular}[t]{c}Prediction\end{tabular}}}}%
    \put(0.49873978,0.24016389){\color[rgb]{0,0,0}\makebox(0,0)[t]{\lineheight{1.25}\smash{\begin{tabular}[t]{c}$\v{S} \ddot{\v{u}}_{n+1}^\text{c} = f_\text{t}(t_{n+1}) \v{f}_\text{x}^\text{c} - \v{K}^\text{c} \v{u}_\text{pred}$ \end{tabular}}}}%
    \put(0.49794037,0.273977){\color[rgb]{0,0,0}\makebox(0,0)[t]{\lineheight{1.25}\smash{\begin{tabular}[t]{c}Acceleration calculation\end{tabular}}}}%
    \put(0,0){\includegraphics[width=\unitlength,page=1]{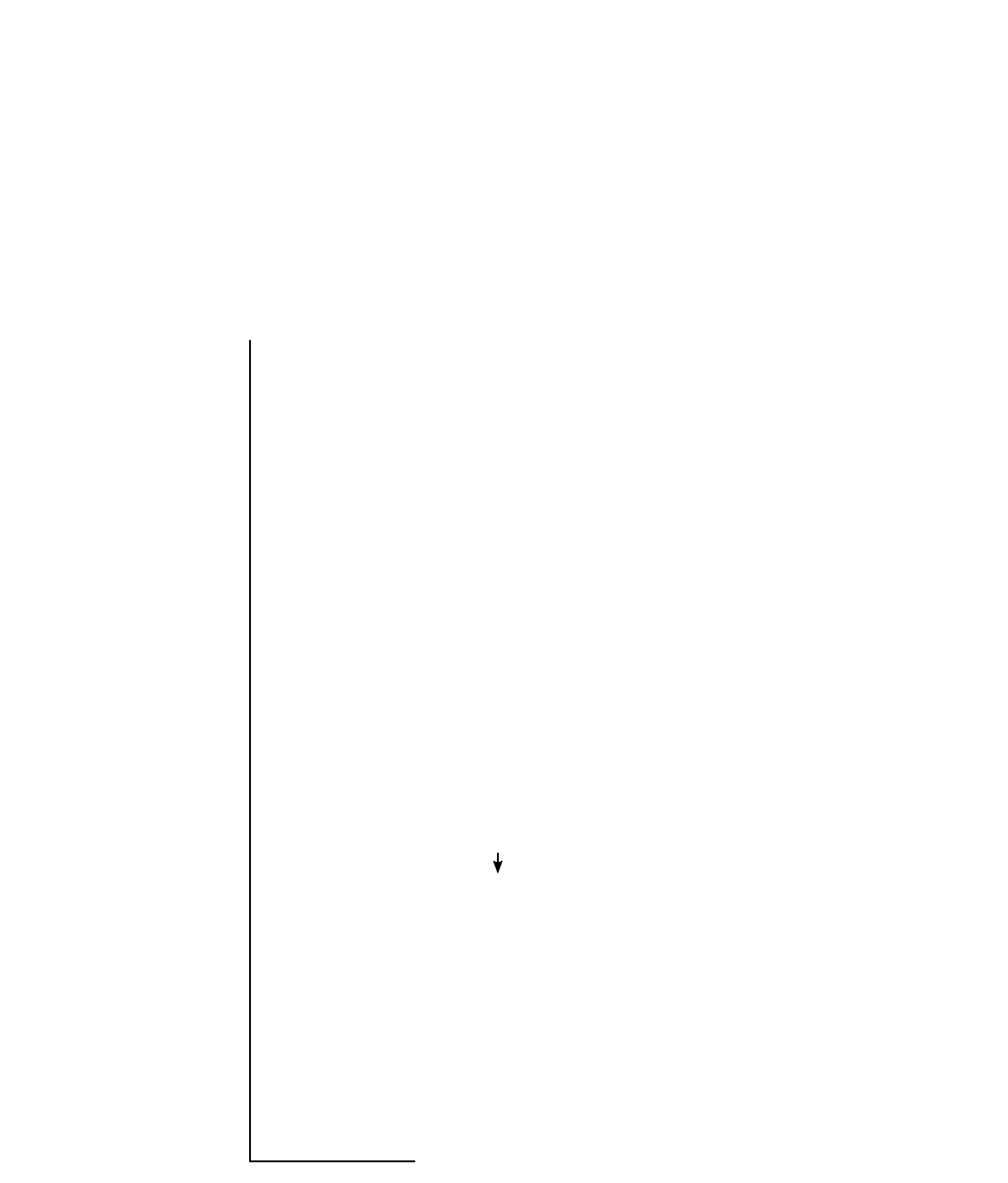}}%
    \put(0.50127344,0.1273111){\color[rgb]{0,0,0}\makebox(0,0)[t]{\lineheight{1.25}\smash{\begin{tabular}[t]{c}$\v{u}_{n+1}^\text{c} = \v{u}_\text{pred}^\text{c} + \beta \Delta t^2 \ddot{\v{u}}_{n+1}^\text{c}$\\$\dot{\v{u}}_{n+1}^\text{c} = \dot{\v{u}}_\text{pred}^\text{c} + \gamma \Delta t \ddot{\v{u}}_{n+1}^\text{c}$\end{tabular}}}}%
    \put(0.50176521,0.15571065){\color[rgb]{0,0,0}\makebox(0,0)[t]{\lineheight{1.25}\smash{\begin{tabular}[t]{c}Correction\end{tabular}}}}%
    \put(0,0){\includegraphics[width=\unitlength,page=2]{figures/CDMTrapImex.pdf}}%
    \put(0.49121163,0.90981615){\color[rgb]{0,0,0}\makebox(0,0)[t]{\lineheight{1.25}\smash{\begin{tabular}[t]{c}$\v{S} = \v{M}^\text{cc} + \beta \Delta t^2 \v{K}^\text{cc}$\end{tabular}}}}%
    \put(0.49170338,0.94128826){\color[rgb]{0,0,0}\makebox(0,0)[t]{\lineheight{1.25}\smash{\begin{tabular}[t]{c}Calculation and factorization of system matrix\end{tabular}}}}%
    \put(0,0){\includegraphics[width=\unitlength,page=3]{figures/CDMTrapImex.pdf}}%
    \put(0.49104896,0.70750757){\color[rgb]{0,0,0}\makebox(0,0)[t]{\lineheight{1.25}\smash{\begin{tabular}[t]{c}$t_{n+1} \leq T?$\\\end{tabular}}}}%
    \put(0,0){\includegraphics[width=\unitlength,page=4]{figures/CDMTrapImex.pdf}}%
    \put(0.50748615,0.63348695){\color[rgb]{0,0,0}\makebox(0,0)[lt]{\lineheight{1.25}\smash{\begin{tabular}[t]{l}yes\end{tabular}}}}%
    \put(0,0){\includegraphics[width=\unitlength,page=5]{figures/CDMTrapImex.pdf}}%
    \put(0.63358804,0.68293107){\color[rgb]{0,0,0}\makebox(0,0)[lt]{\lineheight{1.25}\smash{\begin{tabular}[t]{l}no\end{tabular}}}}%
    \put(0.49396798,0.55919442){\color[rgb]{0,0,0}\makebox(0,0)[t]{\lineheight{1.25}\smash{\begin{tabular}[t]{c}$\ddot{\v{u}}^\text{d}_{n} = (\v{M}^\text{dd})^{-1} (f_\text{t}(t_n) \v{f}_\text{x}^d - \v{K}^\text{d} \v{u}_n  $\\$\v{u}^\text{d}_{n+1} = 2\v{u}^\text{d}_n - \v{u}^\text{d}_{n-1} + \Delta t^2 \ddot{\v{u}}^\text{d}_{n} $\\\end{tabular}}}}%
    \put(0.4987876,0.59567367){\color[rgb]{0,0,0}\makebox(0,0)[t]{\lineheight{1.25}\smash{\begin{tabular}[t]{c}Diagonal CDM step\end{tabular}}}}%
    \put(0,0){\includegraphics[width=\unitlength,page=6]{figures/CDMTrapImex.pdf}}%
    \put(-1.74966306,1.53466131){\color[rgb]{0,0,0}\makebox(0,0)[lt]{\begin{minipage}{0.19334445\unitlength}\centering \end{minipage}}}%
    \put(-1.08210377,-0.45776136){\color[rgb]{0,0,0}\makebox(0,0)[lt]{\begin{minipage}{0.33731597\unitlength}\centering \end{minipage}}}%
    \put(0.70331444,0.70210452){\color[rgb]{0,0,0}\makebox(0,0)[t]{\lineheight{1.25}\smash{\begin{tabular}[t]{c}end\\\end{tabular}}}}%
    \put(0,0){\includegraphics[width=\unitlength,page=7]{figures/CDMTrapImex.pdf}}%
    \put(0.49396798,0.01417313){\color[rgb]{0,0,0}\makebox(0,0)[t]{\lineheight{1.25}\smash{\begin{tabular}[t]{c}$n = n+1$\\\end{tabular}}}}%
    \put(0,0){\includegraphics[width=\unitlength,page=8]{figures/CDMTrapImex.pdf}}%
  \end{picture}%
\endgroup%

%% file: content/lumping.tex
We now introduce a lumped scheme to compare it with our implicit-explicit time integration scheme.
A fully diagonal mass matrix that is favorable for explicit time integration is obtained by additionally lumping the cut elements. In the works of~\cite{Duczek2014, Joulaian2014, Mossaiby2019}, Hinton-Rock-Zienkiewicz (HRZ) lumping~\cite{HRZ76} is employed to approximate the consistent element mass matrices of cut cells using diagonally lumped element mass matrices. In HRZ lumping, the diagonal terms of the cut element mass matrix are scaled to preserve its total mass, while off-diagonal entries are set to \textit{zero}, i.e.,
\begin{equation}
	\tilde{M}_{ii}^e = \frac{m_e}{\sum_{k=1}^{n_e} M_{kk}^e} M_{ii}^e
\end{equation}
and
\begin{equation}
	\tilde{M}_{ij}^e = 0 \qquad \text{if } i \neq j \text{.}
\end{equation}
According to~\cite{DG19}, HRZ lumping leads to a reduction of the convergence rates. On the contrary, unlike row-summing (see~\cite{Hughes2012, Voet2023}), it guarantees positive diagonal components and captures the material distribution in the cell more accurately. In the paper at hand, the CDM is used when HRZ lumping is applied for the cut cells. Nicoli et al.~\cite{NAC22, Nicoli2023} lump the cut cells by using moment fitting to adjust the integration weights while retaining the GLL points as integration points. On this lumped SCM system, they integrate in time using a leap-frog algorithm with smaller time step sizes in the cut cells. Another way to mitigate the time step size constraint of SCM, is the use of eigenvalue stabilization, which is investigated in~\cite{Eisentraeger2023}. A comprehensive comparative study of various lumping techniques applied to the SCM is given in~\cite{K21}. From now on, the SCM with no further lumping on the cut cells is referred to as consistent SCM, and the SCM with additional HRZ lumping on the cut cells is denoted as HRZ-lumped SCM.

%% file: content/SpringMasses.tex
\label{subsec:1D}

\figref{fig:springCoupledMasses} shows a system of ten spring-coupled masses that we analyze in this section to discuss key attributes concerning the accuracy and stability of the Newmark IMEX method. The results transfer to systems obtained from spatial discretizations of the wave equation, such as the SCM.
%
\begin{figure}[H]%
    \centering%
	\def\svgwidth{1.\textwidth}%
	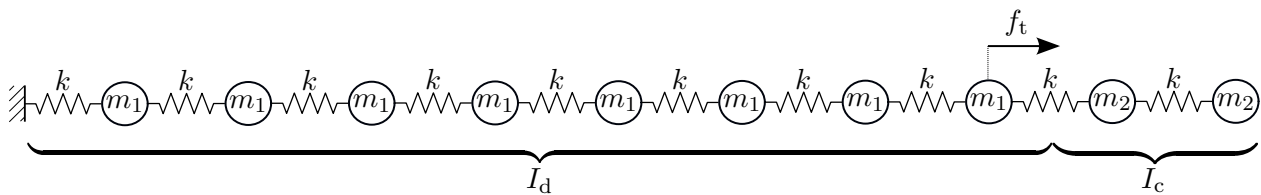%
    \caption{System composed of ten spring-coupled masses}%
	\label{fig:springCoupledMasses}%
\end{figure}%
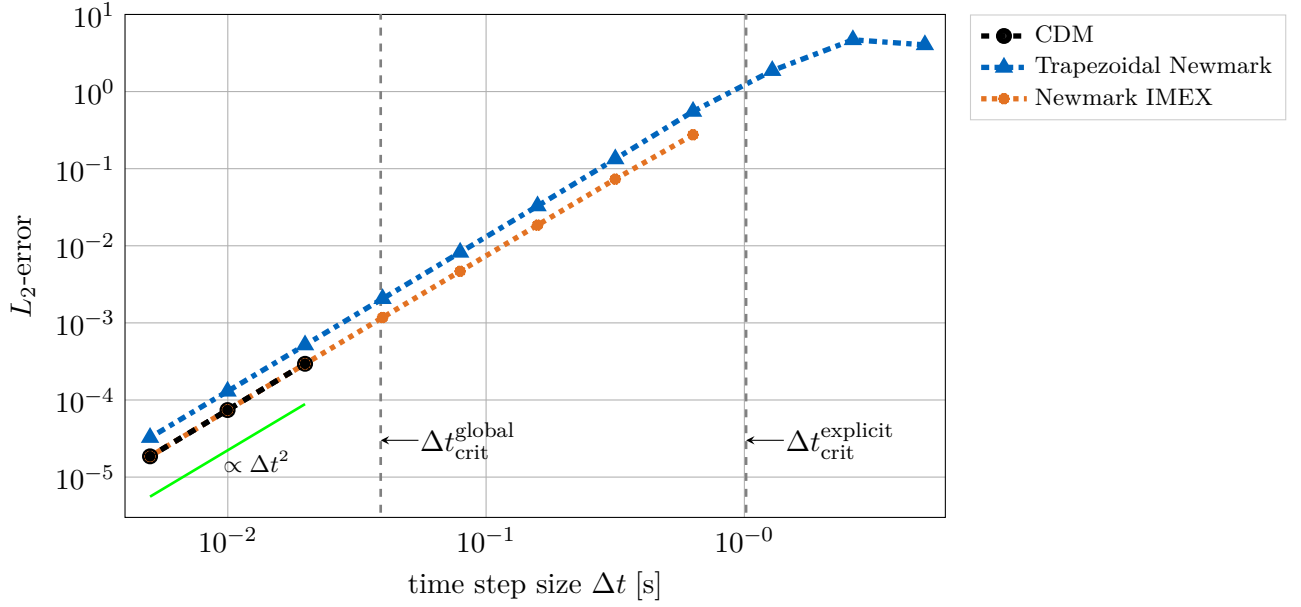
\begin{figure}[t!]
	\centering
		\input{tikzpictures/convergenceSpringMasses.tex}
	\caption{Convergence curves for time integration of the spring-coupled masses with critical time step sizes indicated}
	\label{fig:springCoupledMassesConvergence}
\end{figure}
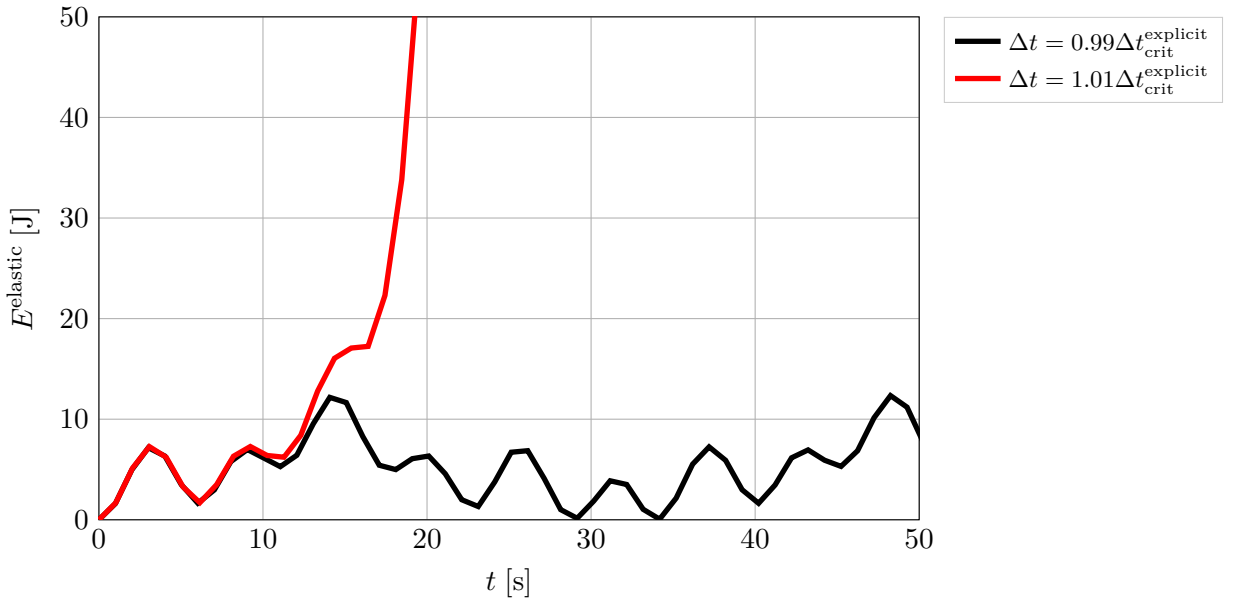
\begin{figure}[b!]
	\centering
		\input{tikzpictures/stabilitySpringMasses.tex}
	\caption{Elastic energy for the Newmark IMEX method with a time step size smaller and higher than the critical time step size of explicit subsystem}
	\label{fig:springCoupledMassesElastic}
\end{figure}
\noindent We assign the first eight masses the value $m_\text{1} = \SI{1}{\kilo\gram}$ and the remaining two the value $m_\text{2} = \SI{e-3}{\kilo\gram}$. The stiffness of all springs is $k = \SI{1}{\newton\per\meter}$. The system is subjected to a harmonic excitation
\begin{equation}
    f_\text{t}(t) = \sin \left( 2 \pi f_\text{s} t \right)
\end{equation}
applied to the third last mass with an excitation frequency of $f_\text{s} = \SI{e-1}{\hertz}$. The equations of motion for all masses can be written as a single second-order system
\begin{equation}\label{eq:springCoupledSystem}
    \v{M} \ddot{\dv{u}} + \v{K} \dv{u} = \v{f} \text{,}
\end{equation}
where $\dv{u}$ are the dofs of the system, representing the displacements of the point masses. The solution of the corresponding Helmholtz equation yields an amplitude vector whose product with the harmonic excitation $f_\text{t}(t)$ forms our time-dependent reference solution. The equation system \eqref{eq:springCoupledSystem} is similar to \eqref{eq:discretizedWaveEquation}, and its structure is identical to a mass-lumped SCM discretization of the wave equation. We emphasize this similarity by referring to the first eight masses as uncut dofs and the remaining two as badly cut dofs. The behavior of the badly cut dofs dominates the stability of the global system.

In the following, we compare the convergence of the explicit CDM, implicit trapezoidal Newmark, and Newmark IMEX time-integration schemes on the time interval $T = \SI{50}{\second}$. The Newmark IMEX method integrates the eight uncut dofs explicitly and the remaining two badly cut dofs implicitly. Our convergence analysis starts with an initial $N_\text{t} = 10$ time steps, resulting in a $\Delta t$ of $\SI{5}{\second}$, and subsequently logarithmically refines the temporal discretization until we reach $N_\text{t} = 10^4$ time steps with a $\Delta t$ of $\SI{5e-3}{\second}$. The initial displacements and velocities are determined from the reference solution outlined above.

\figref{fig:springCoupledMassesConvergence} compares the convergence of the three methods, using the $l_2$-norm of the difference between the computed and reference displacements as an error measure. The green line indicates second-order convergence, the vertical dashed line on the left highlights the critical time step size $\Delta t_\text{crit}^\text{global} = \SI{3.9086e-2}{\second}$ of the global system, and the dashed line on the right highlights the critical time step size $\Delta t_\text{crit}^\text{explicit} = \SI{1.0154}{\second}$ of the explicit subsystem. The CDM and the trapezoidal Newmark method show the expected second-order convergence, with a slightly lower convergence constant for the CDM. The CDM becomes unstable beyond the global critical time step size $\Delta t_\text{crit}^\text{global}$, resulting in no data points to the right of the corresponding dashed line. In contrast, the trapezoidal Newmark method is unconditionally stable. The Newmark IMEX method preserves the convergence rates of the CDM and trapezoidal Newmark methods, which are its basic building blocks. Here, the convergence constant even reaches that of the CDM. As expected, the Newmark IMEX method remains stable up to the critical time step size $\Delta t_\text{crit}^\text{explicit}$ of the explicit subsystem (corresponding to the uncut dofs), which is much less restrictive than $\Delta t_\text{crit}^\text{global}$. 


\figref{fig:springCoupledMassesElastic} shows the evolution of the elastic energy $E^\text{elastic} = \frac{1}{2} \; \dv{u}^\text{T} \v{K} \dv{u}$ for two Newmark IMEX simulations with a time step size slightly below and slightly above the critical time step size of the explicit subsystem. The elastic energy for $\Delta t = 0.99 \; \Delta t_\text{crit}^\text{explicit}$, plotted in black, is stable and remains bounded. However, as the red curve shows, a time step size slightly larger than $\Delta t_\text{crit}^\text{explicit}$ leads to an unstable simulation with an unboundedly growing elastic energy. We refer to~\cite{H78_stability, H78, H81} for further discussion of the theory, derivations, and examples of implicit-explicit time integration.

%% file: figures/1DexampleOverleaf.pdf_tex
\begingroup%
  \makeatletter%
  \providecommand\color[2][]{%
    \errmessage{(Inkscape) Color is used for the text in Inkscape, but the package 'color.sty' is not loaded}%
    \renewcommand\color[2][]{}%
  }%
  \providecommand\transparent[1]{%
    \errmessage{(Inkscape) Transparency is used (non-zero) for the text in Inkscape, but the package 'transparent.sty' is not loaded}%
    \renewcommand\transparent[1]{}%
  }%
  \providecommand\rotatebox[2]{#2}%
  \newcommand*\fsize{\dimexpr\f@size pt\relax}%
  \newcommand*\lineheight[1]{\fontsize{\fsize}{#1\fsize}\selectfont}%
  \ifx\svgwidth\undefined%
    \setlength{\unitlength}{894.41655304bp}%
    \ifx\svgscale\undefined%
      \relax%
    \else%
      \setlength{\unitlength}{\unitlength * \real{\svgscale}}%
    \fi%
  \else%
    \setlength{\unitlength}{\svgwidth}%
  \fi%
  \global\let\svgwidth\undefined%
  \global\let\svgscale\undefined%
  \makeatother%
  \begin{picture}(1,0.14065287)%
    \lineheight{1}%
    \setlength\tabcolsep{0pt}%
    \put(0,0){\includegraphics[width=\unitlength,page=1]{figures/1DexampleOverleaf.pdf}}%
    \put(0.02341216,0.09803207){\color[rgb]{0,0,0}\makebox(0,0)[lt]{\begin{minipage}{0.03896425\unitlength}\centering $k$\end{minipage}}}%
    \put(0,0){\includegraphics[width=\unitlength,page=2]{figures/1DexampleOverleaf.pdf}}%
    \put(0.09255527,0.06753991){\color[rgb]{0,0,0}\makebox(0,0)[t]{\lineheight{1.25}\smash{\begin{tabular}[t]{c}$m_1$\\\end{tabular}}}}%
    \put(3.22258445,-0.42327778){\color[rgb]{0,0,0}\makebox(0,0)[lt]{\begin{minipage}{0.12677072\unitlength}\centering \end{minipage}}}%
    \put(0,0){\includegraphics[width=\unitlength,page=3]{figures/1DexampleOverleaf.pdf}}%
    \put(0.80364699,0.13045942){\color[rgb]{0,0,0}\makebox(0,0)[t]{\lineheight{1.25}\smash{\begin{tabular}[t]{c}$f_\text{t}$\end{tabular}}}}%
    \put(0,0){\includegraphics[width=\unitlength,page=4]{figures/1DexampleOverleaf.pdf}}%
    \put(0.12184094,0.09824761){\color[rgb]{0,0,0}\makebox(0,0)[lt]{\begin{minipage}{0.03896425\unitlength}\centering $k$\end{minipage}}}%
    \put(0,0){\includegraphics[width=\unitlength,page=5]{figures/1DexampleOverleaf.pdf}}%
    \put(0.19098405,0.06775545){\color[rgb]{0,0,0}\makebox(0,0)[t]{\lineheight{1.25}\smash{\begin{tabular}[t]{c}$m_1$\\\end{tabular}}}}%
    \put(0,0){\includegraphics[width=\unitlength,page=6]{figures/1DexampleOverleaf.pdf}}%
    \put(0.22026972,0.09846317){\color[rgb]{0,0,0}\makebox(0,0)[lt]{\begin{minipage}{0.03896425\unitlength}\centering $k$\end{minipage}}}%
    \put(0,0){\includegraphics[width=\unitlength,page=7]{figures/1DexampleOverleaf.pdf}}%
    \put(0.28941282,0.067971){\color[rgb]{0,0,0}\makebox(0,0)[t]{\lineheight{1.25}\smash{\begin{tabular}[t]{c}$m_1$\\\end{tabular}}}}%
    \put(0,0){\includegraphics[width=\unitlength,page=8]{figures/1DexampleOverleaf.pdf}}%
    \put(0.31869849,0.09867871){\color[rgb]{0,0,0}\makebox(0,0)[lt]{\begin{minipage}{0.03896425\unitlength}\centering $k$\end{minipage}}}%
    \put(0,0){\includegraphics[width=\unitlength,page=9]{figures/1DexampleOverleaf.pdf}}%
    \put(0.3878416,0.06818655){\color[rgb]{0,0,0}\makebox(0,0)[t]{\lineheight{1.25}\smash{\begin{tabular}[t]{c}$m_1$\\\end{tabular}}}}%
    \put(0,0){\includegraphics[width=\unitlength,page=10]{figures/1DexampleOverleaf.pdf}}%
    \put(0.41712726,0.09889426){\color[rgb]{0,0,0}\makebox(0,0)[lt]{\begin{minipage}{0.03896425\unitlength}\centering $k$\end{minipage}}}%
    \put(0,0){\includegraphics[width=\unitlength,page=11]{figures/1DexampleOverleaf.pdf}}%
    \put(0.48627038,0.0684021){\color[rgb]{0,0,0}\makebox(0,0)[t]{\lineheight{1.25}\smash{\begin{tabular}[t]{c}$m_1$\\\end{tabular}}}}%
    \put(0,0){\includegraphics[width=\unitlength,page=12]{figures/1DexampleOverleaf.pdf}}%
    \put(0.51555607,0.09910981){\color[rgb]{0,0,0}\makebox(0,0)[lt]{\begin{minipage}{0.03896425\unitlength}\centering $k$\end{minipage}}}%
    \put(0,0){\includegraphics[width=\unitlength,page=13]{figures/1DexampleOverleaf.pdf}}%
    \put(0.58469916,0.06861765){\color[rgb]{0,0,0}\makebox(0,0)[t]{\lineheight{1.25}\smash{\begin{tabular}[t]{c}$m_1$\\\end{tabular}}}}%
    \put(0,0){\includegraphics[width=\unitlength,page=14]{figures/1DexampleOverleaf.pdf}}%
    \put(0.61398482,0.09932536){\color[rgb]{0,0,0}\makebox(0,0)[lt]{\begin{minipage}{0.03896425\unitlength}\centering $k$\end{minipage}}}%
    \put(0,0){\includegraphics[width=\unitlength,page=15]{figures/1DexampleOverleaf.pdf}}%
    \put(0.68312793,0.0688332){\color[rgb]{0,0,0}\makebox(0,0)[t]{\lineheight{1.25}\smash{\begin{tabular}[t]{c}$m_1$\\\end{tabular}}}}%
    \put(0,0){\includegraphics[width=\unitlength,page=16]{figures/1DexampleOverleaf.pdf}}%
    \put(0.71241363,0.09954091){\color[rgb]{0,0,0}\makebox(0,0)[lt]{\begin{minipage}{0.03896425\unitlength}\centering $k$\end{minipage}}}%
    \put(0,0){\includegraphics[width=\unitlength,page=17]{figures/1DexampleOverleaf.pdf}}%
    \put(0.78155671,0.06904875){\color[rgb]{0,0,0}\makebox(0,0)[t]{\lineheight{1.25}\smash{\begin{tabular}[t]{c}$m_1$\\\end{tabular}}}}%
    \put(0,0){\includegraphics[width=\unitlength,page=18]{figures/1DexampleOverleaf.pdf}}%
    \put(0.81084237,0.09975646){\color[rgb]{0,0,0}\makebox(0,0)[lt]{\begin{minipage}{0.03896425\unitlength}\centering $k$\end{minipage}}}%
    \put(0,0){\includegraphics[width=\unitlength,page=19]{figures/1DexampleOverleaf.pdf}}%
    \put(0.87998549,0.0692643){\color[rgb]{0,0,0}\makebox(0,0)[t]{\lineheight{1.25}\smash{\begin{tabular}[t]{c}$m_2$\\\end{tabular}}}}%
    \put(0,0){\includegraphics[width=\unitlength,page=20]{figures/1DexampleOverleaf.pdf}}%
    \put(0.90927115,0.09997201){\color[rgb]{0,0,0}\makebox(0,0)[lt]{\begin{minipage}{0.03896425\unitlength}\centering $k$\end{minipage}}}%
    \put(0,0){\includegraphics[width=\unitlength,page=21]{figures/1DexampleOverleaf.pdf}}%
    \put(0.97841426,0.06947985){\color[rgb]{0,0,0}\makebox(0,0)[t]{\lineheight{1.25}\smash{\begin{tabular}[t]{c}$m_2$\\\end{tabular}}}}%
    \put(0,0){\includegraphics[width=\unitlength,page=22]{figures/1DexampleOverleaf.pdf}}%
    \put(0.42311605,0.00316416){\color[rgb]{0,0,0}\makebox(0,0)[t]{\lineheight{1.25}\smash{\begin{tabular}[t]{c}$I_\text{d}$\end{tabular}}}}%
    \put(0.91259377,0.00316416){\color[rgb]{0,0,0}\makebox(0,0)[t]{\lineheight{1.25}\smash{\begin{tabular}[t]{c}$I_\text{c}$\end{tabular}}}}%
  \end{picture}%
\endgroup%

%% file: tikzpictures/convergenceSpringMasses.tex
\definecolor{TumBlue}{RGB}{0,101,189} 
\definecolor{Orange}{RGB}{227,114,34} 
\definecolor{lightgray204}{RGB}{204,204,204}
\definecolor{darkgray176}{RGB}{176,176,176}
\begin{tikzpicture}
	\begin{loglogaxis}[
		xmin = 4e-3, xmax = 6e0,
		ymin = 3e-6, ymax = 1e1,
        xtick style={color=black},
		xtick={0.01, 0.1, 1},
		xticklabels = {$10^{-2}$, $10^{-1}$, $10^{-0}$},
		ytick={0.00001, 0.0001, 0.001, 0.01, 0.1, 1, 10},
        x grid style={darkgray176},
        y grid style={darkgray176},
        xmajorgrids,
        ymajorgrids,
        xtick style={draw=none},
        ytick style={draw=none},
		width = 0.75\textwidth,
		height = 0.5\textwidth,
		xlabel = {time step size $\Delta t \; [\si{\second}]$},
		ylabel style={align=center}, ylabel=$L_2$-error,
        legend cell align={left},
		legend style={
          fill opacity=0.8,
          draw opacity=1,
          text opacity=1,
          at={(0.97,0.03)},
          anchor=south east,
          draw=lightgray204
        },
		legend pos = outer north east,
        legend style={font=\footnotesize}
		]
		
		\addplot[black, line width = 2.0pt, mark = *, mark size=2, mark options={solid}, dashed] file[] {tikzpictures/data/convergenceCDM.dat};
		\addplot[TumBlue, line width = 2.0pt, mark = triangle*, mark size=2, mark options={solid}, dashdotted] file[] {tikzpictures/data/convergenceTrapezoidal.dat};
		\addplot[Orange, line width = 2.0pt, mark = 10-pointed star, mark size=2, mark options={solid}, dotted] file[] {tikzpictures/data/convergenceIMEX.dat};
		
		\addplot[green, line width=1.0pt] file[] {tikzpictures/data/referenceQuadratic.dat};
		
		\addplot[line width=1pt, gray, dashed] file[] {tikzpictures/data/limitGlobal.dat};
        \draw[{stealth[width=5mm]}-] (0.03908584, 0.00003) -- (0.05472017599999999, 0.00003);
        \node[align=left] at (0.085, 0.00003) {$\Delta t_\text{crit}^\text{global}$};
        
		\addplot[line width=1pt, gray, dashed] file[] {tikzpictures/data/limitExplicit.dat};
        \draw[{stealth[width=5mm]}-] (1.01542661, 0.00003) -- (1.421597254, 0.00003);
        \node[align=left] at (2.35, 0.00003) {$\Delta t_\text{crit}^\text{explicit}$};
        
        \addplot[black, line width = 2.0pt, mark = *, mark size=1, mark options={solid}, dashed] file[] {tikzpictures/data/convergenceCDM.dat};
		
		\legend{CDM, Trapezoidal Newmark, Newmark IMEX},
	\end{loglogaxis}
    \node[] at (0.105\textwidth,0.045\textwidth) {\footnotesize$\propto \Delta t^2$};
\end{tikzpicture}%

%% file: tikzpictures/stabilitySpringMasses.tex
\definecolor{TumBlue}{RGB}{0,101,189} 
\definecolor{Orange}{RGB}{227,114,34} 
\definecolor{lightgray204}{RGB}{204,204,204}
\definecolor{darkgray176}{RGB}{176,176,176}
\begin{tikzpicture}
	\begin{axis}[
		xmin = 0, xmax = 50,
		ymin = 0, ymax = 50,
        xtick style={color=black},
		xtick={0, 10, 20, 30, 40, 50},
		ytick={0, 10, 20, 30, 40, 50},
        x grid style={darkgray176},
        y grid style={darkgray176},
        xmajorgrids,
        ymajorgrids,
        xtick style={draw=none},
        ytick style={draw=none},
		width = 0.75\textwidth,
		height = 0.5\textwidth,
		xlabel = {$t \; [\si{\second}]$},
		ylabel style={align=center}, 
		ylabel={$E^\text{elastic} \; [\si{\joule}]$},
        legend cell align={left},
		legend style={
          fill opacity=0.8,
          draw opacity=1,
          text opacity=1,
          at={(0.97,0.03)},
          anchor=south east,
          draw=lightgray204
        },
		legend pos = outer north east,
        legend style={font=\footnotesize}
		]
		
		\addplot[black, line width=2.0pt] file[] {tikzpictures/data/stableIMEX.dat};
		\addplot[red, line width=2.0pt] file[] {tikzpictures/data/explodingIMEX.dat};
		
		\legend{$\Delta t = 0.99 \Delta t_\text{crit}^\text{explicit}$, $\Delta t = 1.01 \Delta t_\text{crit}^\text{explicit}$},
	\end{axis}
\end{tikzpicture}%

%% file: content/benchmark2D.tex
\label{subsec:2D}

\begin{figure}[t!]
	\centering
    \begin{minipage}{0.85\linewidth}
        \includegraphics[width=\linewidth]{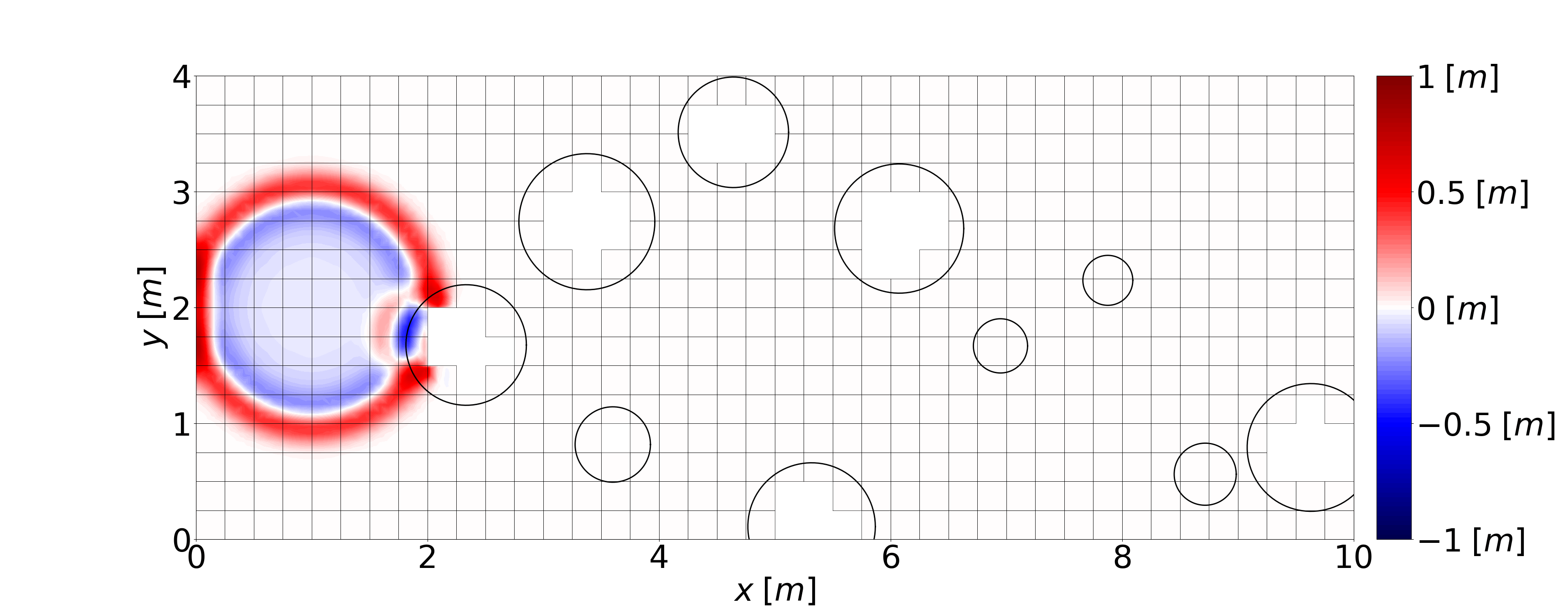}
        \subcaption{$t = \SI{1.5}{\second}$}
    \end{minipage}
    \begin{minipage}{0.85\linewidth}
        \includegraphics[width=\linewidth]{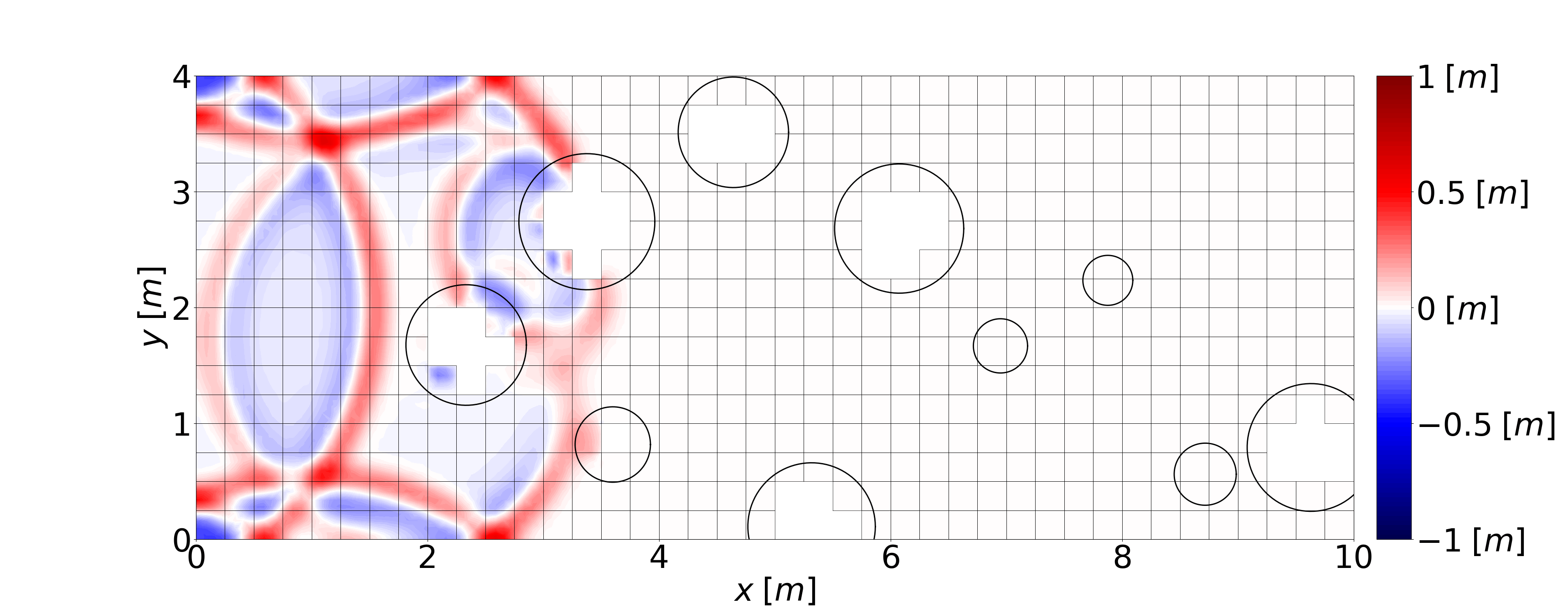}
        \subcaption{$t = \SI{3}{\second}$}
    \end{minipage}
    \begin{minipage}{0.85\linewidth}
        \includegraphics[width=\linewidth]{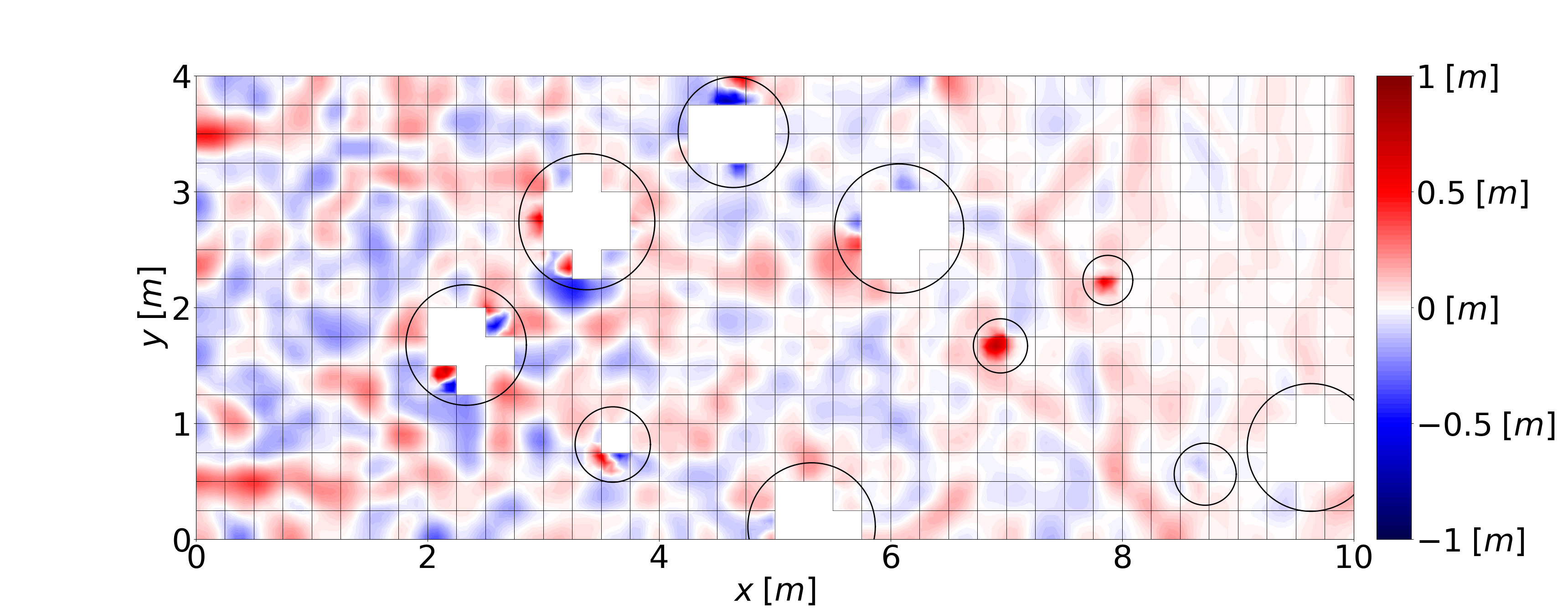}
        \subcaption{$t = T = \SI{10}{\second}$}
    \end{minipage}
	\caption{CDM solution of the perforated plate with $\Delta t = \SI{1}{\milli\second}$}
    \label{fig:RandomHolesForward}
\end{figure}

\begin{figure}[t!]
	\centering
	\input{figures/FillRatiosHistCircles.tex}
	\caption{Histogram of fill ratios $\eta$ for the mesh shown in \figref{fig:RandomHolesForward}}
	\label{fig:fillRatiosCircles}
\end{figure}

We now consider the scalar wave equation to simulate the propagation of acoustic waves in a perforated plate with dimensions $l_x \times l_y = \SI{10}{\meter} \times \SI{4}{\meter}$. We subtract ten circular holes with randomly sampled center points $(x_{c, i}, y_{c, i})$ and radii $r_i$ from uniform distributions:
\begin{equation}
	x_{\text{c}, i} \in \mathcal{U}(\SI{2}{\meter}, \SI{10}{\meter}), \quad
    y_{\text{c}, i} \in \mathcal{U}(\SI{0}{\meter}, \SI{4}{\meter}), \quad
    r_i \in \mathcal{U}(\SI{0.2}{\meter}, \SI{0.6}{\meter}).
\end{equation}
We choose constant values $\rho = \SI{1}{\kg\per\meter\squared}$ and $c = \SI{1}{\meter\per\second}$ for the density and wave speed and scale the density by a factor $\alpha_f = 10^{-6}$ inside the fictitious domain. Starting from the initial condition $u(x, 0) = 0$, we run our simulation until $T = \SI{10}{\second}$, so that waves can travel through the domain exactly once. We assume homogeneous Neumann boundary conditions on all four sides, representing a perfectly reflective behavior. The excitation $f(x, y, t)$ is the product of the spatial function $f_\text{x}(x, y)$ and the temporal function $f_{\text{t}}(t)$. We define $f_\text{t}(t)$ as the derivative of a Gaussian function in time
\begin{equation}
	f_\text{t}(t) = \frac{-(t - t_\text{0})}{\sqrt{2 \pi} \sigma_\text{t}^3} e^{\left(\frac{-(t-t_\text{0})^2}{2 \sigma_\text{t}^2}\right)} ,\label{eq:ft} 
\end{equation}
centered at $t_\text{0} = \frac{1}{f_\text{s}}$, with a temporal width of $\sigma_\text{t} = \frac{1}{2 \pi f_\text{s}}$. The dominant frequency of the excitation signal is set to $f_\text{s} = \SI{2}{\hertz}$, corresponding to a dominant wavelength of $\lambda = \frac{c}{f} = \SI[parse-numbers=false]{\frac{1}{2}}{\meter}$. The spatial distribution $f_\text{x}$ of the excitation is also a Gaussian bell, centered at the position $(x_s, y_s) = (\SI{1}{\meter}, \SI{2}{\meter})$ and a width of $\sigma_s = \SI{6e-2}{\meter}$:
\begin{equation}
	f_\text{x}(x, y) = 10 \cdot e^{\left(-\frac{(x - x_s)^2 + (y - y_s)^2}{2 \sigma_s^2}\right)}\text{.}
	\label{eq:fx}
\end{equation}


\begin{figure}[b!]
	\centering
	\input{figures/tCritsCircles_refined.tex} 
	\caption{Cell-wise critical time step sizes}
	\label{fig:tCritsCircles}
\end{figure}
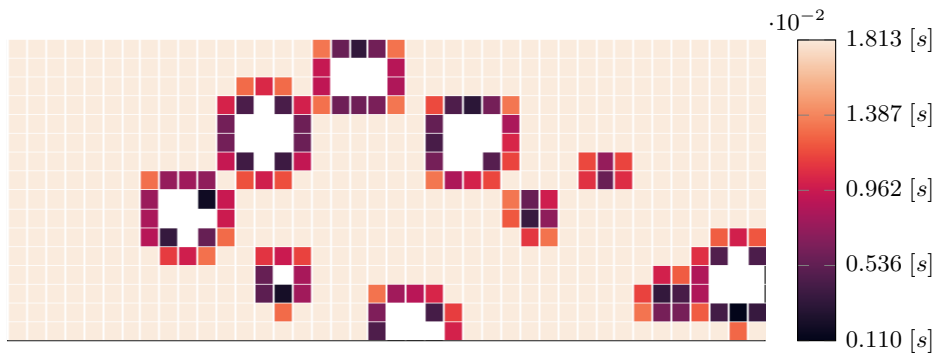

We discretize the domain with a Cartesian SCM grid of $40 \times 16$ cells in the $x$ and $y$ directions and assign a polynomial order of $p = 5$. We integrate cut cells by distributing grids with $p + 1$ Gauss-Legendre points per direction on a quadtree with six levels of refinement towards the domain boundary. We also discard finite cells that are entirely outside the physical domain. \figref{fig:RandomHolesForward} shows the computational mesh with three solution snapshots of the wave spreading from the initial pulse on the left side. There were so many reflections until the third snapshot that no clear wavefront is visible anymore. The high polynomial degree allows the use of a coarse mesh for the wave field while still satisfying the sampling condition. For $p = 5$, the spatial discretization must resolve approximately any wavelength in the excitation signal with at least one element. Our mesh resolves the dominant wavelength with two cells and even wavelengths of the decaying, higher frequency content up to $2 f_\text{s}$ with at least one cell.

\figref{fig:fillRatiosCircles} shows a histogram of the fill ratios of all cells. Approximately \SI{80}{\percent} of the cells are uncut, resulting in $n^\text{d} = 11\,936$ diagonal dofs and $n^\text{c} = 3\,473$ dofs supported by at least one cut cell. We consider cells below the fill ratio threshold $\epsilon = 10^{-10}$ empty. This example's smallest nonzero fill ratio is $\eta_{min} = \num{7.66e-4}$. While only a few cells are in the $[\epsilon, 0.1]$ bucket, they significantly reduce the global critical time step size, as discussed in~\secref{sec:intro}. \figref{fig:tCritsCircles} shows the cell-wise critical time step sizes for the consistent SCM. The values are uniform across uncut cells but significantly lower for cells intersecting the domain boundary.
\tabref{tab:criticaltimestepsizes} compares the global critical time step, the cell-wise critical time step of uncut cells, and the minimum cell-wise critical time step of cut cells. While the minimum cell-wise value predicts the global critical time step well for the consistent and HRZ-lumped SCM, the Newmark IMEX method only depends on the critical time step of uncut cells. We expect this result since the Newmark IMEX method treats cut cells implicitly, and its critical time step only depends on the explicit, i.e., the uncut dofs. Therefore, the cell-wise critical time step for uncut dofs is a good and cheap estimate for the Newmark IMEX critical time step. Interestingly, the comparison of the consistent and HRZ-lumped SCM approaches shows a larger global critical time step for the consistent SCM despite the more restrictive cell-wise minimum that lies over 16 times below the uncut cell estimate. The eigenvalue stabilization approach presented in \cite{Eisentraeger2023} may be able to lift the significant restrictions on the critical time step in an explicit treatment of cut cells in the SCM.

\begin{table}[t!]%
    \vspace{-0.4cm}
    \newcommand{\blap}[1]{{\begin{tabular}[c]{@{}c@{}}#1\end{tabular}}}%
    \renewcommand{\arraystretch}{1.5}%
    \centering%
    \caption{Comparison of the critical time step sizes for different SCM versions}%
    \begin{tabular}{|l|c|c|c|c|}%
        \hline
        & \blap{global system      \\[-0.3em] $\Delta t_\text{crit}^\text{global}$} 
        & \blap{cell-wise, uncut   \\[-0.3em] $\displaystyle\Delta t_\text{crit}^\text{uncut}$} 
        & \blap{cell-wise, minimum \\[-0.3em] $\displaystyle\Delta t_\text{crit}^\text{cut, min}$} 
        & $\displaystyle\frac{\Delta t_\text{crit}^\text{uncut}}{\Delta t_\text{crit}^\text{cut, min}}$ \\ 
        \hline\hline
        CDM, consistent SCM   & \hphantom{1}\SI{2.13}{\milli\second} & \SI{18.13}{\milli\second} & \SI{1.10}{\milli\second} & 16.4 \\ \hline  
        CDM, HRZ-lumped SCM   & \hphantom{1}\SI{1.42}{\milli\second} & ---''---                  & \SI{1.30}{\milli\second} & 13.9 \\ \hline
        Newmark IMEX SCM      & \SI{18.14}{\milli\second}            & ---''---                  & -                        & -       \\ \hline
    \end{tabular}
    \label{tab:criticaltimestepsizes}
\end{table}

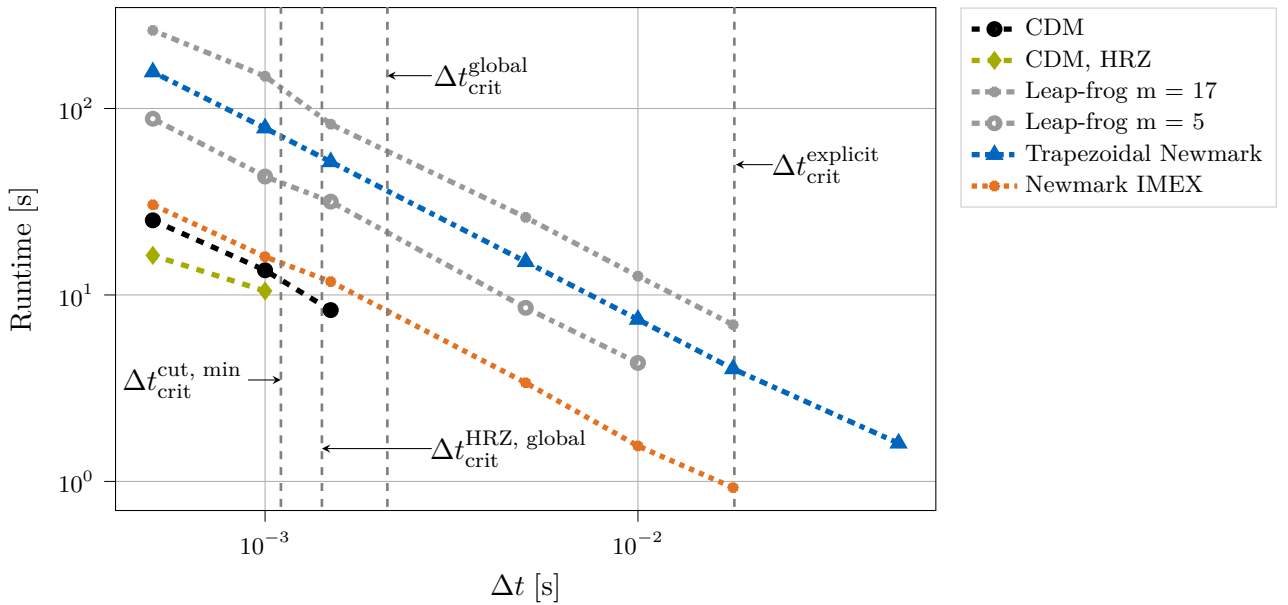
\begin{figure}[b!]
	\centering
	\input{figures/dtRuntime.tex} 
    \vspace{-0.5cm}
	\caption{Runtime versus time step size for the perforated plate}
	\label{fig:dtRuntimeCircles}
\end{figure}

\begin{figure}[t!]
	\centering
	\input{figures/dtL2Overleaf.tex} 
	\caption{$L_2$ error versus time step size for the perforated plate}
	\label{fig:dtL2Circles}
\end{figure}
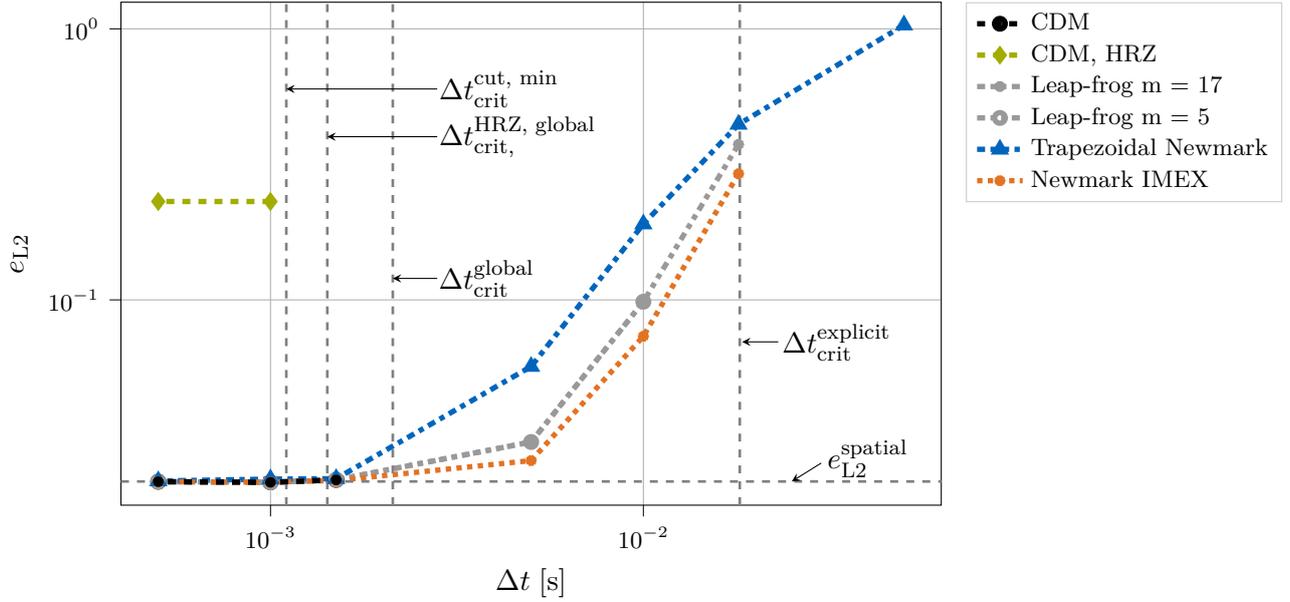

We now investigate the accuracy and runtime of several time integration methods by comparing them to a reference solution computed on a grid of $100 \times 40$ cells with $p = 5$ and integrated in time with $\Delta t = \SI{5e-5}{\second}$. Following~\cite{DG19_1}, we approximate the relative error in the $L_2$-norm summing over randomly sampled coordinates $(x_i, y_i)$ inside the physical domain:
\begin{equation}
	e_{L_2} = \sqrt{\frac{\sum_{i=1}^{n_p} (u(x_i, y_i) - u_{ref}(x_i, y_i))^2}{\sum_{i=1}^{n_p} u_{ref}(x_i, y_i)^2}}
\end{equation}
We first distribute random points in the extended, square domain but only keep the $n_\text{p} = 8572$ points inside the physical domain (i.e., those not inside one of the circular holes).
Our previous analysis of the critical time step sizes considered the CDM using the consistent SCM, the CDM using the HRZ-lumped SCM, and our Newmark IMEX approach that also uses the consistent SCM. We now include in our analysis a fully implicit scheme that integrates all dofs of the consistent SCM with an implicit trapezoidal Newmark method. We also consider a leap-frog algorithm since it is often used for similar problems, e.g., in ~\cite{NAC22}. It subdivides each time step into $m$ smaller intervals to integrate the cut cells with higher accuracy and improved stability. We refer to~\cite{meineMA} for details on the leap-frog formulation and its implementation. We measure the runtime of all simulations on a single Ryzen 7 2700x core. We use the Scipy function \texttt{scipy.sparse.linalg.splu} where necessary to factorize the matrices and solve at each time step.

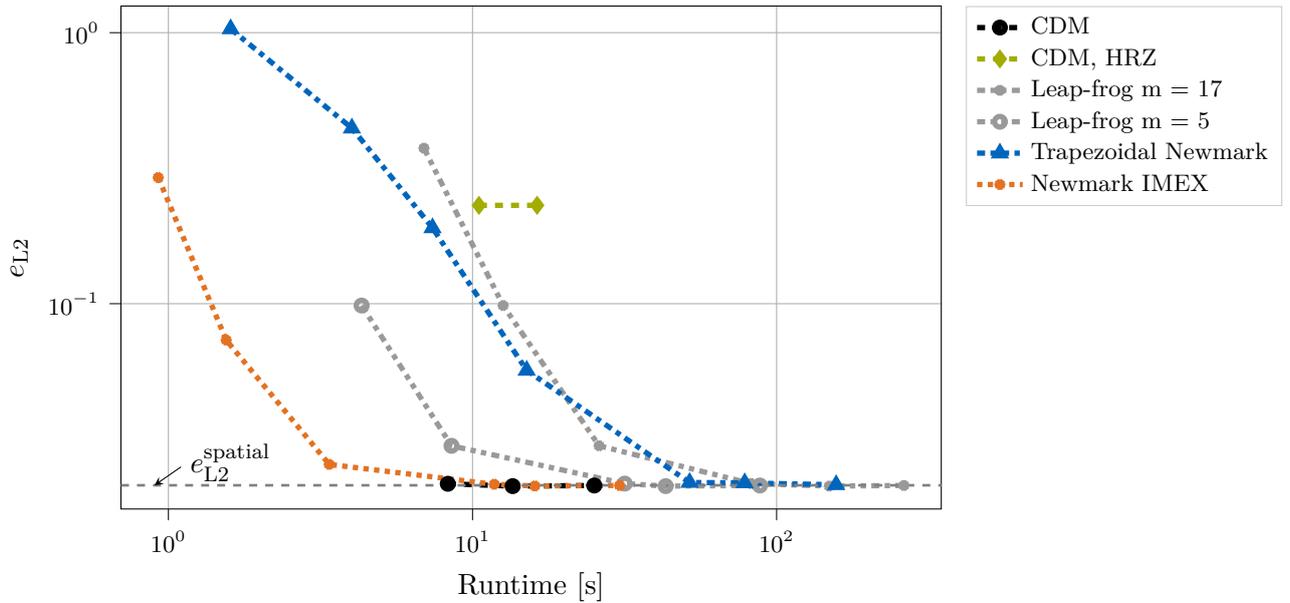
\begin{figure}[b!]
	\centering
	\input{figures/RuntimeL2.tex} 
    \vspace{-0.5cm}
	\caption{$L_2$-error versus runtime for the perforated plate}
	\label{fig:RuntimeL2Circles}
\end{figure}

In the comparison of the runtime for a fixed number of time steps in \figref{fig:dtRuntimeCircles}, the CDM using HRZ-lumped SCM and the CDM using the consistent SCM are most efficient, but they can only use small time step sizes due to their explicit treatment of cut cells. The plots exclude the unstable computations exceeding the critical time step size. Due to the unconditional stability, the implicit treatment of all cells in the trapezoidal Newmark approach is possible for the full range of time step sizes. However, the solution of a triangular matrix system with the factorized system matrix $\mathbf{S}$ and a different force vector $\mathbf{\hat{f}}_n$ (see Section~\ref{sec:timeInt}) at each time step makes it significantly more expensive. In contrast, Newmark IMEX is almost as fast as the CDM while being impacted much less by its stability constraint. This result aligns with the analysis of the spring-coupled masses in the previous section. The leap-frog algorithm trades performance to extend its stability range since finer substeps for the cut cells increase the runtime. The two conditions, $\Delta t_\text{coarse} < \Delta t_\text{crit}^\text{uncut}$ and $\Delta t_\text{fine} = \Delta t_\text{coarse} / m < \Delta t_\text{crit}^\text{cut, min}$, determine the stability limit of the leap-frog algorithm. Therefore, $m = 17$ substeps are more expensive but allow a larger coarse time step than $m = 5$.

\figref{fig:dtL2Circles} compares accuracy depending over the time step size. Except for the CDM applied to HRZ-lumped SCM, all methods converge to the error limit imposed by the chosen spatial discretization. The HRZ lumping of cut cells strongly deteriorates the overall accuracy (see also~\cite{K21}).
The Newmark IMEX method is the most accurate for given time step sizes, followed by the leap-frog algorithms and the implicit trapezoidal Newmark method. 
The two leap-frog algorithms behave similarly since increasing the number of substeps here impacts stability but not so much accuracy. The implicit trapezoidal Newmark method's error is comparatively high, particularly for large time step sizes. \figref{fig:RuntimeL2Circles} combines the information from~\figref{fig:dtRuntimeCircles} and \figref{fig:dtL2Circles} and is crucial for determining the most efficient of the investigated algorithms. Newmark IMEX significantly outperforms the other methods, confirming that our approach can leverage the advantages of implicitly integrating cut cells while explicitly integrating uncut cells without suffering from their disadvantages in this example.

%% file: figures/FillRatiosHistCircles.tex
\begin{tikzpicture}

\definecolor{darkgray176}{RGB}{176,176,176}
\pgfplotsset{every tick label/.append style={font=\footnotesize}}
\pgfplotsset{every label/.append style={font=\footnotesize}}

\begin{axis}[
tick align=outside,
width=1\textwidth,
height = 5cm,
tick pos=left,
ymajorgrids,
x grid style={darkgray176},
xlabel={Fill ratio $\eta$},
xmin=0, xmax=1.1,
xtick style={color=black},
xtick={0.05,0.15,0.25,0.35,0.45,0.55,0.65,0.75,0.85,0.95,1.05},
xticklabels={
  {$[\epsilon, 0.1)$},
  {$[0.1, 0.2)$},
  {$[0.2, 0.3)$},
  {$[0.3, 0.4)$},
  {$[0.4, 0.5)$},
  {$[0.5, 0.6)$},
  {$[0.6, 0.7)$},
  {$[0.7, 0.8)$},
  {$[0.8, 0.9)$},
  {$[0.9, 1)$},
  {$full$}
},
y grid style={darkgray176},
ylabel={Ratio of cells},
ymin=0, ymax=0.844908180300498,
ytick style={color=black}
]
\draw[draw=none,fill=black] (axis cs:0.005,0) rectangle (axis cs:0.095,0.0317195325542571);
\draw[draw=none,fill=black] (axis cs:0.105,0) rectangle (axis cs:0.195,0.00667779632721202);
\draw[draw=none,fill=black] (axis cs:0.205,0) rectangle (axis cs:0.295,0.0200333889816361);
\draw[draw=none,fill=black] (axis cs:0.305,0) rectangle (axis cs:0.395,0.010016694490818);
\draw[draw=none,fill=black] (axis cs:0.405,0) rectangle (axis cs:0.495,0.018363939899833);
\draw[draw=none,fill=black] (axis cs:0.505,0) rectangle (axis cs:0.595,0.00834724540901499);
\draw[draw=none,fill=black] (axis cs:0.605,0) rectangle (axis cs:0.695,0.021702838063439);
\draw[draw=none,fill=black] (axis cs:0.705,0) rectangle (axis cs:0.795,0.020033388981636);
\draw[draw=none,fill=black] (axis cs:0.805,0) rectangle (axis cs:0.895,0.028380634390651);
\draw[draw=none,fill=black] (axis cs:0.904999999995,0) rectangle (axis cs:0.994999999905,0.030050083472454);
\draw[draw=none,fill=black] (axis cs:1.004999999905,0) rectangle (axis cs:1.094999999995,0.804674457429045);
\end{axis}

\end{tikzpicture}

%% file: figures/TCritsCircles_refined.tex
\begin{tikzpicture}[every tick label/.append style={font=\footnotesize},
	every label/.append style={font=\footnotesize}]

\definecolor{darkgray176}{RGB}{176,176,176}

\begin{axis}[
	width=0.7\textwidth,
colorbar,
colorbar style={ytick={0.0011026750137449,0.00535983065899446,0.00961698630424401,0.0138741419494936,0.0181312975947431},y tick label style={/pgf/number format/.cd,
	fixed,
	fixed zerofill,
	precision=5,
	/tikz/.cd} ,yticklabels={$0.110\; [s]$,$0.536\; [s]$,$0.962\; [s]$,$1.387\; [s]$,$1.813\; [s]$},ylabel={}},
colormap={mymap}{[1pt]
 rgb(0pt)=(0.01060815,0.01808215,0.10018654);
  rgb(1pt)=(0.01428972,0.02048237,0.10374486);
  rgb(2pt)=(0.01831941,0.0229766,0.10738511);
  rgb(3pt)=(0.02275049,0.02554464,0.11108639);
  rgb(4pt)=(0.02759119,0.02818316,0.11483751);
  rgb(5pt)=(0.03285175,0.03088792,0.11863035);
  rgb(6pt)=(0.03853466,0.03365771,0.12245873);
  rgb(7pt)=(0.04447016,0.03648425,0.12631831);
  rgb(8pt)=(0.05032105,0.03936808,0.13020508);
  rgb(9pt)=(0.05611171,0.04224835,0.13411624);
  rgb(10pt)=(0.0618531,0.04504866,0.13804929);
  rgb(11pt)=(0.06755457,0.04778179,0.14200206);
  rgb(12pt)=(0.0732236,0.05045047,0.14597263);
  rgb(13pt)=(0.0788708,0.05305461,0.14995981);
  rgb(14pt)=(0.08450105,0.05559631,0.15396203);
  rgb(15pt)=(0.09011319,0.05808059,0.15797687);
  rgb(16pt)=(0.09572396,0.06050127,0.16200507);
  rgb(17pt)=(0.10132312,0.06286782,0.16604287);
  rgb(18pt)=(0.10692823,0.06517224,0.17009175);
  rgb(19pt)=(0.1125315,0.06742194,0.17414848);
  rgb(20pt)=(0.11813947,0.06961499,0.17821272);
  rgb(21pt)=(0.12375803,0.07174938,0.18228425);
  rgb(22pt)=(0.12938228,0.07383015,0.18636053);
  rgb(23pt)=(0.13501631,0.07585609,0.19044109);
  rgb(24pt)=(0.14066867,0.0778224,0.19452676);
  rgb(25pt)=(0.14633406,0.07973393,0.1986151);
  rgb(26pt)=(0.15201338,0.08159108,0.20270523);
  rgb(27pt)=(0.15770877,0.08339312,0.20679668);
  rgb(28pt)=(0.16342174,0.0851396,0.21088893);
  rgb(29pt)=(0.16915387,0.08682996,0.21498104);
  rgb(30pt)=(0.17489524,0.08848235,0.2190294);
  rgb(31pt)=(0.18065495,0.09009031,0.22303512);
  rgb(32pt)=(0.18643324,0.09165431,0.22699705);
  rgb(33pt)=(0.19223028,0.09317479,0.23091409);
  rgb(34pt)=(0.19804623,0.09465217,0.23478512);
  rgb(35pt)=(0.20388117,0.09608689,0.23860907);
  rgb(36pt)=(0.20973515,0.09747934,0.24238489);
  rgb(37pt)=(0.21560818,0.09882993,0.24611154);
  rgb(38pt)=(0.22150014,0.10013944,0.2497868);
  rgb(39pt)=(0.22741085,0.10140876,0.25340813);
  rgb(40pt)=(0.23334047,0.10263737,0.25697736);
  rgb(41pt)=(0.23928891,0.10382562,0.2604936);
  rgb(42pt)=(0.24525608,0.10497384,0.26395596);
  rgb(43pt)=(0.25124182,0.10608236,0.26736359);
  rgb(44pt)=(0.25724602,0.10715148,0.27071569);
  rgb(45pt)=(0.26326851,0.1081815,0.27401148);
  rgb(46pt)=(0.26930915,0.1091727,0.2772502);
  rgb(47pt)=(0.27536766,0.11012568,0.28043021);
  rgb(48pt)=(0.28144375,0.11104133,0.2835489);
  rgb(49pt)=(0.2875374,0.11191896,0.28660853);
  rgb(50pt)=(0.29364846,0.11275876,0.2896085);
  rgb(51pt)=(0.29977678,0.11356089,0.29254823);
  rgb(52pt)=(0.30592213,0.11432553,0.29542718);
  rgb(53pt)=(0.31208435,0.11505284,0.29824485);
  rgb(54pt)=(0.31826327,0.1157429,0.30100076);
  rgb(55pt)=(0.32445869,0.11639585,0.30369448);
  rgb(56pt)=(0.33067031,0.11701189,0.30632563);
  rgb(57pt)=(0.33689808,0.11759095,0.3088938);
  rgb(58pt)=(0.34314168,0.11813362,0.31139721);
  rgb(59pt)=(0.34940101,0.11863987,0.3138355);
  rgb(60pt)=(0.355676,0.11910909,0.31620996);
  rgb(61pt)=(0.36196644,0.1195413,0.31852037);
  rgb(62pt)=(0.36827206,0.11993653,0.32076656);
  rgb(63pt)=(0.37459292,0.12029443,0.32294825);
  rgb(64pt)=(0.38092887,0.12061482,0.32506528);
  rgb(65pt)=(0.38727975,0.12089756,0.3271175);
  rgb(66pt)=(0.39364518,0.12114272,0.32910494);
  rgb(67pt)=(0.40002537,0.12134964,0.33102734);
  rgb(68pt)=(0.40642019,0.12151801,0.33288464);
  rgb(69pt)=(0.41282936,0.12164769,0.33467689);
  rgb(70pt)=(0.41925278,0.12173833,0.33640407);
  rgb(71pt)=(0.42569057,0.12178916,0.33806605);
  rgb(72pt)=(0.43214263,0.12179973,0.33966284);
  rgb(73pt)=(0.43860848,0.12177004,0.34119475);
  rgb(74pt)=(0.44508855,0.12169883,0.34266151);
  rgb(75pt)=(0.45158266,0.12158557,0.34406324);
  rgb(76pt)=(0.45809049,0.12142996,0.34540024);
  rgb(77pt)=(0.46461238,0.12123063,0.34667231);
  rgb(78pt)=(0.47114798,0.12098721,0.34787978);
  rgb(79pt)=(0.47769736,0.12069864,0.34902273);
  rgb(80pt)=(0.48426077,0.12036349,0.35010104);
  rgb(81pt)=(0.49083761,0.11998161,0.35111537);
  rgb(82pt)=(0.49742847,0.11955087,0.35206533);
  rgb(83pt)=(0.50403286,0.11907081,0.35295152);
  rgb(84pt)=(0.51065109,0.11853959,0.35377385);
  rgb(85pt)=(0.51728314,0.1179558,0.35453252);
  rgb(86pt)=(0.52392883,0.11731817,0.35522789);
  rgb(87pt)=(0.53058853,0.11662445,0.35585982);
  rgb(88pt)=(0.53726173,0.11587369,0.35642903);
  rgb(89pt)=(0.54394898,0.11506307,0.35693521);
  rgb(90pt)=(0.5506426,0.11420757,0.35737863);
  rgb(91pt)=(0.55734473,0.11330456,0.35775059);
  rgb(92pt)=(0.56405586,0.11235265,0.35804813);
  rgb(93pt)=(0.57077365,0.11135597,0.35827146);
  rgb(94pt)=(0.5774991,0.11031233,0.35841679);
  rgb(95pt)=(0.58422945,0.10922707,0.35848469);
  rgb(96pt)=(0.59096382,0.10810205,0.35847347);
  rgb(97pt)=(0.59770215,0.10693774,0.35838029);
  rgb(98pt)=(0.60444226,0.10573912,0.35820487);
  rgb(99pt)=(0.61118304,0.10450943,0.35794557);
  rgb(100pt)=(0.61792306,0.10325288,0.35760108);
  rgb(101pt)=(0.62466162,0.10197244,0.35716891);
  rgb(102pt)=(0.63139686,0.10067417,0.35664819);
  rgb(103pt)=(0.63812122,0.09938212,0.35603757);
  rgb(104pt)=(0.64483795,0.0980891,0.35533555);
  rgb(105pt)=(0.65154562,0.09680192,0.35454107);
  rgb(106pt)=(0.65824241,0.09552918,0.3536529);
  rgb(107pt)=(0.66492652,0.09428017,0.3526697);
  rgb(108pt)=(0.67159578,0.09306598,0.35159077);
  rgb(109pt)=(0.67824099,0.09192342,0.3504148);
  rgb(110pt)=(0.684863,0.09085633,0.34914061);
  rgb(111pt)=(0.69146268,0.0898675,0.34776864);
  rgb(112pt)=(0.69803757,0.08897226,0.3462986);
  rgb(113pt)=(0.70457834,0.0882129,0.34473046);
  rgb(114pt)=(0.71108138,0.08761223,0.3430635);
  rgb(115pt)=(0.7175507,0.08716212,0.34129974);
  rgb(116pt)=(0.72398193,0.08688725,0.33943958);
  rgb(117pt)=(0.73035829,0.0868623,0.33748452);
  rgb(118pt)=(0.73669146,0.08704683,0.33543669);
  rgb(119pt)=(0.74297501,0.08747196,0.33329799);
  rgb(120pt)=(0.74919318,0.08820542,0.33107204);
  rgb(121pt)=(0.75535825,0.08919792,0.32876184);
  rgb(122pt)=(0.76145589,0.09050716,0.32637117);
  rgb(123pt)=(0.76748424,0.09213602,0.32390525);
  rgb(124pt)=(0.77344838,0.09405684,0.32136808);
  rgb(125pt)=(0.77932641,0.09634794,0.31876642);
  rgb(126pt)=(0.78513609,0.09892473,0.31610488);
  rgb(127pt)=(0.79085854,0.10184672,0.313391);
  rgb(128pt)=(0.7965014,0.10506637,0.31063031);
  rgb(129pt)=(0.80205987,0.10858333,0.30783);
  rgb(130pt)=(0.80752799,0.11239964,0.30499738);
  rgb(131pt)=(0.81291606,0.11645784,0.30213802);
  rgb(132pt)=(0.81820481,0.12080606,0.29926105);
  rgb(133pt)=(0.82341472,0.12535343,0.2963705);
  rgb(134pt)=(0.82852822,0.13014118,0.29347474);
  rgb(135pt)=(0.83355779,0.13511035,0.29057852);
  rgb(136pt)=(0.83850183,0.14025098,0.2876878);
  rgb(137pt)=(0.84335441,0.14556683,0.28480819);
  rgb(138pt)=(0.84813096,0.15099892,0.281943);
  rgb(139pt)=(0.85281737,0.15657772,0.27909826);
  rgb(140pt)=(0.85742602,0.1622583,0.27627462);
  rgb(141pt)=(0.86196552,0.16801239,0.27346473);
  rgb(142pt)=(0.86641628,0.17387796,0.27070818);
  rgb(143pt)=(0.87079129,0.17982114,0.26797378);
  rgb(144pt)=(0.87507281,0.18587368,0.26529697);
  rgb(145pt)=(0.87925878,0.19203259,0.26268136);
  rgb(146pt)=(0.8833417,0.19830556,0.26014181);
  rgb(147pt)=(0.88731387,0.20469941,0.25769539);
  rgb(148pt)=(0.89116859,0.21121788,0.2553592);
  rgb(149pt)=(0.89490337,0.21785614,0.25314362);
  rgb(150pt)=(0.8985026,0.22463251,0.25108745);
  rgb(151pt)=(0.90197527,0.23152063,0.24918223);
  rgb(152pt)=(0.90530097,0.23854541,0.24748098);
  rgb(153pt)=(0.90848638,0.24568473,0.24598324);
  rgb(154pt)=(0.911533,0.25292623,0.24470258);
  rgb(155pt)=(0.9144225,0.26028902,0.24369359);
  rgb(156pt)=(0.91717106,0.26773821,0.24294137);
  rgb(157pt)=(0.91978131,0.27526191,0.24245973);
  rgb(158pt)=(0.92223947,0.28287251,0.24229568);
  rgb(159pt)=(0.92456587,0.29053388,0.24242622);
  rgb(160pt)=(0.92676657,0.29823282,0.24285536);
  rgb(161pt)=(0.92882964,0.30598085,0.24362274);
  rgb(162pt)=(0.93078135,0.31373977,0.24468803);
  rgb(163pt)=(0.93262051,0.3215093,0.24606461);
  rgb(164pt)=(0.93435067,0.32928362,0.24775328);
  rgb(165pt)=(0.93599076,0.33703942,0.24972157);
  rgb(166pt)=(0.93752831,0.34479177,0.25199928);
  rgb(167pt)=(0.93899289,0.35250734,0.25452808);
  rgb(168pt)=(0.94036561,0.36020899,0.25734661);
  rgb(169pt)=(0.94167588,0.36786594,0.2603949);
  rgb(170pt)=(0.94291042,0.37549479,0.26369821);
  rgb(171pt)=(0.94408513,0.3830811,0.26722004);
  rgb(172pt)=(0.94520419,0.39062329,0.27094924);
  rgb(173pt)=(0.94625977,0.39813168,0.27489742);
  rgb(174pt)=(0.94727016,0.4055909,0.27902322);
  rgb(175pt)=(0.94823505,0.41300424,0.28332283);
  rgb(176pt)=(0.94914549,0.42038251,0.28780969);
  rgb(177pt)=(0.95001704,0.42771398,0.29244728);
  rgb(178pt)=(0.95085121,0.43500005,0.29722817);
  rgb(179pt)=(0.95165009,0.44224144,0.30214494);
  rgb(180pt)=(0.9524044,0.44944853,0.3072105);
  rgb(181pt)=(0.95312556,0.45661389,0.31239776);
  rgb(182pt)=(0.95381595,0.46373781,0.31769923);
  rgb(183pt)=(0.95447591,0.47082238,0.32310953);
  rgb(184pt)=(0.95510255,0.47787236,0.32862553);
  rgb(185pt)=(0.95569679,0.48489115,0.33421404);
  rgb(186pt)=(0.95626788,0.49187351,0.33985601);
  rgb(187pt)=(0.95681685,0.49882008,0.34555431);
  rgb(188pt)=(0.9573439,0.50573243,0.35130912);
  rgb(189pt)=(0.95784842,0.51261283,0.35711942);
  rgb(190pt)=(0.95833051,0.51946267,0.36298589);
  rgb(191pt)=(0.95879054,0.52628305,0.36890904);
  rgb(192pt)=(0.95922872,0.53307513,0.3748895);
  rgb(193pt)=(0.95964538,0.53983991,0.38092784);
  rgb(194pt)=(0.96004345,0.54657593,0.3870292);
  rgb(195pt)=(0.96042097,0.55328624,0.39319057);
  rgb(196pt)=(0.96077819,0.55997184,0.39941173);
  rgb(197pt)=(0.9611152,0.5666337,0.40569343);
  rgb(198pt)=(0.96143273,0.57327231,0.41203603);
  rgb(199pt)=(0.96173392,0.57988594,0.41844491);
  rgb(200pt)=(0.96201757,0.58647675,0.42491751);
  rgb(201pt)=(0.96228344,0.59304598,0.43145271);
  rgb(202pt)=(0.96253168,0.5995944,0.43805131);
  rgb(203pt)=(0.96276513,0.60612062,0.44471698);
  rgb(204pt)=(0.96298491,0.6126247,0.45145074);
  rgb(205pt)=(0.96318967,0.61910879,0.45824902);
  rgb(206pt)=(0.96337949,0.6255736,0.46511271);
  rgb(207pt)=(0.96355923,0.63201624,0.47204746);
  rgb(208pt)=(0.96372785,0.63843852,0.47905028);
  rgb(209pt)=(0.96388426,0.64484214,0.4861196);
  rgb(210pt)=(0.96403203,0.65122535,0.4932578);
  rgb(211pt)=(0.96417332,0.65758729,0.50046894);
  rgb(212pt)=(0.9643063,0.66393045,0.5077467);
  rgb(213pt)=(0.96443322,0.67025402,0.51509334);
  rgb(214pt)=(0.96455845,0.67655564,0.52251447);
  rgb(215pt)=(0.96467922,0.68283846,0.53000231);
  rgb(216pt)=(0.96479861,0.68910113,0.53756026);
  rgb(217pt)=(0.96492035,0.69534192,0.5451917);
  rgb(218pt)=(0.96504223,0.7015636,0.5528892);
  rgb(219pt)=(0.96516917,0.70776351,0.5606593);
  rgb(220pt)=(0.96530224,0.71394212,0.56849894);
  rgb(221pt)=(0.96544032,0.72010124,0.57640375);
  rgb(222pt)=(0.96559206,0.72623592,0.58438387);
  rgb(223pt)=(0.96575293,0.73235058,0.59242739);
  rgb(224pt)=(0.96592829,0.73844258,0.60053991);
  rgb(225pt)=(0.96612013,0.74451182,0.60871954);
  rgb(226pt)=(0.96632832,0.75055966,0.61696136);
  rgb(227pt)=(0.96656022,0.75658231,0.62527295);
  rgb(228pt)=(0.96681185,0.76258381,0.63364277);
  rgb(229pt)=(0.96709183,0.76855969,0.64207921);
  rgb(230pt)=(0.96739773,0.77451297,0.65057302);
  rgb(231pt)=(0.96773482,0.78044149,0.65912731);
  rgb(232pt)=(0.96810471,0.78634563,0.66773889);
  rgb(233pt)=(0.96850919,0.79222565,0.6764046);
  rgb(234pt)=(0.96893132,0.79809112,0.68512266);
  rgb(235pt)=(0.96935926,0.80395415,0.69383201);
  rgb(236pt)=(0.9698028,0.80981139,0.70252255);
  rgb(237pt)=(0.97025511,0.81566605,0.71120296);
  rgb(238pt)=(0.97071849,0.82151775,0.71987163);
  rgb(239pt)=(0.97120159,0.82736371,0.72851999);
  rgb(240pt)=(0.97169389,0.83320847,0.73716071);
  rgb(241pt)=(0.97220061,0.83905052,0.74578903);
  rgb(242pt)=(0.97272597,0.84488881,0.75440141);
  rgb(243pt)=(0.97327085,0.85072354,0.76299805);
  rgb(244pt)=(0.97383206,0.85655639,0.77158353);
  rgb(245pt)=(0.97441222,0.86238689,0.78015619);
  rgb(246pt)=(0.97501782,0.86821321,0.78871034);
  rgb(247pt)=(0.97564391,0.87403763,0.79725261);
  rgb(248pt)=(0.97628674,0.87986189,0.8057883);
  rgb(249pt)=(0.97696114,0.88568129,0.81430324);
  rgb(250pt)=(0.97765722,0.89149971,0.82280948);
  rgb(251pt)=(0.97837585,0.89731727,0.83130786);
  rgb(252pt)=(0.97912374,0.90313207,0.83979337);
  rgb(253pt)=(0.979891,0.90894778,0.84827858);
  rgb(254pt)=(0.98067764,0.91476465,0.85676611);
  rgb(255pt)=(0.98137749,0.92061729,0.86536915)
},
point meta max=0.0181312975947431,
point meta min=0.0011026750137449,
tick align=outside,
unit vector ratio*=1 1 1, 
tick pos=left,
x grid style={darkgray176},
xmin=0, xmax=40,
xtick style={draw=none},
y dir=reverse,
y grid style={darkgray176},
ymin=0, ymax=16,
ytick style={draw=none},
yticklabels=\empty,
xticklabels=\empty,
]
\addplot graphics [includegraphics cmd=\pgfimage,xmin=0, xmax=40, ymin=16, ymax=0] {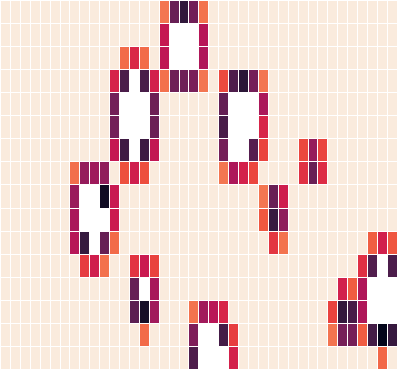};
\end{axis}

\end{tikzpicture}

%% file: figures/dtRuntime.tex
\begin{tikzpicture}[]

\definecolor{crimson2143940}{RGB}{214,39,40}
\definecolor{darkgray176}{RGB}{176,176,176}
\definecolor{darkorange25512714}{RGB}{255,127,14}
\definecolor{forestgreen4416044}{RGB}{44,160,44}
\definecolor{gray127}{RGB}{127,127,127}
\definecolor{lightgray204}{RGB}{204,204,204}
\definecolor{mediumpurple148103189}{RGB}{148,103,189}
\definecolor{orchid227119194}{RGB}{227,119,194}
\definecolor{sienna1408675}{RGB}{140,86,75}
\definecolor{steelblue31119180}{RGB}{31,119,180}

\definecolor{Black}{RGB}{0, 0, 0} 
\definecolor{TumBlue}{RGB}{0,101,189} 
\definecolor{Gray}{RGB}{153,153,153} 
\definecolor{TumColor2}{RGB}{0,82,147} 
\definecolor{TumColor2}{RGB}{152,198,234} 
\definecolor{Orange}{RGB}{227,114,34} 
\definecolor{Green}{RGB}{162,173,0} 
\definecolor{TumColor2}{RGB}{0,101,189}

\pgfplotsset{every tick label/.append style={font=\footnotesize}}
\pgfplotsset{every label/.append style={font=\footnotesize}}

\begin{axis}[
legend cell align={left},
width=0.75\textwidth,
height=0.5\textwidth,
legend style={
  fill opacity=0.8,
  draw opacity=1,
  text opacity=1,
  at={(0.03,0.97)},
  anchor=north west,
  draw=lightgray204
},
log basis x={10},
log basis y={10},
legend pos= outer north east,
legend style={font=\footnotesize},
tick align=outside,
tick pos=left,
x grid style={darkgray176},
xlabel={$\Delta t \; [\si{\second}]$},
xmajorgrids,
xmin=0.000397164117362141, xmax=0.0629462705897084,
xminorgrids,
xmode=log,
xtick style={color=black},
xtick={1e-05,0.0001,0.001,0.01,0.1,1},
xticklabels={
  \(\displaystyle {10^{-5}}\),
  \(\displaystyle {10^{-4}}\),
  \(\displaystyle {10^{-3}}\),
  \(\displaystyle {10^{-2}}\),
  \(\displaystyle {10^{-1}}\),
  \(\displaystyle {10^{0}}\)
},
y grid style={darkgray176},
ylabel={Runtime $[\si{\second}]$},
ymajorgrids,
ymin=0.698230819818063, ymax=347.860070133297,
yminorgrids,
ymode=log,
ytick style={color=black},
ytick={0.01,0.1,1,10,100,1000,10000},
yticklabels={
  \(\displaystyle {10^{-2}}\),
  \(\displaystyle {10^{-1}}\),
  \(\displaystyle {10^{0}}\),
  \(\displaystyle {10^{1}}\),
  \(\displaystyle {10^{2}}\),
  \(\displaystyle {10^{3}}\),
  \(\displaystyle {10^{4}}\)
}
]

\addplot [line width=2pt, Black, mark=*, mark size=2, mark options={solid}, dashed]
table {%
	0.0015 8.29394054412842
	0.001 13.553920507431
	0.0005 25.1387796401978
};
\addlegendentry{CDM}
\addplot [line width=2pt, Green, mark=diamond*, mark size=2, mark options={solid}, dashed]
table {%
0.001 10.4988887310028
0.0005 16.3078768253326
};
\addlegendentry{CDM, HRZ}
\addplot [line width=2pt, Gray, mark=star, mark size=2, mark options={solid}, dashdotdotted]
table {%
0.018 6.92195343971252
0.01 12.6109111309052
0.005 26.0707900524139
0.0015 82.4153189659119
0.001 148.949785709381
0.0005 262.29815530777
};
\addlegendentry{Leap-frog m = 17}
\addplot [line width=2pt, Gray, mark=o, mark size=2, mark options={solid}, dashdotdotted]
table {%
0.01 4.31896638870239
0.005 8.53093075752258
0.0015 31.6661343574524
0.001 43.1707127094269
0.0005 88.1124360561371
};
\addlegendentry{Leap-frog m = 5}

\addplot [line width=2pt, TumBlue, mark=triangle*, mark size=2, mark options={solid}, dashdotted]
table {%
0.05 1.60199284553528
0.018 4.01198244094849
0.01 7.38696956634521
0.005 15.0709342956543
0.0015 51.746776342392
0.001 78.5276608467102
0.0005 156.95432472229
};
\addlegendentry{Trapezoidal Newmark}
\addplot [line width=2pt, Orange, mark=10-pointed star, mark size=2, dotted]
table {%
0.018 0.925994396209717
0.01 1.54899501800537
0.005 3.37998533248901
0.0015 11.7889494895935
0.001 16.0259330272675
0.0005 30.4628674983978
};
\addlegendentry{Newmark IMEX}

\draw[{stealth[width=5mm]}-] (0.01831,50)   -- (0.023,50);
\node[align=left] at (0.032, 50) {$\Delta t_\text{crit}^\text{explicit}$};

\draw[{stealth[width=5mm]}-] (0.0021282,150)   -- (0.0028,150);
\node[align=left] at (0.0038, 150) {$\Delta t_\text{crit}^\text{global}$};

\draw[{stealth[width=5mm]}-] (0.0014205,1.5)   -- (0.0028,1.5);
\node[align=left] at (0.0045, 1.5) {$\Delta t_\text{crit}^\text{HRZ, global}$};

\draw[-{stealth[width=5mm]}] (0.0009,3.5)   -- (0.0011027,3.5);
\node[align=left] at (0.0006, 3.5) {$\Delta t_\text{crit}^\text{cut, min}$};

\addplot [line width=1pt, gray, dashed]
table {%
	0.018140 0.6
	0.018140 400
};
\addplot [line width=1pt, gray, dashed]
table {%
	0.0021282 0.6
	0.0021282 400
};
\addplot [line width=1pt, gray, dashed]
table {%
	0.0014205 0.6
	0.0014205 400
};

\addplot [line width=1pt, gray, dashed]
table {%
	0.0011027 0.6
	0.0011027 400
};
\end{axis}

\end{tikzpicture}

%% file: figures/dtL2Overleaf.tex

\begin{tikzpicture}[]
\definecolor{crimson2143940}{RGB}{214,39,40}
\definecolor{darkgray176}{RGB}{176,176,176}
\definecolor{darkorange25512714}{RGB}{255,127,14}
\definecolor{forestgreen4416044}{RGB}{44,160,44}
\definecolor{gray127}{RGB}{127,127,127}
\definecolor{lightgray204}{RGB}{204,204,204}
\definecolor{mediumpurple148103189}{RGB}{148,103,189}
\definecolor{orchid227119194}{RGB}{227,119,194}
\definecolor{sienna1408675}{RGB}{140,86,75}
\definecolor{steelblue31119180}{RGB}{31,119,180}

\definecolor{Black}{RGB}{0, 0, 0} 
\definecolor{TumBlue}{RGB}{0,101,189} 
\definecolor{Gray}{RGB}{153,153,153} 
\definecolor{TumColor2}{RGB}{0,82,147} 
\definecolor{TumColor2}{RGB}{152,198,234} 
\definecolor{Orange}{RGB}{227,114,34} 
\definecolor{Green}{RGB}{162,173,0} 
\definecolor{TumColor2}{RGB}{0,101,189}

\pgfplotsset{every tick label/.append style={font=\footnotesize}}
\pgfplotsset{every label/.append style={font=\footnotesize}}
\begin{axis}[
legend cell align={left},
width=0.75\textwidth,
height=0.5\textwidth,
legend style={
  fill opacity=0.8,
  draw opacity=1,
  text opacity=1,
  at={(0.97,0.03)},
  anchor=south east,
  draw=lightgray204
},
legend pos= outer north east,
legend style={font=\footnotesize},
log basis x={10},
log basis y={10},
tick align=outside,
tick pos=left,
x grid style={darkgray176},
xlabel={$\Delta t \; [\si{\second}]$},
xmajorgrids,
xmin=0.000397164117362141, xmax=0.0629462705897084,
xminorgrids,
xmode=log,
xtick style={color=black},
xtick={1e-05,0.0001,0.001,0.01,0.1,1},
xticklabels={
  \(\displaystyle {10^{-5}}\),
  \(\displaystyle {10^{-4}}\),
  \(\displaystyle {10^{-3}}\),
  \(\displaystyle {10^{-2}}\),
  \(\displaystyle {10^{-1}}\),
  \(\displaystyle {10^{0}}\)
},
y grid style={darkgray176},
ylabel={$e_\text{L2}$},
ymajorgrids,
ymin=0.0174814137894623, ymax=1.25387336731297,
yminorgrids,
ymode=log,
ytick style={color=black},
ytick={0.001,0.01,0.1,1,10,100},
yticklabels={
  \(\displaystyle {10^{-3}}\),
  \(\displaystyle {10^{-2}}\),
  \(\displaystyle {10^{-1}}\),
  \(\displaystyle {10^{0}}\),
  \(\displaystyle {10^{1}}\),
  \(\displaystyle {10^{2}}\)
}
]

\addplot [line width=2pt, Black, mark=*, mark size=2, mark options={solid}, dashed]
table {%
	0.0015 0.0216566371141543
	0.001 0.0212288033004938
	0.0005 0.0213390821676393
};
\addlegendentry{CDM}
\addplot [line width=2pt, Green, mark=diamond*, mark size=2, mark options={solid}, dashed]
table {%
	0.001 0.230596450456585
	0.0005 0.23067392951055
};
\addlegendentry{CDM, HRZ}
\addplot [line width=2pt, Gray, mark=star, mark size=2, mark options={solid}, dashdotdotted]
table {%
	0.018 0.374835298223246
	0.01 0.0984745067732454
	0.005 0.0299041988804622
	0.0015 0.0216893334044662
	0.001 0.0212345806963219
	0.0005 0.0213404907028305
};
\addlegendentry{Leap-frog m = 17}
\addplot [line width=2pt, Gray, mark=o, mark size=2, mark options={solid}, dashdotdotted]
table {%
	0.01 0.0984597001325326
	0.005 0.0298953568129395
	0.0015 0.0216880957711176
	0.001 0.021234360178657
	0.0005 0.0213404385121496
};
\addlegendentry{Leap-frog m = 5}

\addplot [line width=2pt, TumBlue, mark=triangle*, mark size=2, mark options={solid}, dashdotted]
table {%
	0.05 1.03253484726925
	0.018 0.44428426558302
	0.01 0.190474962022801
	0.005 0.0567678622736857
	0.0015 0.021889498109562
	0.001 0.0218180056984379
	0.0005 0.021474377006394
};
\addlegendentry{Trapezoidal Newmark}
\addplot [line width=2pt, Orange, mark=10-pointed star, mark size=2, dotted]
table {%
	0.018 0.292065535125105
	0.01 0.0734138617456025
	0.005 0.025530743625991
	0.0015 0.0215582515621646
	0.001 0.0212544334748173
	0.0005 0.0213473443429327
};
\addlegendentry{Newmark IMEX}

\addplot [line width=1pt, gray, dashed]
table {%
	0.018140 0.0174814137894623
	0.018140 1.25387336731297
};
\addplot [line width=1pt, gray, dashed]
table {%
	0.0021282 0.0174814137894623
	0.0021282 1.25387336731297
};
\addplot [line width=1pt, gray, dashed]
table {%
	0.0014205 0.0174814137894623
	0.0014205 1.25387336731297
};

\addplot [line width=1pt, gray, dashed]
table {%
	0.0011027 0.0174814137894623
	0.0011027 1.25387336731297
};
\addplot [line width=1pt, gray, dashed]
table {%
	0.000397164117362141 0.02138
	0.0629462705897084 0.02138
};

\draw[{stealth[width=5mm]}-] (0.025,0.02138)   -- (0.03,0.025);
\node[align=left] at (0.040, 0.026) {$e_\text{L2}^\text{spatial}$};

\draw[{stealth[width=5mm]}-] (0.01840,0.07)   -- (0.023,0.07);
\node[align=left] at (0.033, 0.07) {$\Delta t_\text{crit}^\text{explicit}$};

\draw[{stealth[width=5mm]}-] (0.0021282,0.12)   -- (0.0028,0.12);
\node[align=left] at (0.0038, 0.12) {$\Delta t_\text{crit}^\text{global}$};

\draw[{stealth[width=5mm]}-] (0.0014205,0.4)   -- (0.0028,0.4);
\node[align=left] at (0.0046, 0.4) {$\Delta t_\text{crit, }^\text{HRZ, global}$};

\draw[{stealth[width=5mm]}-] (0.0011027,0.6)   -- (0.0028,0.6);
\node[align=left] at (0.0041, 0.6) {$\Delta t_\text{crit}^\text{cut, min}$};

\addplot [line width=2pt, Black, mark=*, mark size=1, mark options={solid}, dashed]
table {%
	0.0015 0.0216566371141543
	0.001 0.0212288033004938
	0.0005 0.0213390821676393
};

\end{axis}
\end{tikzpicture}

%% file: figures/RuntimeL2.tex
\begin{tikzpicture}[]

\definecolor{crimson2143940}{RGB}{214,39,40}
\definecolor{darkgray176}{RGB}{176,176,176}
\definecolor{darkorange25512714}{RGB}{255,127,14}
\definecolor{forestgreen4416044}{RGB}{44,160,44}
\definecolor{gray127}{RGB}{127,127,127}
\definecolor{lightgray204}{RGB}{204,204,204}
\definecolor{mediumpurple148103189}{RGB}{148,103,189}
\definecolor{orchid227119194}{RGB}{227,119,194}
\definecolor{sienna1408675}{RGB}{140,86,75}
\definecolor{steelblue31119180}{RGB}{31,119,180}
\definecolor{Black}{RGB}{0, 0, 0} 
\definecolor{TumBlue}{RGB}{0,101,189} 
\definecolor{Gray}{RGB}{153,153,153} 
\definecolor{TumColor2}{RGB}{0,82,147} 
\definecolor{TumColor2}{RGB}{152,198,234} 
\definecolor{Orange}{RGB}{227,114,34} 
\definecolor{Green}{RGB}{162,173,0} 
\definecolor{TumColor2}{RGB}{0,101,189}

\pgfplotsset{every tick label/.append style={font=\footnotesize}}
\pgfplotsset{every label/.append style={font=\footnotesize}}

\begin{axis}[
legend cell align={left},
width=0.75\textwidth,
height=0.5\textwidth,
legend style={
  fill opacity=0.8,
  draw opacity=1,
  text opacity=1,
  at={(0.97,0.03)},
  anchor=south east,
  draw=lightgray204
},
log basis x={10},
log basis y={10},
tick align=outside,
legend style={font=\footnotesize},
tick pos=left,
x grid style={darkgray176},
xlabel={Runtime $[\si{\second}]$},
xmajorgrids,
legend pos= outer north east,
xmin=0.698230819818063, xmax=347.860070133297,
xminorgrids,
xmode=log,
xtick style={color=black},
xtick={0.01,0.1,1,10,100,1000,10000},
xticklabels={
  \(\displaystyle {10^{-2}}\),
  \(\displaystyle {10^{-1}}\),
  \(\displaystyle {10^{0}}\),
  \(\displaystyle {10^{1}}\),
  \(\displaystyle {10^{2}}\),
  \(\displaystyle {10^{3}}\),
  \(\displaystyle {10^{4}}\)
},
y grid style={darkgray176},
ylabel={$e_\text{L2}$},
ymajorgrids,
ymin=0.0174814137894623, ymax=1.25387336731297,
yminorgrids,
ymode=log,
ytick style={color=black},
ytick={0.001,0.01,0.1,1,10,100},
yticklabels={
  \(\displaystyle {10^{-3}}\),
  \(\displaystyle {10^{-2}}\),
  \(\displaystyle {10^{-1}}\),
  \(\displaystyle {10^{0}}\),
  \(\displaystyle {10^{1}}\),
  \(\displaystyle {10^{2}}\)
}
]

\addplot [line width=2pt, Black, mark=*, mark size=2, mark options={solid}, dashed]
table {%
	8.29394054412842 0.0216566371141543
	13.553920507431 0.0212288033004938
	25.1387796401978 0.0213390821676393
};
\addlegendentry{CDM}
\addplot [line width=2pt, Green, mark=diamond*, mark size=2, mark options={solid}, dashed]
table {%
10.4988887310028 0.230596450456585
16.3078768253326 0.23067392951055
};
\addlegendentry{CDM, HRZ}
\addplot [line width=2pt, Gray, mark=star, mark size=2, mark options={solid}, dashdotdotted]
table {%
6.92195343971252 0.374835298223246
12.6109111309052 0.0984745067732454
26.0707900524139 0.0299041988804622
82.4153189659119 0.0216893334044662
148.949785709381 0.0212345806963219
262.29815530777 0.0213404907028305
};
\addlegendentry{Leap-frog m = 17}
\addplot [line width=2pt, Gray, mark=o, mark size=2, mark options={solid}, dashdotdotted]
table {%
4.31896638870239 0.0984597001325326
8.53093075752258 0.0298953568129395
31.6661343574524 0.0216880957711176
43.1707127094269 0.021234360178657
88.1124360561371 0.0213404385121496
};
\addlegendentry{Leap-frog m = 5}

\addplot [line width=2pt, TumBlue, mark=triangle*, mark size=2, mark options={solid}, dashdotted]
table {%
1.60199284553528 1.03253484726925
4.01198244094849 0.44428426558302
7.38696956634521 0.190474962022801
15.0709342956543 0.0567678622736857
51.746776342392 0.021889498109562
78.5276608467102 0.0218180056984379
156.95432472229 0.021474377006394
};
\addlegendentry{Trapezoidal Newmark}
\addplot [line width=2pt, Orange, mark=10-pointed star, mark size=2, dotted]
table {%
0.925994396209717 0.292065535125105
1.54899501800537 0.0734138617456025
3.37998533248901 0.025530743625991
11.7889494895935 0.0215582515621646
16.0259330272675 0.0212544334748173
30.4628674983978 0.0213473443429327
};
\addlegendentry{Newmark IMEX}

\addplot [line width=1pt, gray, dashed]
table {%
	0.698230819818063 0.02138
	347.860070133297 0.02138
};

\draw[{stealth[width=5mm]}-] (0.9,0.02138)   -- (1.1,0.025);
\node[align=left] at (1.6, 0.026) {$e_\text{L2}^\text{spatial}$};
\end{axis}

\end{tikzpicture}

%% file: content/example3D.tex
\label{subsec:3D}

\begin{figure}[t!]
	\centering
	\begin{subfigure}[t]{0.7\textwidth}
	    \centering
    	\includegraphics[width=\textwidth]{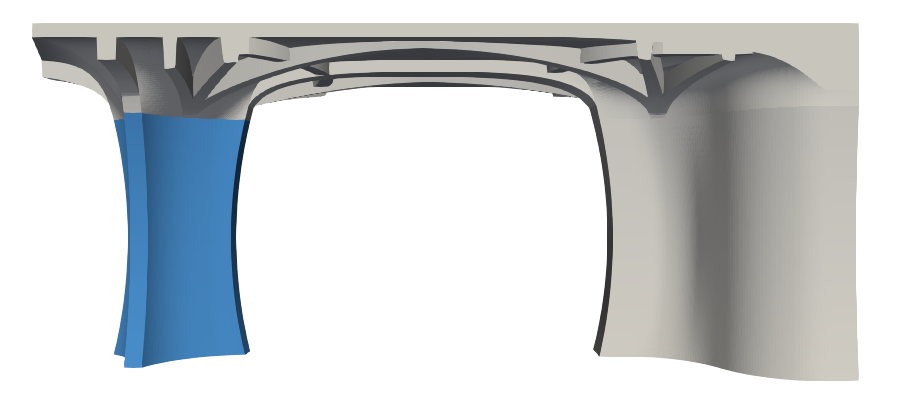}
    	\caption{Whole structure}
	\end{subfigure}
	\hfill
	\begin{subfigure}[t]{0.25\textwidth}
	    \centering
		\input{tikzpictures/pillarWithSourceAndReceivers.tex}
    	\caption{Pillar with source and receiver positions}
	\label{fig:3Dstructure_sr}
	\end{subfigure}
	\caption{3D example~\cite{kloft2023, BKK23}}
	\label{fig:3Dstructure}
\end{figure}

\begin{table}[b!]
	\centering
	\caption{Time step sizes and computation times}
	\renewcommand{\arraystretch}{1.5}
	\begin{tabular}{|l|c|c|c|}
		\hline
		& time step size $\Delta t$ & critical time step $\Delta t_\text{crit}$ & computation time \\
		\hline
        \hline
        CDM, consistent SCM & $\SI{2e-7}{\second}$ & $\SI{2.0001e-7}{\second}$ & $\SI{510.99}{\second}$ \\
        \hline
        CDM, HRZ-lumped SCM & $\SI{2e-7}{\second}$ & $\SI{2.0015e-7}{\second}$ & $\SI{326.17}{\second}$ \\
        \hline
        Trapezoidal Newmark SCM & $\SI{1e-6}{\second}$ & $\infty$ & $\SI{278.4}{\second}$ \\
        \hline 
        Newmark IMEX SCM& $\SI{1e-6}{\second}$ & $\SI{1.8551e-6}{\second}$ & $\SI{116.78}{\second}$ \\
		\hline
	\end{tabular}
	\label{Tab:times3D}
\end{table}

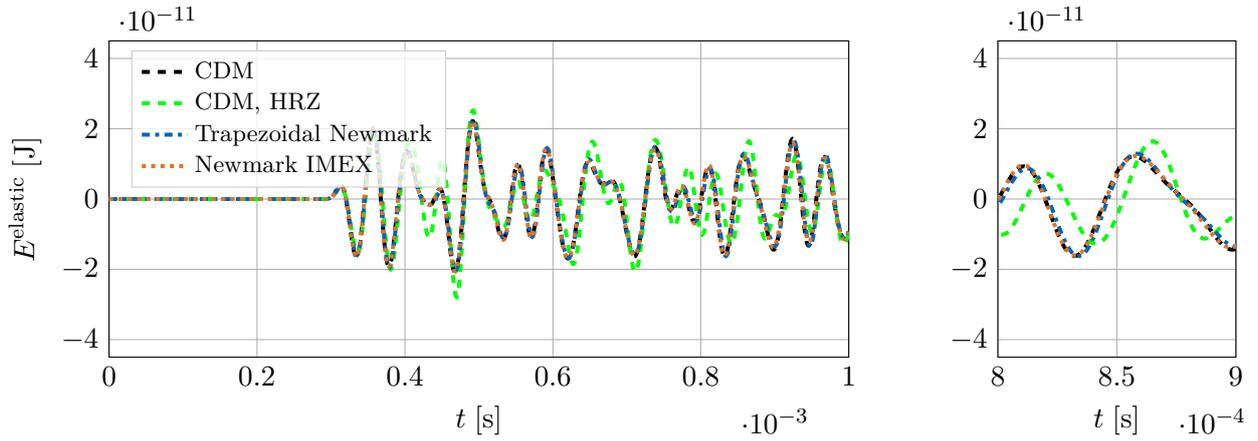
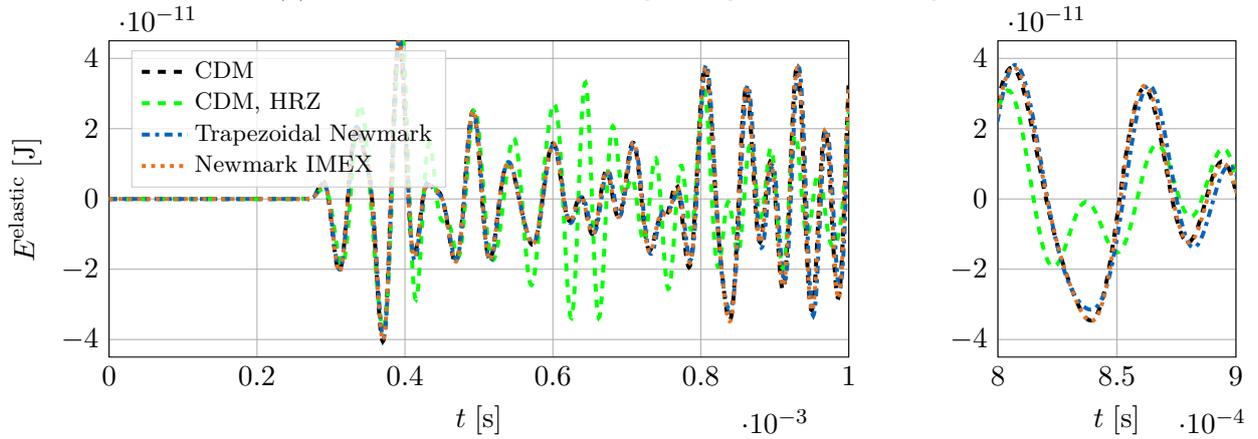
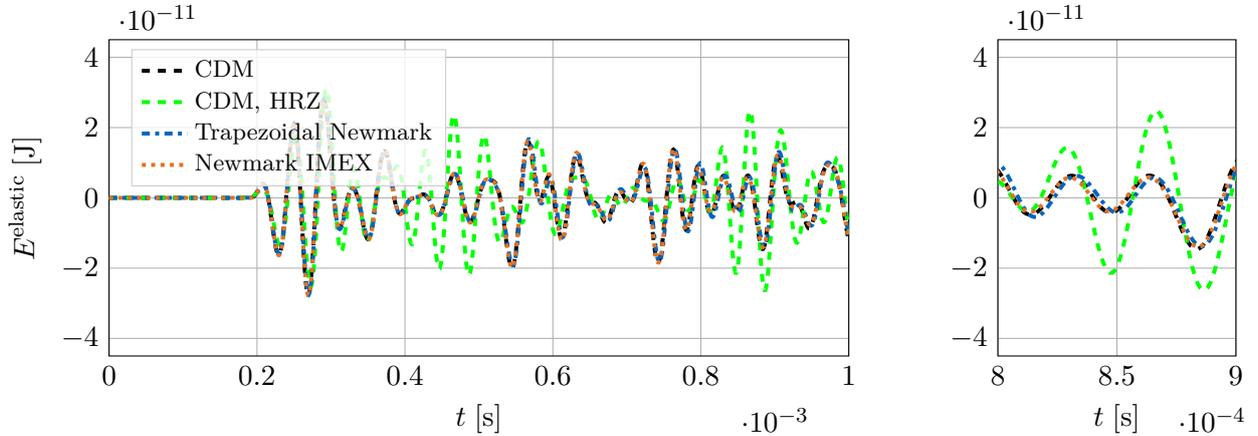
\begin{figure}[t!]
	\centering
	\begin{subfigure}[b]{1.0\textwidth}
		\centering
	    \input{tikzpictures/signalR0_whole} 
		\hfill
	    \input{tikzpictures/signalR0_part} 
		\caption{Receiver $R0$ -- left: whole signal; right: zoomed-in signal}
	\end{subfigure}
	\begin{subfigure}[b]{1.0\textwidth}
		\centering
	    \input{tikzpictures/signalR1_whole} 
		\hfill
	    \input{tikzpictures/signalR1_part} 
		\caption{Receiver $R1$ -- left: whole signal; right: zoomed-in signal}
	\end{subfigure}
	\begin{subfigure}[b]{1.0\textwidth}
		\centering
	    \input{tikzpictures/signalR2_whole} 
		\hfill
	    \input{tikzpictures/signalR2_part} 
		\caption{Receiver $R2$ -- left: whole signal; right: zoomed-in signal}
	\end{subfigure}
	\caption{Solutions of the 3D example: wave signals at the three receiver positions}
	\label{fig:solutions3D}
\end{figure}

Efficient methods for simulating wave propagation are essential for solving inverse problems in non-destructive testing, where many objective function and gradient evaluations with forward model evaluations are needed. We now analyze the geometry of the three-dimensional full waveform inversion performed in~\cite{BKK23}. \figref{fig:3Dstructure} illustrates the geometry and highlights the left pillar in blue that we extract and simulate in our performance analysis. The geometry is defined by a surface triangulation in the STL format. We define homogeneous Neumann (i.e., perfectly reflecting) boundary conditions on the entire surface. The left pillar of the model is embedded within an extended domain of size $\SI{2}{\meter} \times \SI{1.25}{\meter} \times \SI{1.25}{\meter}$. The material is concrete with a density of $\rho = \SI{2400}{\kg\per\meter\cubed}$ and a wave speed of $c = \SI{3000}{\meter\per\second}$. The density is scaled by $\alpha_f = 10^{-5}$ inside the fictitious domain. We position a source at $S0 = (\SI{1,04066}{\meter}, \SI{0.542926}{\meter}, \SI{1.0}{\meter})$ and three receivers at $R0 = (\SI{0.27966}{\meter}, \SI{0.39788}{\meter}, \SI{0.57757}{\meter})$, $R1 = (\SI{0.34066}{\meter}, \SI{0.18043}{\meter}, \SI{1.0}{\meter})$, and $R2 = (\SI{0.78096}{\meter}, \SI{0.47348}{\meter}, \SI{1.4967}{\meter})$. \figref{fig:3Dstructure_sr} marks the source with a green cross and the receivers with red circles, where $R0$ is the lowest receiver, $R1$ is the middle receiver, and $R2$ is the highest receiver. The structure is excited at $S0$ with a two-cycle sine burst with a central frequency of $f_\text{s} = \SI{20}{\kilo\hertz}$
\begin{equation}
    f_\text{t}(t) = \begin{cases}
        \sin{(2 \pi f_\text{s} t)} \sin^2{\left(\frac{\pi f_\text{s} t}{n_\text{c}}\right)} &\qquad \text{, } t \leq \frac{n_\text{c}}{f_\text{s}} \\
        0 &\qquad \text{, else}
    \end{cases},
\end{equation}
where $n_\text{c} = 2$ denotes the number of cycles. Beyond $2 \times f_\text{s}$, the frequency content of the two-cycle sine burst decreases to almost zero. Thus, $2 \times f_\text{s}$ is considered as the maximum frequency of interest. We use $u, \dot{u} = 0$ as initial condition and simulate until $T = \SI{e-3}{\second}$. 

Following \secref{sec:theory}, we discretize the computational domain using $20 \times 20 \times 32$ cells of polynomial order $p = 4$ in the $x$-, $y$-, and $z$-directions. This mesh resolves the central wavelength with $2.4$ cells and the maximum wavelength with $1.2$ cells. We integrate cut cells on a quadtree with five levels of refinement towards the domain boundary. The SCM discretization of the wave field results in $n^\text{dof} = 325\,573$ basis functions. Given the thin structure of the pillar, only $n^\text{d} = 158\,053$ basis functions are supported exclusively by uncut cells, while $n^\text{c} = 167\,520$ basis functions are supported by at least one cut cell. We compare the performance of the explicit CDM, the trapezoidal Newmark method (treating all dofs implicitly), and our Newmark IMEX method, all three using the consistent SCM and the CDM using HRZ-lumped SCM. \tabref{Tab:times3D} shows the selected time step sizes, the critical time step sizes, and the computational times for the different time integration methods. The runtimes are measured from single-threaded computations on an Intel i9-9900K processor and include only the time integration with an initial matrix factorization, if necessary. The matrix factorization and triangular solution at each time step use \texttt{pardiso\_64} from Intel's oneMKL library. The relatively coarse time step for the Newmark IMEX and trapezoidal Newmark methods leads to $50$ time steps per period of the central frequency and $25$ time steps per period of the maximum frequency. The CDM methods with consistent SCM and HRZ-lumped SCM use a smaller time step size only slightly below their critical time step, resulting in $250$ time steps per period of the central frequency  and $125$ time steps per period of the maximum frequency.

\figref{fig:solutions3D} compares the time signals at the receiver positions. The solutions of the CDM, the Newmark IMEX method, and the trapezoidal Newmark method are almost identical. The plots on the right side of \figref{fig:solutions3D} magnify a time interval towards the end of the simulation. We use the CDM as a reference since it uses more time steps and is the most accurate.
While the Newmark IMEX looks identical to the CDM, the trapezoidal Newmark method shows minor deviations in amplitude and phase. The much higher number of time steps makes the CDM the most expensive method. In contrast, Newmark IMEX is the fastest of all methods despite the unfavorable ratio between uncut and cut dofs. The trapezoidal Newmark method requires more than twice the computational time of Newmark IMEX. When applying HRZ lumping to the cut cells, the accuracy of the simulation deteriorates significantly. The waves are similar at first, but the difference increases after several reflections because of the error introduced by the HRZ lumping of cells intersecting the boundary. Despite the fully diagonal mass matrix, the computation time is still almost three times higher than that of Newmark IMEX because it requires significantly more time steps to maintain stability.

%% file: tikzpictures/pillarWithSourceAndReceivers.tex
\begin{tikzpicture}
	\centering
	\draw (0.0\textwidth, 0.0\textwidth) node[inner sep=0] {\includegraphics[width=0.65\textwidth]{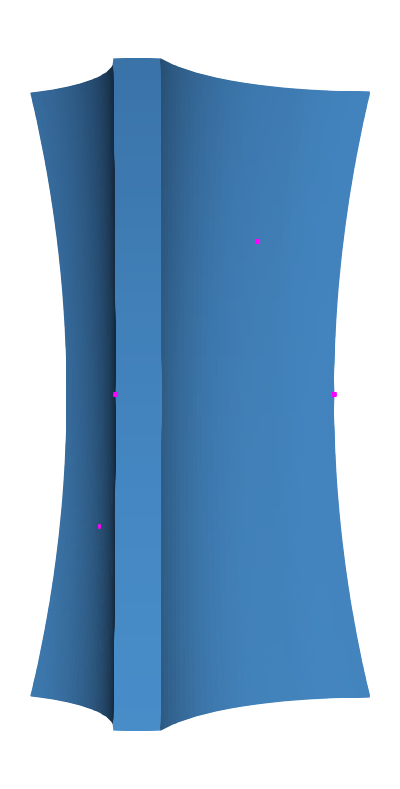}};
	\draw (0.22\textwidth,0.0\textwidth) node[cross=5pt, green, line width=0.3mm]{};
	\draw[red, fill=red] (-0.135\textwidth,0.0\textwidth) circle (2.5pt);
	\draw[red, fill=red] (-0.162\textwidth,-0.218\textwidth) circle (2.5pt);
	\draw[red, fill=red] (0.095\textwidth,0.245\textwidth) circle (2.5pt);
\end{tikzpicture}

%% file: tikzpictures/signalR0_whole.tex
\definecolor{TumBlue}{RGB}{0,101,189} 
\definecolor{Orange}{RGB}{227,114,34} 
\definecolor{lightgray204}{RGB}{204,204,204}
\definecolor{darkgray176}{RGB}{176,176,176}
\begin{tikzpicture}
	\begin{axis}[
		xmin = 0, xmax = 0.001,
		ymin = -4.5e-11, ymax = 4.5e-11,
        xtick style={color=black},
		xtick={0, 0.0002, 0.0004, 0.0006, 0.0008, 0.001},
		ytick={-4e-11, -2e-11, 0, 2e-11,4e-11},
        x grid style={darkgray176},
        y grid style={darkgray176},
        xmajorgrids,
        ymajorgrids,
        xtick style={draw=none},
        ytick style={draw=none},
		width = 0.68571\textwidth,
		height = 0.35\textwidth,
		xlabel = {$t \; [\si{\second}]$},
		ylabel style={align=center}, 
		ylabel={$E^\text{elastic} \; [\si{\joule}]$},
        legend cell align={left},
		legend style={
          fill opacity=0.8,
          draw opacity=1,
          text opacity=1,
          at={(0.97,0.03)},
          anchor=south east,
          draw=lightgray204
        },
		legend pos = north west,
        legend style={font=\footnotesize}
		]
		
		\addplot[black, line width = 1.5pt, dashed] file[] {tikzpictures/data/CDM_P1.dat};
		\addplot[line width=1.5pt, green, dashed] file[] {tikzpictures/data/CDM_HRZ_P1.dat};
		\addplot[TumBlue, line width = 1.5pt, dashdotted] file[] {tikzpictures/data/Trapezoidal_P1.dat};
		\addplot[Orange, line width = 1.5pt, dotted] file[] {tikzpictures/data/IMEX_P1.dat};
		
		\legend{CDM, {CDM, HRZ}, Trapezoidal Newmark, Newmark IMEX},
	\end{axis}
\end{tikzpicture}%

%% file: tikzpictures/signalR0_part.tex
\definecolor{TumBlue}{RGB}{0,101,189} 
\definecolor{Orange}{RGB}{227,114,34} 
\definecolor{lightgray204}{RGB}{204,204,204}
\definecolor{darkgray176}{RGB}{176,176,176}
\begin{tikzpicture}
	\begin{axis}[
		xmin = 0.0008, xmax = 0.0009,
		ymin = -4.5e-11, ymax = 4.5e-11,
        xtick style={color=black},
		xtick={0.0008, 0.00085, 0.0009},
		ytick={-4e-11, -2e-11, 0, 2e-11,4e-11},
        x grid style={darkgray176},
        y grid style={darkgray176},
        xmajorgrids,
        ymajorgrids,
        xtick style={draw=none},
        ytick style={draw=none},
		width = 0.28571\textwidth,
		height = 0.35\textwidth,
		xlabel = {$t \; [\si{\second}]$},
		ylabel style={align=center}, 
        legend cell align={left},
		legend style={
          fill opacity=0.8,
          draw opacity=1,
          text opacity=1,
          at={(0.97,0.03)},
          anchor=south east,
          draw=lightgray204
        },
		legend pos = north west,
        legend style={font=\footnotesize}
		]
		
		\addplot[black, line width = 1.5pt, dashed] file[] {tikzpictures/data/CDM_P1.dat};
		\addplot[line width=1.5pt, green, dashed] file[] {tikzpictures/data/CDM_HRZ_P1.dat};
		\addplot[TumBlue, line width = 1.5pt, dashdotted] file[] {tikzpictures/data/Trapezoidal_P1.dat};
		\addplot[Orange, line width = 1.5pt, dotted] file[] {tikzpictures/data/IMEX_P1.dat};
		
	\end{axis}
\end{tikzpicture}%

%% file: tikzpictures/signalR1_whole.tex
\definecolor{TumBlue}{RGB}{0,101,189} 
\definecolor{Orange}{RGB}{227,114,34} 
\definecolor{lightgray204}{RGB}{204,204,204}
\definecolor{darkgray176}{RGB}{176,176,176}
\begin{tikzpicture}
	\begin{axis}[
		xmin = 0, xmax = 0.001,
		ymin = -4.5e-11, ymax = 4.5e-11,
        xtick style={color=black},
		xtick={0, 0.0002, 0.0004, 0.0006, 0.0008, 0.001},
		ytick={-4e-11, -2e-11, 0, 2e-11,4e-11},
        x grid style={darkgray176},
        y grid style={darkgray176},
        xmajorgrids,
        ymajorgrids,
        xtick style={draw=none},
        ytick style={draw=none},
		width = 0.68571\textwidth,
		height = 0.35\textwidth,
		xlabel = {$t \; [\si{\second}]$},
		ylabel style={align=center}, 
		ylabel={$E^\text{elastic} \; [\si{\joule}]$},
        legend cell align={left},
		legend style={
          fill opacity=0.8,
          draw opacity=1,
          text opacity=1,
          at={(0.97,0.03)},
          anchor=south east,
          draw=lightgray204
        },
		legend pos = north west,
        legend style={font=\footnotesize}
		]
		
		\addplot[black, line width = 1.5pt, dashed] file[] {tikzpictures/data/CDM_P2.dat};
		\addplot[line width=1.5pt, green, dashed] file[] {tikzpictures/data/CDM_HRZ_P2.dat};
		\addplot[TumBlue, line width = 1.5pt, dashdotted] file[] {tikzpictures/data/Trapezoidal_P2.dat};
		\addplot[Orange, line width = 1.5pt, dotted] file[] {tikzpictures/data/IMEX_P2.dat};
		
		\legend{CDM, {CDM, HRZ}, Trapezoidal Newmark, Newmark IMEX},
	\end{axis}
\end{tikzpicture}%

%% file: tikzpictures/signalR1_part.tex
\definecolor{TumBlue}{RGB}{0,101,189} 
\definecolor{Orange}{RGB}{227,114,34} 
\definecolor{lightgray204}{RGB}{204,204,204}
\definecolor{darkgray176}{RGB}{176,176,176}
\begin{tikzpicture}
	\begin{axis}[
		xmin = 0.0008, xmax = 0.0009,
		ymin = -4.5e-11, ymax = 4.5e-11,
        xtick style={color=black},
		xtick={0.0008, 0.00085, 0.0009},
		ytick={-4e-11, -2e-11, 0, 2e-11,4e-11},
        x grid style={darkgray176},
        y grid style={darkgray176},
        xmajorgrids,
        ymajorgrids,
        xtick style={draw=none},
        ytick style={draw=none},
		width = 0.28571\textwidth,
		height = 0.35\textwidth,
		xlabel = {$t \; [\si{\second}]$},
		ylabel style={align=center}, 
        legend cell align={left},
		legend style={
          fill opacity=0.8,
          draw opacity=1,
          text opacity=1,
          at={(0.97,0.03)},
          anchor=south east,
          draw=lightgray204
        },
		legend pos = north west,
        legend style={font=\footnotesize}
		]
		
		\addplot[black, line width = 1.5pt, dashed] file[] {tikzpictures/data/CDM_P2.dat};
		\addplot[line width=1.5pt, green, dashed] file[] {tikzpictures/data/CDM_HRZ_P2.dat};
		\addplot[TumBlue, line width = 1.5pt, dashdotted] file[] {tikzpictures/data/Trapezoidal_P2.dat};
		\addplot[Orange, line width = 1.5pt, dotted] file[] {tikzpictures/data/IMEX_P2.dat};
		
	\end{axis}
\end{tikzpicture}%

%% file: tikzpictures/signalR2_whole.tex
\definecolor{TumBlue}{RGB}{0,101,189} 
\definecolor{Orange}{RGB}{227,114,34} 
\definecolor{lightgray204}{RGB}{204,204,204}
\definecolor{darkgray176}{RGB}{176,176,176}
\begin{tikzpicture}
	\begin{axis}[
		xmin = 0, xmax = 0.001,
		ymin = -4.5e-11, ymax = 4.5e-11,
        xtick style={color=black},
		xtick={0, 0.0002, 0.0004, 0.0006, 0.0008, 0.001},
		ytick={-4e-11, -2e-11, 0, 2e-11,4e-11},
        x grid style={darkgray176},
        y grid style={darkgray176},
        xmajorgrids,
        ymajorgrids,
        xtick style={draw=none},
        ytick style={draw=none},
		width = 0.68571\textwidth,
		height = 0.35\textwidth,
		xlabel = {$t \; [\si{\second}]$},
		ylabel style={align=center}, 
		ylabel={$E^\text{elastic} \; [\si{\joule}]$},
        legend cell align={left},
		legend style={
          fill opacity=0.8,
          draw opacity=1,
          text opacity=1,
          at={(0.97,0.03)},
          anchor=south east,
          draw=lightgray204
        },
		legend pos = north west,
        legend style={font=\footnotesize}
		]
		
		\addplot[black, line width = 1.5pt, dashed] file[] {tikzpictures/data/CDM_P3.dat};
		\addplot[line width=1.5pt, green, dashed] file[] {tikzpictures/data/CDM_HRZ_P3.dat};
		\addplot[TumBlue, line width = 1.5pt, dashdotted] file[] {tikzpictures/data/Trapezoidal_P3.dat};
		\addplot[Orange, line width = 1.5pt, dotted] file[] {tikzpictures/data/IMEX_P3.dat};
		
		\legend{CDM, {CDM, HRZ}, Trapezoidal Newmark, Newmark IMEX},
	\end{axis}
\end{tikzpicture}%

%% file: tikzpictures/signalR2_part.tex
\definecolor{TumBlue}{RGB}{0,101,189} 
\definecolor{Orange}{RGB}{227,114,34} 
\definecolor{lightgray204}{RGB}{204,204,204}
\definecolor{darkgray176}{RGB}{176,176,176}
\begin{tikzpicture}
	\begin{axis}[
		xmin = 0.0008, xmax = 0.0009,
		ymin = -4.5e-11, ymax = 4.5e-11,
        xtick style={color=black},
		xtick={0.0008, 0.00085, 0.0009},
		ytick={-4e-11, -2e-11, 0, 2e-11,4e-11},
        x grid style={darkgray176},
        y grid style={darkgray176},
        xmajorgrids,
        ymajorgrids,
        xtick style={draw=none},
        ytick style={draw=none},
		width = 0.28571\textwidth,
		height = 0.35\textwidth,
		xlabel = {$t \; [\si{\second}]$},
		ylabel style={align=center}, 
        legend cell align={left},
		legend style={
          fill opacity=0.8,
          draw opacity=1,
          text opacity=1,
          at={(0.97,0.03)},
          anchor=south east,
          draw=lightgray204
        },
		legend pos = north west,
        legend style={font=\footnotesize}
		]
		
		\addplot[black, line width = 1.5pt, dashed] file[] {tikzpictures/data/CDM_P3.dat};
		\addplot[line width=1.5pt, green, dashed] file[] {tikzpictures/data/CDM_HRZ_P3.dat};
		\addplot[TumBlue, line width = 1.5pt, dashdotted] file[] {tikzpictures/data/Trapezoidal_P3.dat};
		\addplot[Orange, line width = 1.5pt, dotted] file[] {tikzpictures/data/IMEX_P3.dat};
		
	\end{axis}
\end{tikzpicture}%

%% file: content/conclusion.tex
We present an implicit-explicit~(IMEX) time integration approach for the immersed scalar wave equation. The spatial discretization by the spectral cell method incorporates complex geometries by an extended domain. In the presented hybrid time integration, called the immersed Newmark IMEX method, we split the system obtained from the spatial discretization into two parts: one corresponding to the basis functions with support on at least one cell intersecting the domain boundary and another consisting of the remaining dofs whose basis function are supported only by uncut cells. The cut dofs are integrated in time with the implicit trapezoidal Newmark method to gain independence on their restrictive critical time step size. The uncut dofs are integrated with explicit second-order central differences. We use nodal lumping to obtain a diagonal mass matrix of the explicit subsystem.

We demonstrate that immersed Newmark IMEX eliminates the severe time step size restrictions imposed by the cut cells by determining the highest eigenfrequency of a cut cell when varying the cut ratios.
The analysis of a simplified system of ten spring-coupled masses shows that Newmark IMEX also preserves second-order convergence. Its stability depends on the explicit subsystem of the uncut dofs. Our two- and three-dimensional examples with significant geometric complexity confirm these results. Immersed Newmark IMEX outperforms other methods significantly when comparing runtime per accuracy since it benefits from the efficiency of the spectral element method  for uncut cells. The direct linear solver for the explicit subsystem costs less than using significantly more time steps. Even the leap-frog algorithm that limits the additional time steps to the explicit subsystem is more expensive per accuracy and time step size. Newmark IMEX performs best even in the unfavorable case of many cut cells, as our three-dimensional example shows. It is four times faster than the explicitly integrated spectral cell method with a consistent mass matrix and more than two times faster with a higher accuracy than the trapezoidal Newmark method. 
We also observe that an SCM with an HRZ lumping of cut cells significantly reduces the accuracy while suffering from the same critical time step constraints.

While immersed Newmark IMEX appears viable for a broad range of problems, limitations exist. The implicit integration of cut cells requires solving a system of equations at each time step that we accelerate by computing a factorization in advance. 
On the one hand, the memory consumption of direct solvers may limit the number of cut cells for computations on a single machine. The explicit schemes we consider do not store an inverse and can even be implemented matrix-free. On the other hand, larger problems may require a parallel computation on a distributed memory system to reduce the runtime. The involvement of a direct solver significantly complicates the necessary implementations, while methods like the HRZ-lumped spectral cell method are simple to parallelize. Finding ways to upscale the immersed Newmark IMEX method towards massively parallel computations remains a challenge for future investigation.